\documentclass[a4paper, 10pt]{article}
\usepackage{amssymb} % that is for real numbers etc.
\usepackage{mathtools} % replaces amsmath
\usepackage{amsthm} % enables to number all with 1 single counter
\usepackage[pdftex]{graphicx}
\usepackage[pdftex]{color}
\usepackage{natbib}
\usepackage{url}
\usepackage{subcaption}
\usepackage{multirow}
\usepackage[margin=1cm]{caption}
\usepackage{dsfont}
\usepackage{pgfplots}
\usepackage{tikz}
\usepackage{filecontents}
\usepackage[shortlabels]{enumitem}
\usepackage{geometry}
\usepackage[titletoc,title]{appendix}
\usepackage{float}
\usepackage{comment}

\newcount\fromtop
\newfont{\smcal}{cmu10 scaled 1200}
\newfont{\handw}{cmmi10 scaled 1200}
\newfont{\handws}{cmmi10 scaled 800}
\newtheorem{Prop}{Proposition}[section]
\newtheorem{Lem}[Prop]{Lemma}
\newtheorem{As}[Prop]{Assumption}
\newtheorem{Ag}[Prop]{Agreement}

\newtheorem{Th}[Prop]{Theorem}
\newtheorem{Rm}[Prop]{Remark}
\newtheorem{Def}[Prop]{Definition}

\newtheorem{Test}[Prop]{Test}
\newtheorem{Tests}[Prop]{Tests}
\newtheorem{Cor}[Prop]{Corollary}

\newtheorem{Ex}[Prop]{Example}

\newcommand{\grad}{\mbox{\rm grad}}
\newcommand{\cov}{\mbox{\rm Cov}}

\newcommand{\var}{\mbox{\rm Var}}

\newcommand{\cD}{{\mathcal{D}}}

\newcommand{\cN}{{\mathcal{N}}}

\newcommand{\EE}{\mathbb E}
\newcommand{\RR}{\mathbb R}
\newcommand{\NN}{\mathbb N}

\newcommand{\BL}{{\textup{BL}_1(\mathbb{R})}}
\newcommand{\lrBrack}[1]{\left[#1\right]}
\newcommand{\lrCurlyBrack}[1]{\left(#1\right)}

\newcommand{\SSS}{{{\mathbb S}^1}}
\newcommand{\TTT}{{{\mathbb T}^m}}

\newcommand{\Prb}{\mathbb P}

\DeclareMathOperator*{\argmin}{\mbox{\rm argmin}}

\newcommand{\iid}{\operatorname{\stackrel{i.i.d.}{\sim}}}

\newcommand{\inD}{\operatorname{\stackrel{\cD}{\to}}}
\newcommand{\eps}{\varepsilon}

\DeclareMathOperator{\sign}{sign}

\newcommand{\fm}{\mathfrak{m}}
\newcommand{\fmnhat}{\widehat{\mathfrak{m}}_n^*}

\newcommand{\konvPStar}{\xrightarrow{\;\;\mathbb{P}^*\;}}

\newcommand\commentout[1]{}

\makeatletter
\renewcommand{\paragraph}{%
  \@startsection{paragraph}{4}%
  {\z@}{3.25ex \@plus 1ex \@minus .2ex}{-0.5em}%
  {\normalfont\normalsize\bfseries}%
}
\makeatother

\begin{document}
\title{Finite Sample Smeariness of Fr\'echet Means \\ and Application to Climate}
\author{Shayan Hundrieser\footnote{
Institute for Mathematical Stochastics,
% Felix-Bernstein-Institut f\"ur Mathematische Statistik in den Biowissenschaften, 
Georg-August-Universit\"at G\"ottingen
}\,\,\footnote{Cluster of Excellence "Multiscale Bioimaging: from Molecular Machines to Networks of Excitable Cells" (MBExC), University of G\"ottingen, Germany},\,  Benjamin Eltzner\footnote{Max Planck Institute for Biophysical Chemistry, G\"ottingen, Germany}\,\,
and Stephan F. Huckemann$^*$} 
\date{\today}
\maketitle

\begin{abstract} 

Fr\'echet means on non-Euclidean spaces may exhibit nonstandard asymptotic rates rendering quantile-based asymptotic inference inapplicable. We show here that this affects, among others, all circular distributions whose support exceeds a half circle. We exhaustively describe this phenomenon and introduce a new concept which we call finite samples smeariness (FSS). In the presence of FSS, it turns out that quantile-based tests for equality of Fr\'echet means 
systematically feature effective levels higher than their nominal level which perseveres asymptotically in case of Type I FSS. In contrast, suitable bootstrap-based tests correct for FSS and asymptotically attain the correct level. 
For illustration of the relevance of FSS in real data, we apply our method to directional wind data from two European cities. It turns out that quantile based tests, not correcting for FSS, find a multitude of significant wind changes. This multitude condenses to a few years featuring significant wind changes, when our bootstrap tests are applied, correcting for FSS.

\end{abstract}

\par
\vspace{9pt}

\par
\vspace{9pt}Figure 
\noindent {\it AMS 2000 Subject Classification:} \begin{minipage}[t]{6cm}
Primary 62 H 11\\ Secondary 60 F 05, 62 G 10 
 \end{minipage}
\par

\section{Introduction}\label{scn:intro}

Considering random variables $X_1,\ldots,X_n \iid X$ on a metric space $(M,d)$, generalizing the concept of the \emph{expected value}, \cite{F48} proposed to consider minimizers of the expected squared distance:
\begin{eqnarray}\label{eq:population-mean}  \argmin_{p\in M}F(p) &\mbox{ where}& F(p) =\EE\left[d(p,X)^2\right]\,.\end{eqnarray}
 Under mild assumptions, \cite{Z77} derived strong set-wise asymptotic consistency for
\begin{eqnarray}\label{eq:sample-mean}  \argmin_{p\in M}F_n(p) &\mbox{ where}& F_n(p) =\frac{1}{n}d(p,X_j)^2\end{eqnarray}
Under stronger conditions, among others that $M$ is a finite dimensional manifold, and under uniqueness of the minimizer $\mu$ of (\ref{eq:population-mean}), called a \emph{Fr\'echet population mean}, \cite{BP05} derived, in a local chart $\phi:U \to \RR^m$ near $\mu \in U \subset M$, a central limit theorem for a measurable selection $\widehat{\mu}_n$ of (\ref{eq:sample-mean}),  called a \emph{Fr\'echet sample mean}, with a Gaussian limiting distribution and the usual rate of $n^{-1/2}$:
\begin{eqnarray}\label{eq:BP-CLT}
 \sqrt{n}\left(\phi(\widehat{\mu}_n) - \phi(\mu)\right) \inD \cN(0,\Sigma)\,.
\end{eqnarray}
While the covariance matrix $\Sigma$ above reflects the asymptotic rescaled covariance of $\phi(\widehat{\mu}_n)$, it is usually approximated by the covariance matrix $\widehat{\Sigma}_n$ of the sample $\phi(X_1),\ldots, \phi(X_n)$, giving rise to the quantile-based one- and two-sample tests proposed by \cite{BP05}, see also e.g. \cite{BL17}. These tests rest on the approximation 
\begin{eqnarray}\label{eq:BP-approx}
\left(\phi(\widehat{\mu}_n) - \phi(\mu)\right)^T \widehat{\Sigma}_n^{-1}\left(\phi(\widehat{\mu}_n) - \phi(\mu)\right) &\inD& \chi^2_m\,.
\end{eqnarray}
In order to assess the validity of this approximation, we consider the suitably rescaled quotient of Fr\'echet variances
\begin{eqnarray}\label{eq:modulation}
\fm_n := \frac{n\EE[d(\mu,\widehat{\mu}_n)^2]}{\EE[d(\mu,X)^2]}\,,
\end{eqnarray}
which has been called by \cite{Pennec19} the \emph{modulation of the rate of convergence of the variance} for sample size $n$, which we will abbreviate as \emph{variance modulation}. The following phenomena have been observed in the literature:
\begin{itemize}
 \item[(A)] $\fm_n = 1$ for all $n \in \NN$,
 \item[(B)] $\fm_n=0$ for all $n>N$ where $N$ is a suitable random positive integer (stickiness),
 \item[(C)] $\lim_{n\to \infty}\fm_n = \infty$ (smeariness).
 \end{itemize}
Phenomenon (A) is the case on Euclidean spaces $(M,d)$ whenever second moments are finite. As we will show here, it is also the case if $(M,d)$ is a flat torus with sufficiently concentrated $X$. For some nontrivial random variables on non-manifolds, using not a local chart but a suitable embedding, Phenomenon (B) has been observed by \cite{BLO13,BLO18} on the BHV spaces of \cite{BHV01} for phylogenetic trees and it has been observed on related spaces by \cite{HHMMN13,H_Mattingly_Miller_Nolen2015}. Furthermore, Phenomenon (C) has been observed on the circle by \cite{HH15} with $\fm_n$ of rate $n^{\frac{\gamma}{\gamma+1}}$ with arbitrary $1\leq \gamma \in \NN$  and on spheres of arbitrary dimension by \cite{EH19} with $\fm_n$ of rate $n^{2/3}$.

Notably, all but (A) render the approximation (\ref{eq:BP-approx}) invalid. At this point we remark that stickiness (B) is conceptually different from the opposite of smeariness (C), namely
\begin{itemize}
 \item[(D)] $\fm_n >0$ for all $n\in \NN$ and $\lim_{n\to \infty}\fm_n = 0$,
\end{itemize}
which has been observed by \cite{schotz2019arbitrary}.

In this contribution we bring to attention two new phenomena:
\begin{itemize}
 \item[(E)] $\fm_n > 1$ for all $n \in \NN$ and $1 < \lim_{n\to \infty}\fm_n < \infty$
 \item[(F)] $\fm_n > 1$ for all $n \in \NN$ and  $\lim_{n\to \infty}\fm_n = 1$,
\end{itemize}
and investigate them on the circle. Phenomenon (E) affects all (!) parametric distributions on the circle with nowhere vanishing density  
like the Fisher-von-Mises, the wrapped Gaussian, etc. In particular, it may render the approximation (\ref{eq:BP-approx}) invalid as dramatically illustrated in the left display of Table \ref{tab:FSS-size}.

\begin{table}[h]
\centering 
\begin{minipage}{0.45\textwidth}
\begin{center} quantile-based
\end{center}
$$
\boxed{ 
\begin{array}{r|cccc}
\lambda  & 0     & 1/4   & 1/2   & 3/4 \\ \hline
n = 100  & 0.320 & 0.447 & 0.582 & 0.689 \\
n = 1000 & 0.330 & 0.474 & 0.656 & 0.818 \\
n = 10000 & 0.331 & 0.477 & 0.666 & 0.876
 \end{array}
 }$$
\end{minipage}
\begin{minipage}{0.45\textwidth}
\begin{center}bootstrap-based\end{center}
$$
\boxed{ 
\begin{array}{r|cccc}
\lambda & 0     & 1/4   & 1/2   & 3/4   \\ \hline
n = 100  & 0.045 & 0.041 & 0.039 & 0.035 \\
n = 1000 & 0.046 & 0.045 & 0.044 & 0.042 \\
n = 10000 & 0.050 & 0.049 & 0.049 & 0.051
 \end{array}
 }$$
\end{minipage}\\[0.4cm]
\caption{\it Empirical rejection probability based on 100\,000 simulation runs of two-sample tests for equality of Fr\'echet means with nominal significance level $0.05$ under the null hypothesis of a mixture of two equally weighted ($\beta=1/2$) von-Mises distributions on the circle with antipodal means (the first with concentration parameter $\kappa = 3$ and the second with varying concentration parameter $\lambda$) as defined in (\ref{eq:von-Mises}). Left: the quantile-based Test \ref{tests:BP}(ii) resting on (\ref{eq:BP-approx}), Right: the bootstrap-based Test \ref{tests:bootstrap} with $B = 1000$.  \label{tab:FSS-size}} 
\end{table}
Phenomenon (F) affects all distributions on the circle whose support strictly exceeds a closed half circle as long as a neighborhood of the antipodal of their Fr\'echet mean carries no probability mass. While this phenomenon renders the approximation (\ref{eq:BP-approx}) asymptotically valid, for surprisingly high sample sizes, the approximation may still be far off. 

In simulations, we see that in cases (E) and (F), $\fm_n$ initially has a rate comparable to smeariness as in Phenomenon (C), in particular, the rate of $\EE[d(\widehat{\mu_n},\mu)^2]$ is strictly lower than the classical $n^{-1}$. Equivalently, $n\mapsto \fm_n$ starts off nonhorizontal as illustrated in Figures  \ref{fig:fss-circle} and \ref{fig:wiggly-FSS}. Since in these cases, $n\mapsto \fm_n$ is only asymptotically horizontal, we give these new phenomena the name \emph{finite sample smeariness} (FSS) and distinguish between Type I  (E) and Type II (F). Alternatively for these phenomena, the term \emph{lethargic means} has been proposed.

While under FSS, empirical levels of quantile-based tests can deviate strongly from their nominal level, we show for the circle that suitably designed bootstrap tests approximately keep their level, as illustrated in the right display of Table \ref{tab:FSS-size}. In fact,  we show in this work that the deviation of the quantile-based test perseveres as the sample size $n$ tends to infinity  whereas the bootstrap-based test asymptotically attains the correct level. In addition to simulations, in application to wind direction data we see that asymptotic quantile based two-sample tests, on the circle and a torus, due to FSS present in the data, give a wrong impression of the extent of extreme wind change events for two continental European cities. Using bootstrap tests which preserve the nominal level, extreme wind changes reduce to a few concise events, which in particular make the years 2003, 2005 and 2017, 2018 exceptional (the latter strongly), hinting towards an effect of recent climate change.

In the following Section  \ref{scn:FSS} we introduce FSS on  arbitrary metric spaces and give a test for the presence of FSS in data. Further, we explore FSS exhaustively on the circle and on tori. In Section \ref{scn:smeary} we investigate the asymptotics of circular Fr\'echet sample means and prove a consistency result for bootstrap Fr\'echet sample means under FSS. For tests to follow, we also verify that all moments of properly centered and scaled Fr\'echet sample means converge to the respective moment of the limit distribution. Section \ref{scn:tests} explores the quantile based test by \cite{BP05} and an implementation of the bootstrap test, which was also suggested by \cite{BP05}. This is followed by simulations in Section \ref{scn:simulations} and the application to wind direction data in Section \ref{scn:wind}.
We conclude in Section \ref{scn:outlook} with an outlook to higher dimensions and list open problems that arise from our findings.

All of the more elaborate proofs are deferred to the supplement.

\section{Finite Sample Smeariness}\label{scn:FSS}

In the presence of finite sample smeariness (FSS) for finite sample sizes $n$, often considerably large, the rate of $\phi(\widehat{\mu}_n) - \phi(\mu)$, in the notation from \eqref{eq:BP-CLT}, is not of order $n^{-1/2}$ but similar to smeary rates as have been observed by \cite{HH15} on circles and by \cite{EH19} on spheres of arbitrary dimension. In case of FSS, however, asymptotically the rate returns to $n^{-1/2}$ but, possibly, with unanticipated asymptotic variance. In this section, we first give a general definition. Then, we describe this new phenomenon in detail on the circle and on flat tori and finally we propose a test for the presence of FSS in general data.

\subsection{Definition of Finite Sample Smeariness}
The following general definition assumes a metric space $(M,d)$ isometrically embedded in a finite dimensional Euclidean space, such that for every subset of $M$, being compact is equivalent to being closed and bounded. Then Fr\'echet mean sets defined by \eqref{eq:population-mean} are compact because they are closed by continuity and bounded due to
$$ d(\mu_1,\mu_2) \leq \EE[d(\mu_1,X)] + \EE[d(\mu_2,X)] \leq \sqrt{\EE[d(\mu_1,X)^2]}+\sqrt{\EE[d(\mu_2,X)^2]}\,.$$
%\begin{Def} 
If the set $K$ of Fr\'echet means is the union of disjoint manifolds of same dimension, the \emph{uniform measure on} $K$ is the normed measure obtained from the Euclidean measure conditioned to $K$. Thus, we can agree on the following. 

\begin{Ag}\label{ag:unifom}
If sample Fr\'echet means defined by \eqref{eq:sample-mean} are not unique and the mean set is sufficiently well behaved admitting a uniform measure, we agree that a measurable selection $\widehat{\mu}_n$ is drawn independently of $X_1,\ldots,X_n$ from that uniform measure on the Fr\'echet sample mean set. Under uniqueness of the Fr\'echet population mean $\mu$, $\EE[d(\widehat{\mu}_n,\mu)^2]$ is then well defined and so is the variance modulation $\fm_n$ from \eqref{eq:modulation}.
\end{Ag}

\begin{Def}[Finite sample smeariness]\label{def:FSS}
  A random variable $X$ on  a metric space $(M,d)$ with unique  Fr\'echet population mean $\mu$  is \emph{finite sample smeary} (FSS) if  $1<\sup_{n\in \NN}\fm_n< \infty$. We speak of \emph{Type I} FSS if $\lim_{n\to \infty}\fm_n > 1$ and of \emph{Type II} FSS if $\lim_{n\to \infty}\fm_n = 1$.
\end{Def}

\subsection{Finite Sample Smeariness on Circles and Tori}
The \emph{circle} is the space $\SSS = [-\pi,\pi)$ with $-\pi$ and $\pi$ identified and the usual arc length distance
$$d_\SSS(x,y) = \min\big\{|x-y|, 2\pi -|x-y|\big\}\,.$$
On the circle, we have the following characterization of the variance modulation $\fm_n$.

\begin{Th}\label{thm:CirclS-FSS} For arbitrary $n\geq 2$ suppose that $X$ is a random variable on $\SSS$ with unique Fr\'echet mean $\mu$ and support $J \subseteq \SSS$. Then $\fm_n>1$ for all $n \in \NN$ under any of the two following conditions 
\begin{itemize}
 \item[(i)] the interior of $J$ contains a closed half circle,
 \item[(ii)] $J$ contains two antipodal points, each of which is assumed by $X$ with positive probability. 
\end{itemize}
Moreover, $\fm_n = 1$ for all $n \in \NN$ under any of the two following conditions 
\begin{itemize}
 \item[(iii)] $J$ is strictly contained in a closed half circle,
 \item[(iv)] $J$ is a  closed half circle and one of its end points is assumed by $X$ with zero probability.
\end{itemize}
Finally, suppose that $X$ has a continuous density $f$ near the antipode $\overline{\mu}$ of $\mu$.  
\begin{itemize}
 \item[(v)] If $f(\overline{\mu}) =0$ then $\lim_{n\to \infty} \fm_n =1$,
 \item[(vi)] if $0<f(\overline{\mu})<\frac{1}{2\pi}$ then $\lim_{n\to \infty} \fm_n =\frac{1}{(1-f(\overline{\mu}) 2\pi)^2} >1$.
 \end{itemize}
\end{Th}

\begin{proof} The cases $(i) - (iv)$ follow at once from reducing to one of the cases of Lemma \ref{lem:FSS-foundation2} in Supplement~\ref{app:ProofModulationCurve}. 
For the cases $(v)$ and $(vi)$ we note that $\sup_{n\in \NN}\EE[d(\widehat{\mu}_n, \mu)^2]< \infty$ (Proposition \ref{prop:UniformIntegrability}) which asserts  by \cite[Corollary to Theorem 25.12]{Billingsley1995}, that $n\EE[d(\widehat{\mu}_n, \mu)^2]$ converges to the variance of the limiting distribution in the first assertion of Theorem \ref{them:CLT_Circle}.
\end{proof}

As a consequence, we derive the following characterization of FSS on the circle. 
\begin{Cor}\label{cor:ConnectionFSSAndDensity}
 Let $X$ be a random variable on $\SSS$ which has a continuous density near the antipode $\overline \mu$ of the Fr\'echet population mean $\mu$. 
\begin{itemize}
 \item[(i)] Then, $\mu$ is FSS of type I if and only if $0<f(\overline \mu ) < 1/(2\pi)$ . 
 \item[(ii)] Further, $\mu$ is FSS of type II if and only if  $f(\overline \mu) = 0$ and  condition (i) or (ii) of Theorem \ref{thm:CirclS-FSS} holds.  \end{itemize}
\end{Cor}

\begin{Rm}
 In the above case of a continuous density $f$ near the antipode $\overline{\mu}$ of the Fr\'echet mean, $f(\overline{\mu}) = (2\pi)^{-1}$ leads to smeariness as detailed in \cite{HH15} and  $f(\overline{\mu}) > (2\pi)^{-1}$ is not possible, cf. \cite[Theorem 3.1(ii)]{HH15}.
\end{Rm}

Simple cases of FSS on the circle are illustrated in the following. 

\begin{figure}[H]
\includegraphics[width=0.95\textwidth, trim = 0 0 0 0, clip]{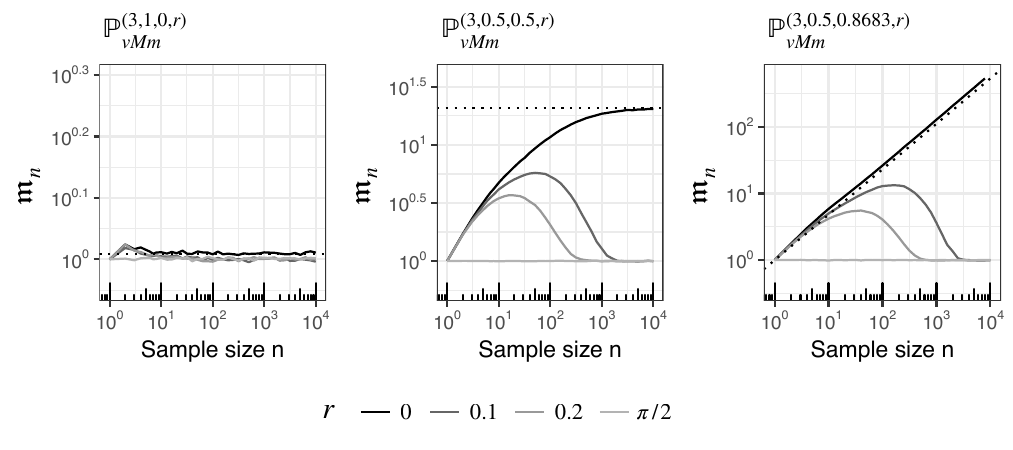}\vspace{-0.4cm}
  \caption{\it
  Log-log plots of variance modulation curves $n\mapsto \fm_n$, for varying von Mises mixture distributions (vMmds) defined in (\ref{eq:von-Mises}) in black ($r=0$) and vMmds with a disk cut out of radius $r=0.1, 0.2,\pi/2$ (in fading gray)  about the antipode of the mean, as defined in (\ref{eq:FvM-modified}) based on 100\,000 simulation runs for each sample size. 
  In the left and middle display, the vMmds (black) feature Type I FSS, with the dotted line giving the asymptotic scaled variance, the two vMmds with an interval cut out (gray curves) of radius $0<r<\pi/2$ feature Type II FSS. In the right display, the vMmd (black) features smeariness, with theoretical scaled variance (dotted), the vMmds with an interval cut out  (gray curves) of radius $0<r<\pi/2$ feature Type II FSS. Only the vMmds with an interval cut out of radius $r=\pi/2$ (light gray) features no FSS at~all.
  }
  \label{fig:fss-circle}
\end{figure}

\begin{Ex}[Von Mises mixtures] On $\SSS$ we consider von Mises mixtures with antipodal modes with respect to arc length measure 
\begin{equation}\label{eq:von-Mises}
d\Prb^{\kappa, \beta, \lambda}_{vMm}(x) \coloneqq \beta\, \frac{\exp\left( \kappa \cos(x) \right) }{2 \pi I_0(\kappa)}\,dx + (1- \beta)\frac{\exp\left( \lambda \cos(x+\pi) \right) }{2 \pi I_0(\lambda)}\,dx \quad  \text{ for }\kappa,\lambda \geq  0, \beta \in [0,1],
\end{equation}
where $I_0(\cdot)$ is the modified Bessel function of the first kind of order 0, e.g.,  \citet[p. 36]{MJ00}. By symmetry, a von Mises mixture $d\Prb_{vMm}^{\kappa,\beta, \lambda}$ attains either a unique  mean at $0$ or $\pi$, or nonunique means at $\{-t,t\}$ for some $t\in(0, \pi)$. 
Furthermore, we define for $r\geq 0$ the function cutting out and mirroring a disk of radius $r$ about $-\pi$: $$\zeta^r\colon \SSS \rightarrow \SSS,\quad  p \mapsto \begin{cases} p & \text{ if } p \in [-\pi+r, \pi-r)\\
p+\pi & \text{ if } p \in [-\pi, -\pi+r)\\
p-\pi & \text{ if } p \in [\pi-r, \pi)
\end{cases}$$
For the von Mises mixture $\Prb_{vMm}^{\kappa, \beta, \lambda}$ we then denote the push-forward measure under $\zeta^r$ by 
\begin{equation}\label{eq:FvM-modified}
\Prb_{vMm}^{\kappa, \beta, \lambda,r} \coloneqq \zeta^r_* \,\Prb_{vMm}^{\kappa, \beta, \lambda}\,,
\end{equation}
which preserves all mass except for that in the disk which is mirrored. Recall that by Theorem \ref{thm:CirclS-FSS}, all of the $\Prb_{vMm}^{\kappa, \beta, \lambda, r}$ with unique mean $\mu=0$ and $r < \pi/2 - \eps$ for some $\eps>0$ feature FSS (if they are not smeary themselves), which is of Type I if $r =0$ and Type II if $r >0$. For the parameters $(\kappa, \beta, \lambda) \in \{(3,1,0), (3,0.5,0.5), (3,0.5,0.8683)\}$ and varying choices for the parameter $r\in \{0, 0.1,0.2,\pi\}$ the respective variance modulation curves are depicted in Figure \ref{fig:fss-circle}. Notably, in case of Type I FSS, the curve  $\fm_n$ rises from $1$ to  $(1-2\pi f(\overline \mu))^{-2}$ whereas under Type II FSS, the curve first rises from  $1$ and eventually drops down to $1$.
\end{Ex}

\begin{figure}[t]
  \includegraphics[width=0.95\textwidth]{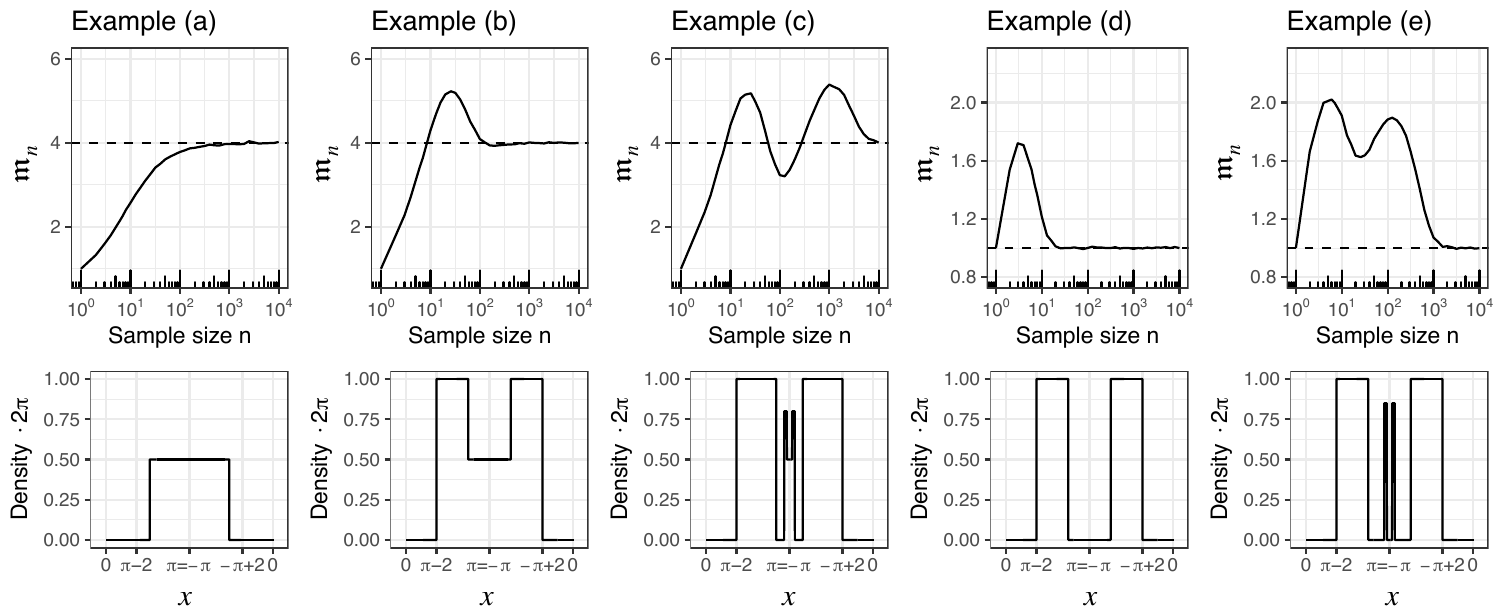}
  \caption{\it Top: Variance modulation curves as in Figure 
  \ref{fig:fss-circle} (here the vertical is not in log-scale), for each distribution in Table \ref{tab:valuesForWigglySmearyiness} of Example \ref{ex:intrigued-FSS} based on 100\,000 simulations for each sample size.
  Bottom: Density part of the corresponding distributions (that comprise a $\delta$-measure at the origin).
  }\label{fig:wiggly-FSS}
\end{figure}

Depending on the probability distribution near $\overline \mu$, also more complicated versions of increase and decrease may occur, as Example \ref{ex:intrigued-FSS} and Figure \ref{fig:wiggly-FSS} teach: every pair of bumps of the density near the antipode may result in a bump of $\fm_n$. As before, however, it starts at $1$ and eventually settles at $(1-2\pi f(\overline \mu))^{-2}$, producing Type I FSS if $f(\overline \mu)>0$ and Type II FSS else.

\begin{Ex}[Relating antipodal density to variance modulation]\label{ex:intrigued-FSS}
To investigate the  relationship between the variance modulation of intrinsic sample means $n \mapsto \fm_n$ and the density $f^{(t,w)}$ of $X$ near the antipode of the intrinsic population mean $\mu=0$ we consider a family of distributions for which the density near the antipode $\overline \mu$ is piecewise constant. Let $l\in \NN$, $w=(w_1, \dots, w_l) \in [0,1]^l$, $t=(t_1, \dots, t_l)\in [0,\pi)^l$ with $t_0 \coloneqq 0 < t_1 < \dots < t_l< \pi$ and define the distribution $\Prb^{(t,w)}_U$ by
\begin{align*}& \quad \quad \quad  d\Prb^{(t,w)}_U(x)\coloneqq  k \cdot d \delta_0(x) + \frac{1}{2\pi}\cdot f^{(t,w)}(x)dx \quad \text{ with } 
 \\
 f^{(t,w)}(x) \coloneqq & \begin{cases}
 	w_i & \text{ if } \;\;	x \in [-\pi +t_{i-1}, -\pi + t_{i})\cup(\pi -t_{i}, \pi - t_{i-1}] \text{ for some } i \in \{1, \dots, l\}\\
 	0 & \text{ if } \;\; x \in [-\pi+t_l,\pi-t_l], \end{cases}
\end{align*}
where $k=k(t,w)>0$ is a normalization constant to ensure that $\Prb^{(t,w)}_U$ is a probability measure. In Table \ref{tab:valuesForWigglySmearyiness} we list cases (a) -- (e) of parameter choices considered.

\begin{table}[H]
\centering
% \boxed{
\begin{tabular}{c||c|c|c|c|c}
 & $(a)$ & $(b)$ & $(c)$ & $(d)$ & $(e)$\\[0.1cm]
\hline 
$l$ & 1 & 2 &4 & 2 & 4\\
$t$ & 1.5 & (0.8, 2) & (0.1, 0.2, 0.5, 2) & (0.8, 2) & (0.1,  0.2, 0.5, 2)\\
$w$ & 0.5&(0.5, 1)&(0.5, 0.8, 0.0, 1)& (0.0, 1)& (0.0, 0.85, 0.0, 1)\\
\hline
Type of FSS & I & I & I & II & II
\end{tabular}
% }
\caption{Selected values for the parameters $l,t,w$ and their resulting type of FSS.}
\label{tab:valuesForWigglySmearyiness}
\end{table}

In all cases the population mean of  $\Prb^{(t,w)}_U$  is unique and located at $\mu = 0$. Whenever the density at the antipode is strictly between zero and $1/2\pi$ we have FSS of Type I. Regardless of the type of FSS, every pair of bumps in the density near the antipode corresponds to a single bump in the variance modulation curve. Thus, in case of Type I FSS the rescaled Fr\'echet sample variance $n \EE[d(\widehat{\mu}_n,\mu)^2]$ approaches asymptotically a value strictly above the Euclidean variance $\EE[d(X,\mu)^2]$, in case of Type II FSS it approaches asymptotically the Euclidean variance.

\end{Ex}

\begin{Rm}
As Example \ref{ex:intrigued-FSS} and Figure \ref{fig:wiggly-FSS} teach, the variance modulation $\fm_n$ may be non-monotonic exhibiting different behaviours for different regimes of sample size.
\end{Rm}

The following example illustrates Theorem \ref{thm:CirclS-FSS} with two point masses at or beyond the equator (case (ii)), or before (case (i)), with a possibly (in case (iii)) non-unique sample mean set, of which uniformly at random, a sample mean is selected. 
 
\begin{figure}[!h]
  \centering
  \includegraphics[width=0.95\textwidth, trim = 0 0 0 0, clip]{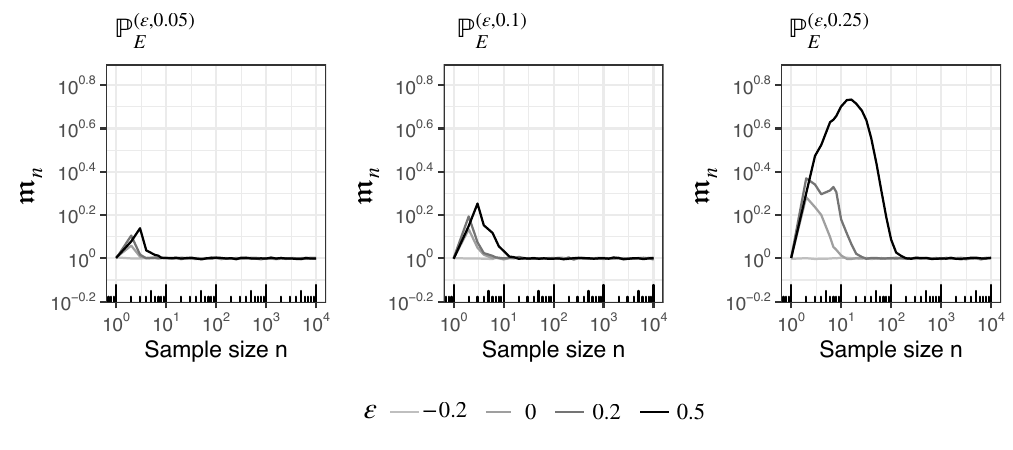}
  \caption{\it  Variance modulation curves, as in Figure 
  \ref{fig:fss-circle}, for the three types of distributions $\Prb^{(\eps, w)}_{E}$ from Example \ref{ex:TriplePointFSS} with weight $w$ of point mass at $\pm (\pi/2 +\eps)$ based on 100\,000 simulation runs for each sample size. Black: $\eps = 0.5$, dark gray: $\eps = 0.2$, gray: $\eps = 0$, light gray: $\eps = -0.2$. We have Type II FSS for all values of $\eps \geq 0$ and no FSS (constant scaled variance) for $\eps <0$.
  \label{fig:weightAtEquator}}
\end{figure}

\begin{Ex}\label{ex:TriplePointFSS}
	Letting $\eps \in [-\pi/2, \pi/2)$ and  $w\in (0,1/4]$, define the circular distribution $\Prb^{(\eps, w)}_{E}$ by \[d\Prb^{(\eps, w)}_{E}(x) \coloneqq (1 - 2w)\cdot \mathds{1}_{[-1/2,1/2]}(x)dx + w\, d\delta_{\pi/2 + \eps}(x) + w\, d\delta_{-\pi/2 - \eps}(x)\,,\]
	which assigns at least half the mass to $[-1/2,1/2]$ and the rest is evenly distributed close to the equator at  $\pi/2 + \eps$ and $-\pi/2 - \eps$. 
	\begin{itemize}
	\item[(i)]
	For $\eps<0$ these distribution are supported in $(-\pi/2,\pi/2)$ and thus by Theorem \ref{thm:CirclS-FSS} (i) feature no FSS. 
	\item[(ii)]
	For $\eps>0$, in contrast by Theorem \ref{thm:CirclS-FSS} (iii), they always feature FSS, which is, by Theorem \ref{them:CLT_Circle} (i) of Type II. 
	\item[(iii)] 
	For $\eps =0$, by the same argument, they also feature FSS of Type II. Indeed, samples featuring only points at $\pm \pi/2$ and no others, occurring with positive probability, have nonunique sample means, namely one closer to $0$ and one closer to $-\pi$. As agreed, of these with probability $1/2$ the one closer to $-\pi$ is chosen which accounts for a higher variance than the Euclidean. 
	\end{itemize}
	Each panel in Figure \ref{fig:weightAtEquator} illustrates the three cases, with larger effect on FSS the larger the weight $w$ of each of the point masses.
\end{Ex}

While we defined FSS for arbitrary metric spaces, we have investigated it only for the circle. These results extend at once to the $m$-torus $\TTT = \big(\SSS\big)^m = \bigtimes_{i=1}^m [-\pi,\pi)$, $m\in \NN$, equipped with the canonical product metric
$$ d_{\TTT} (x,y) = \sqrt{\sum_{i=1}^m d(x^{(i)},y^{(i)})^2},\quad x=\begin{pmatrix}x^{(1)}\\ \vdots \\x^{(m)}\end{pmatrix},~y=\begin{pmatrix}y^{(1)}\\ \vdots \\y^{(m)}\end{pmatrix}\in \TTT\,.$$

Indeed,  $\mu \in \TTT$ is a minimizer of the population Fr\'echet function (\ref{eq:population-mean}) of a random variable $X$ on $\TTT$ if and only if all of its coordinates $\mu^{(i)}$ are minimizers of the population Fr\'echet functions of the marginals $X^{(i)}$ on the $i$-th circle, $i\in \{1,\ldots,m\}$.

\begin{figure}[h!]
  \centering
  \includegraphics[width=0.5\textwidth]{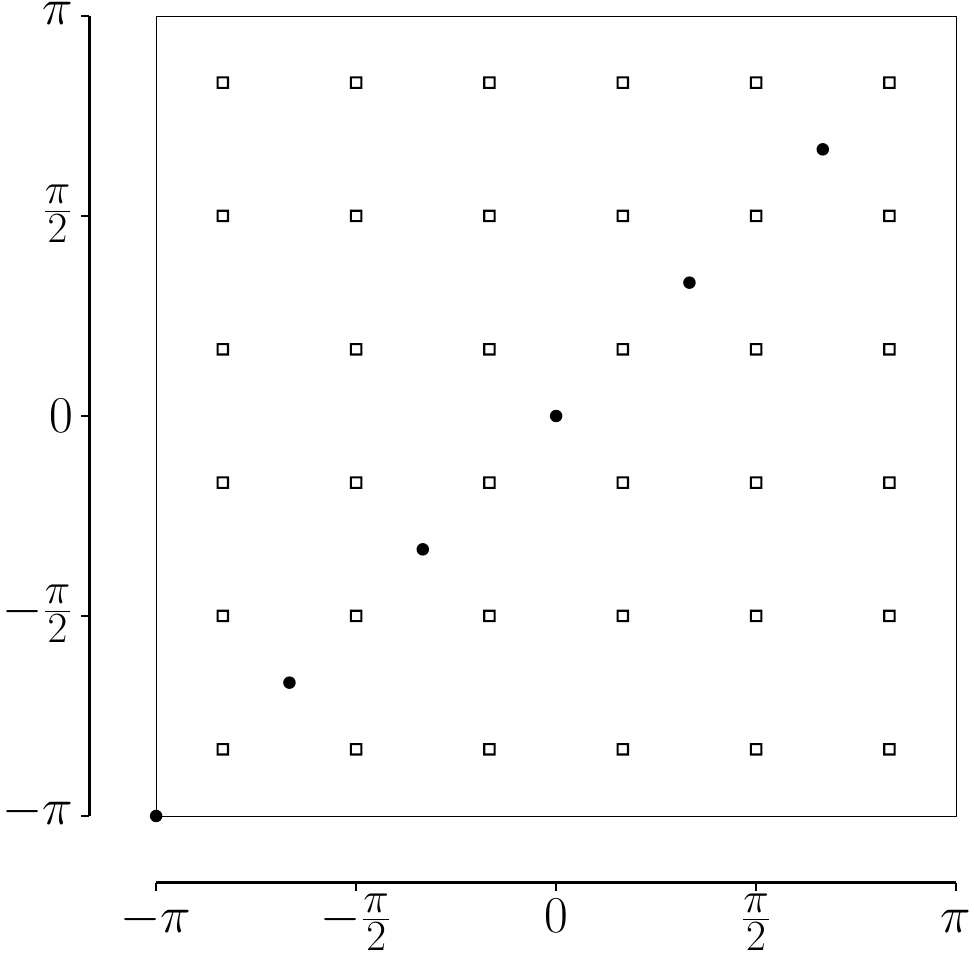}
  \caption{\it Six sample points (black circles) on the diagonal of the two-torus $\mathbb{T}^2$ and their equally spaced $6^2 = 36$ Fr\'echet sample means (empty squares), cf. Remark \ref{rm:torus}.}
  \label{fig:sample-mean}
\end{figure}

\begin{Rm}
In consequence, due to \citet[Corollary 3]{HH15}, for a sample $X_1,\ldots,X_n$ on the $m$-torus $\TTT$ there are at most $n^m$ minimizers of the sample Fr\'echet function (\ref{eq:sample-mean}). For instance every one of the $n^m$ grid points
$$\left(\left(-\pi + \frac{2\pi (2j_1-1)}{2n}\right),\ldots,\left(-\pi + \frac{2\pi (2j_m-1)}{2n}\right)\right),\quad j_1,\ldots,j_m \in \{1,\ldots,n\}  $$
is a minimizer of the sample Fr\'echet function of $X_j = \left(-\pi + \frac{2\pi (j-1)}{n}\right)  (1,\ldots,1)^T \in \TTT$, $j\in \{1,\ldots,n\}$, cf. Figure \ref{fig:sample-mean}. Indeed, their $i$-th marginal $(1\leq i \leq m)$ is given by the sample  $X^{(i)}_j = \left(-\pi + \frac{2\pi (j-1)}{n}\right)$, $j\in \{1,\ldots,n\}$ which has exactly $n$  minimizers of its sample Fr\'echet function which are equally spaced between the sample points.
\end{Rm}

In particular, Theorem \ref{thm:CirclS-FSS} yields the following characterization of FSS and the variance modulation for random variables on $\TTT$.
    
\begin{Cor}\label{Cor:TorS-FSS} For arbitrary $n\geq 2$ and $m \in \NN$ let $X=(X^{(1)},\ldots,X^{(m)})$ be a random variable on the torus $\TTT$ with unique Fr\'echet population mean $\mu$. For $i \in \{1, \dots m\}$ let the support of $X^{(k)}$ be $J^{(i)}\subseteq \SSS$. Then $\fm_n>1$ for all $n \in \NN$  under any of the following two conditions
\begin{itemize}
    \item[(i)] there exists $i\in \{1, \dots m\}$ such that the interior of $J^{(i)}$ contains a closed half circle,
    \item[(ii)] there exists $i\in \{1, \dots m\}$ such that $J^{(i)}$ contains two antipodal points, each of which are attained by $X^{(i)}$ with positive probability
\end{itemize}
Moreover, $\fm_n=1$ for all $n \in \NN$ under any of the following two conditions 
\begin{itemize}
    \item[(iii)] for each $i \in \{1, \dots, m\}$  the support $J^{(i)}$ is strictly contained in a closed half circle,
    \item[(iv)] for each $i \in \{1, \dots, m\}$  the support $J^{(i)}$ is contained in a closed half circle where one of the end points is a.s. not attained by $X$.
\end{itemize}
Finally, suppose that each component $X^{(i)}$ for $i \in \{1, \dots, m\}$ has near the antipode $\overline{\mu}^{(i)}$ of $\mu^{(i)}$ a continuous density $f^{(i)}$.  
\begin{itemize}
  \item[(v)] If $f^{(i)}(\overline{\mu}^{(i)}) =0$ for all $i \in \{1, \dots, m \}$, then $\lim_{n\to \infty} \fm_n =1$,
  \item[(vi)] if $0\leq f^{(i)}(\overline{\mu}^{(i)})<\frac{1}{2\pi}$ for all $i \in \{1, \dots, m \}$ with $f^{(i)}(\overline{\mu}^{(i)})>0$ for at least one component then $\lim_{n\to \infty} \fm_n = \left(\sum_{k = 1}^m \frac{ \EE[d^2(X^{(k)},\mu^{(i)})] }{(1-f^{(i)}(\overline{\mu}^{(i)}) 2\pi)^2} \right) / \left(\sum_{i = 1}^m  \EE[d^2(X^{(i)},\mu^{(i)})]  \right) >1$.
\end{itemize}
\end{Cor}

As a consequence, we obtain a similar characterization as in Corollary \ref{cor:ConnectionFSSAndDensity} for FSS on the torus. 

\begin{Cor}
Suppose Let $X=(X^{(1)},\ldots,X^{(m)})$ be a random variable on $\TTT$ and suppose for each $i \in \{1, \dots, m\}$ there exists a  continuous density near the antipode $\overline \mu^{(i)}$ of the respective coordinate's Fr\'echet population mean $\mu^{(i)}$. 
\begin{itemize}
 	\item[(i)] Then, $\mu$ is FSS of type I if and only if  $0\leq f^{(i)}(\overline{\mu}^{(i)})<\frac{1}{2\pi}$ for all  $i \in \{1, \dots, m \}$  with $f^{(i)}(\overline{\mu}^{(i)})>0$ for at least one component. 
	\item[(ii)] Further, $\mu$ is FSS of type II if and only if $f^{(i)}(\overline{\mu}^{(i)}) =0$  for all  $i \in \{1, \dots, m \}$ and  condition (i) or (ii) of Corollary \ref{Cor:TorS-FSS} holds. 
 \end{itemize}
\end{Cor}

For other spaces, an investigation of FSS is beyond the scope of this paper. Already for spheres, finding examples of smeary distributions is rather involved, and to date, only two-smeariness could be confirmed by \cite{EH19}. It is, however, known for rather general manifolds that ``smeariness begets FSS'' as \cite{DHH-GSI21} constructed from every smeary random variable a random variable featuring FSS of Type I. On the other hand, for manifolds featuring a simply connected submanifold of constant positive sectional curvature, not exceeded by any sectional curvatures of the manifold, \cite{DHH-GSI21} have given examples of directional smeary random variables.  In particular, \cite{EHH-GSI21} showed that every non trivial random variable on a sphere is FSS of Type I. In the light of \emph{geometrical smeariness} introduced by \cite{E19} we conjecture  that Type I FSS is present in all nondegenerate random variables on positively curved spaces. For such spaces, \cite{Afsari09} has shown that the Hessian of the Fr\'echet function is smaller than its Euclidean equivalent, which is twice the identity matrix. Moreover, for very small sample sizes ($n$ approx. less than $10$) of highly concentrated random variables, \cite{Pennec19} explicitly derived $ \fm_n>1$ but well below what is predicted asymptotically by the CLT. We allow considerably larger sample sizes with no concentration constraints, however. 

\subsection{Testing for Finite Sample Smeariness on Metric Spaces}
\begin{figure}[hbt]
  \centering
  \includegraphics[width=1\textwidth, trim = 0 0 0 0, clip]{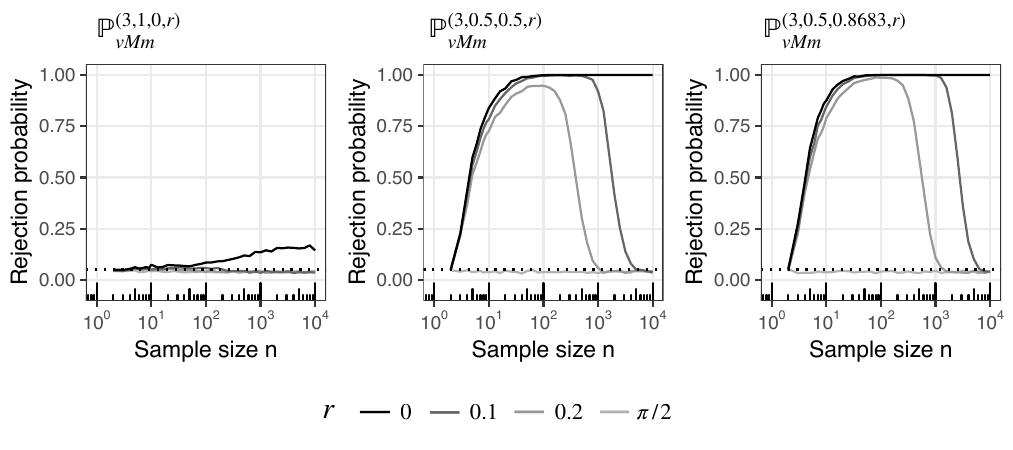}
  \caption{\it Empirical rejection probability curves for Test \ref{test:FSS} with nominal size $\alpha = 0.05$ (dotted horizontal) for the vMmds from Figure \ref{fig:fss-circle} based on 5000 simulation runs for each sample size and $B = 1000$. For the vMmds with a disk cut out of size $r=\pi/2$ about the antipode of its mean, which feature no FSS, all of the curves  (light gray) keep the level. For Type II FSS (gray: $r=0.2$, dark gray: $r=0.1$) rejection probabilities are almost one within the regime of FSS; beyond, where there is no FSS any more, they drop back to the nominal size (see Figure \ref{fig:fss-circle}). Also for the vMmd consisting of only one vMd (left panel, black) with almost no FSS visible, rejection probabilities increase visibly with sample size. For Type I FSS (middle, black) and smeariness (right, black), the rejection probabilities remain one also for higher sample sizes.
  \label{fig:FSS_ScalePlot_3Point}}
\end{figure}
\noindent In order to heuristically assess the presence of FSS for a given sample $X_1,\ldots,X_n\iid X$ on a metric space $(M,d)$ consider random variables  with unique population mean $\mu$ and unique sample mean $\widehat{\mu}_n$ (in case of well behaved nonuniqueness, take a uniformly random selection as in Agreement \ref{ag:unifom}) and set
\begin{eqnarray}\label{variances:eq} V := F(\mu),~\widehat{V}_n := F_n(\widehat{\mu}_n),~W:= \EE[d(\mu,X)^4],~\widehat{W}_n := \frac{1}{n}\sum_{j=1}^n d(\widehat{\mu}_n,X_j)^4\,,\end{eqnarray}
where we assume that the 4-th moment exists.
Then, under mild additional assumptions the  CLT for the Fr\'echet variance by \cite{dubey2019frechet} teaches that
$$ \sqrt{n} (\widehat{V}_n - V) \inD \cN\left(0,W - V^2\right)\,,$$
where $\widehat{W}_n - \widehat{V}_n^2$ is an asymptotically unbiased estimator for $W - V^2$, that also satisfies $\sqrt{n}$ asymptotic normality.

Further, let $B>0$ be a large integer and $\mu^*_b$ be a mean of an $n$-out-of-$n$ bootstrap sample of $X_1,\ldots,X_n$ for $b=1,\ldots,B$. Further, let $\widehat{\mu}^*$ be the mean of the $\mu^*_1,\ldots,\mu^*_B$ and set 
\begin{eqnarray}\label{bt-variances:eq}\widehat{V}^* := \frac{1}{B} \sum_{b=1}^Bd(\widehat\mu^*,\mu^*_b)^2,~ \widehat{W}^* := \frac{1}{B} \sum_{b=1}^Bd(\widehat\mu^*,\mu^*_b)^4\,.\end{eqnarray}
Then, under absence of smeariness and absence of FSS, $\sqrt{B} (\widehat{V}^* - V_n/n)$ is approximately normal with zero expectation and its variance can be estimated by $\widehat{W}^* - (\widehat{V}^*)^2$. Hence, in this case, with the standard-normal $\alpha$-quantile $\phi_\alpha$ we expect that
$$ \Prb\left\{ \widehat{V}^* > \frac{\widehat{V}_n}{n} + \phi_{1-\alpha} \,\sqrt{\frac{\widehat{W}^* - (\widehat{V}^*)^2}{B}}\right\} \approx \alpha\,.$$

\begin{Test}[For the Presence of FSS]\label{test:FSS} For random variables $X_1,\ldots,X_n$ on a metric space $(M,d)$ with sample mean $\widehat\mu_n$, bootstrap means $\mu^*_1,\ldots,\mu^*_B$, and their mean $\widehat{\mu}^*$, as above and the notation from (\ref{variances:eq}) and (\ref{bt-variances:eq}),
\begin{equation}
  \fmnhat := \frac{n\widehat{V}^*}{\widehat{V}_n}\label{eq:ScaleFSS}
\end{equation}
denotes the empirical (bootstrap-based) variance modulation for $X_1,\ldots,X_n$. With the standard normal $\alpha$-quantile $\phi_\alpha$ for $\alpha \in (0,1)$ \emph{reject} the absence of FSS  for sample size $n$ at nominal level $\alpha$ if
$$ \fmnhat - 1 > h_{n,\alpha}\mbox{ where } h_{n,\alpha}= \frac{n\phi_\alpha}{\sqrt{B}} \,\frac{\sqrt{\widehat{W}^* - (\widehat{V}^*)^2}}{\widehat{V}_n}\,.$$
\end{Test}
  
Figure \ref{fig:FSS_ScalePlot_3Point} shows simulations for distributions $\Prb_{vMm}^{(3,\beta, \lambda,r)}$ featuring Type I FSS for $r=0$ as defined in (\ref{eq:von-Mises}) and Type II FSS for $r>0$ as defined in (\ref{eq:FvM-modified}).
Indeed for $r = \pi/2$, the null hypothesis of absence of FSS is true and Test \ref{test:FSS} keeps the level $\alpha = 0.05$. 
For Type II FSS, rejection probabilities are almost one within the regime of FSS (compare with Figure \ref{fig:fss-circle}). Beyond that regime the null hypothesis is true again and detected with the correct size. For the von Mises mixture distribution (vMmd) consisting of only one von Mises distribution ($\beta = 1$, left panel, black) with almost no FSS visible in Figure \ref{fig:fss-circle}, rejection probabilities increase visibly with sample size.  As expected, the power of Test \ref{test:FSS} increases with proximity to a smeary distribution ($\beta = 0.5, \lambda = 0.8683, r=0$, right panel, black), and specifically in case of Type II FSS also with smaller hole size and proximity to a distribution featuring Type I FSS.

\section{Asymptotics of Fr\'echet Means on the Circle and the Torus}\label{scn:smeary}

In this section, we give with an exposition on the asymptotic behavior of Fr\'echet sample means on the circle. We then proceed with a proof on consistency of the bootstrap and afterwards show under the presence of FSS that moments of the Fr\'echet sample means and their bootstrapped versions  converge to the respective moment of the limit distribution. 

\subsection{Central Limit Theorem and Bootstrap Consistency for Fr\'echet Means} 
We begin with the central limit theorem for Fr\'echet sample means on the circle  under the following assumption.

\begin{As}\label{ass:antipodalDensityBounded}
	Let $X$ be a random element on $\SSS$ 
	\begin{enumerate} 
		\item[(i)]	with unique Fr\'echet population mean $\mu=0$ and 
		\item[(ii)]	that features a continuous density $f$ on $(\pi - \delta, \pi)\cup[-\pi,-\pi + \delta)$ with respect to the arc length measure for some $\delta>0$ such that $$
  f(-\pi) = \lim_{x \searrow -\pi} f(x) = \lim_{x \nearrow \pi} f(x) < \frac{1}{2\pi}.$$

	\end{enumerate}
\end{As}

\begin{Th}[Central Limit Theorem on the Circle by \cite{MQC12,HH15}]\label{them:CLT_Circle}
	Under Assumption \ref{ass:antipodalDensityBounded} let $n\in \NN$ and consider i.i.d. random elements $X_1,\ldots,X_n\sim X$ on $\SSS$ with a measurable selection $\widehat{\mu}_n$ of Fr\'echet sample means. Then, for $n\to \infty$, we have that
	$$    \sqrt{n} \,\widehat{\mu}_n  \xrightarrow{\mathcal{D}} \mathcal{N}\left(0,\frac{\EE[X^2]}{\big(1-2\pi f(-\pi)\big)^2}  \right),$$
    where $\EE[X^2]$ denotes Euclidean variance.
\end{Th}	
    
\begin{Rm} The case $f(-\pi) = \frac{1}{2\pi}$ where the density $f$ is $(k+1)$-times continuously differentiable near $-\pi$ for $k \in \NN$, such that the first $k$ derivatives vanish at $-\pi$ whereas the derivative of order $k+1$ at $-\pi$ does not, has been investigated by \cite{HH15}. In this setting, the convergence rate for Fr\'echet sample means turns out to be of order $n^{\frac{-1}{2(k+1)}}$ which is strictly slower than the standard $n^{-1/2}$-rate. The phenomenon of a slower convergence rate is known as \emph{smeariness}. Moreover, it is not possible that the antipode of the Fr\'echet population mean exhibits a point mass or fulfills $f(-\pi) > \frac{1}{2\pi}$, cf. \citet[Theorem 1]{HH15}.
\end{Rm}

Notably, the limit distribution of the Fr\'echet sample mean depends on the behavior of the density near the antipode of the Fr\'echet population mean. In particular, in case the density near the antipode of the population mean does not vanish, approximating the distribution of $\sqrt{n} \widehat\mu_n$ by a centered Gaussian with an estimated variance $\frac{1}{n}\sum_{i = 1}^{n} d(\hat \mu_n,X_i)$ is not suited. 

An alternative approach to imitate the law of scaled Fr\'echet means is by means of bootstrap methods. 
In the following, we show that the na\"ive $n$-out-of-$n$ bootstrap is indeed (asymptotically) consistent in approximating the distribution of  Fr\'echet sample means. For this purpose, we denote by $\BL$ the space of bounded Lipschitz functions on $\RR$ which are bounded by one and have Lipschitz modulus at most one. Further, we denote by $\konvPStar$ the convergence in outer probability, cf. \cite{vdVW96}.

\begin{Th}[Consistency of Bootstrap]\label{them:BS_CLT_Circle}
  Under Assumption \ref{ass:antipodalDensityBounded} let $n\in \NN$ and consider i.i.d. random elements $X_1,\ldots,X_n\sim X$ on $\SSS$ with a measurable selection $\widehat{\mu}_n$ of Fr\'echet sample means. 
  Further, given $X_1, \dots, X_n$ consider a bootstrap sample $X_1^*, \dots, X_n^*\iid \frac{1}{n}\sum_{i = 1}^{n}\delta_{X_i}$ with measurable selection $\widehat \mu_{n}^*$ of its Fr\'echet sample mean. Then, for $n\to \infty$ we have
  $$ \sup_{h \in \BL}\bigg|	\EE\lrBrack{h\lrCurlyBrack{\sqrt{n}\lrCurlyBrack{\widehat\mu_n^* - \widehat \mu_n}}\big| X_1, \dots, X_n}- \EE\lrBrack{h\lrCurlyBrack{\sqrt{n}\,\widehat \mu_n}}\bigg|\konvPStar 0.
  $$
\end{Th}	

\begin{proof}
	For $x \geq 0$ the Fr\'echet function of the bootstrap realization is given by 
	\begin{align*}
		  F_n^*(x) &= \frac{1}{n}\sum_{X_j^* \in [x -\pi,\pi)} (X_j ^*-x)^2 + \frac{1}{n}\sum_{X_j^* <x-\pi} (X_j^*+2\pi - x)^2 \\
            &= \frac{1}{n}\sum_{j=1}^n(X_j^* - x)^2 +  \frac{4\pi}{n} \sum_{X_j^* < x-\pi} (X_j^* - x + \pi).	
	\end{align*}
    With  the Euclidean mean $\overline X^*_n = \frac{1}{n}\sum_{i = 1}^{n}X_i^*$ of the bootstrap realizations we have for any $x \geq 0$ with $x - \pi \not \in \{X_1^*, \dots, X_n^*\}$ that 
	 \begin{align}\label{eq:proof-optimalityBS_FrechetFunction}
	  \frac{1}{2}\,\grad\, F_n^*(x)  = x-\overline{X}^*_n -  \frac{2\pi}{n} \sum_{X_j^* \leq  x-\pi} 1\,.
    \end{align}
   We rewrite the sum in \eqref{eq:proof-optimalityBS_FrechetFunction} as follows
    \begin{align*}
    	 -\frac{1}{n} \sum_{X_j^* \leq x-\pi} 1 &= -f(-\pi)x +\underbrace{ \lrBrack{ f(-\pi)x - \mathbb{P}(X \leq x-\pi) } }_{\eqqcolon A(x)}  \\
    	 &+ \underbrace{ \lrBrack{ \mathbb{P}(X \leq x-\pi)  -\frac{1}{n} \lrCurlyBrack{\sum_{X_j \leq x-\pi} 1} }}_{\eqqcolon B_n(x)}+ \underbrace{\lrBrack{ \frac{1}{n}\lrCurlyBrack{\sum_{X_j \leq x-\pi} 1} - \frac{1}{n} \lrCurlyBrack{ \sum_{X_j^* \leq x-\pi}1}}}_{\eqqcolon C_n(x)}. 
    \end{align*}
	Note that the function $A(x)$ is independent from samples $X_1, \dots, X_n$ and bootstrap samples $X_1^*, \dots, X_n^*$. 
	Similarly we obtain for $x<0$ with $x + \pi \not \in \{X_1^*, \dots, X_n^*\}$ that 
	\begin{equation}\label{eq:proof-optimalityBS_FrechetFunction2}
  	  \frac{1}{2}\,\grad\, F_n^*(x) = x-\overline{X}^*_n +  \frac{2\pi}{n} \sum_{X_j^* > x+\pi} 1
	\end{equation}
	and we similarly rewrite  
	\begin{align*}
    	 \frac{1}{n} \sum_{X_j^* > x+\pi} 1 &=
    	 - f(-\pi)x  +
    	 \underbrace{\lrBrack{f(-\pi)x + \mathbb{P}(X> x +\pi)}}_{\eqqcolon A(x)}
    	\\
    	 & + \underbrace{ \lrBrack{ - \mathbb{P}(X >x+\pi)  +\frac{1}{n} \lrCurlyBrack{\sum_{X_j > x+\pi} 1} }}_{\eqqcolon B_n(x)}+ \underbrace{\lrBrack{ - \frac{1}{n}\lrCurlyBrack{\sum_{X_j > x+\pi} 1}+ \frac{1}{n} \lrCurlyBrack{ \sum_{X_j^* > x+\pi}1}}}_{\eqqcolon C_n(x)} 
    \end{align*}	
    Note, we consider $x \pm \pi \neq X_j^*$, thus the condition ``$X_j^*< x + \pi$'' (resp. ``$X_j^*> x- \pi$'') is equivalent to ``$X_j^*\leq x + \pi$'' (resp. ``$X_j^*\geq x- \pi$'') and likewise for analogous conditions with $X_j$. The reason for our definition of $B_n$ and $C_n$ is to ensure that these functions are c\` adl\` ag on $[-\pi, \pi)$, i.e. right-continuous and limits from left exist. 
    Plugging in $x = \widehat \mu_n^*$ into \eqref{eq:proof-optimalityBS_FrechetFunction} and \eqref{eq:proof-optimalityBS_FrechetFunction2}, which is well-defined since $\widehat \mu_n^* \pm \pi \neq X_j^*$ for each $j = 1, \dots, n$ \citep[Theorem 1(i)]{HH15}, asserts
    \begin{eqnarray}\label{eq:ConnectionBS_SampleMeanFrechetMean}
	  \overline X^*_n &=&  \big(1-2\pi f(-\pi)\big)\widehat \mu_n^*+ A(\widehat \mu_n^*) +  B_n(\widehat \mu_n^*) + C_n(\widehat \mu_n^*).
	\end{eqnarray}
	Likewise, it follows for the Fr\'echet sample mean $\widehat \mu_n$ of $X_1, \dots, X_n$ that  
	\begin{eqnarray}\label{eq:Connection_SampleMeanFrechetMean}
	  \overline X_n &=&  \big(1-2\pi f(-\pi)\big)\widehat \mu_n+ A(\widehat \mu_n) +  B_n(\widehat \mu_n), 
	\end{eqnarray}
    which yields
    \begin{equation}\begin{aligned}
	  \sqrt{n}\lrCurlyBrack{  \overline X^*_n - \overline X_n} =&  \sqrt{n}\Big( 1-2\pi f(-\pi)\Big)(\widehat \mu_n^* - \widehat \mu_n)+ \sqrt{n}A(\widehat \mu_n^*) - \sqrt{n}A(\widehat \mu_n)\\
	  & + \sqrt{n}B_n(\widehat \mu_n^*) - \sqrt{n} B_n(\widehat \mu_n) + \sqrt{n}C_n(\widehat \mu_n^*).  \end{aligned}\label{eq:BootstrapEmpiricalScaledQuantity}
	\end{equation}
	By \cite[Theorem 23.4]{vdV00} the bootstrap is consistent for the Euclidean sample mean, i.e.,  for $n \rightarrow \infty$ we have 
	\begin{equation}\label{eq:proof-clt_bootstrap_euclidean}
      \sup_{h \in \BL}\bigg|	\EE\lrBrack{h\lrCurlyBrack{\sqrt{n}\lrCurlyBrack{\overline X^*_n - \overline X_n}}\big| X_1, \dots, X_n}- \EE\lrBrack{h\lrCurlyBrack{
	  \sqrt{n}\,\overline X_n}}\bigg| \konvPStar 0. 
    \end{equation}
    Upon defining for $t \in \RR$ the function $ H(t) \coloneqq \min(2, |t|)$ we note for each $h \in \BL$ and any $t,t' \in \RR$ that the inequality $|h(t+t')-h(t)| \leq  H(t')$ holds since $h$ is bounded by one and Lipschitz with modulus one. Further, it holds that  $H(t + t') \leq H(t) + H(t')$. Hence, we assert by  \eqref{eq:Connection_SampleMeanFrechetMean} and \eqref{eq:BootstrapEmpiricalScaledQuantity} that
    \begin{align*}
    \bigg| h\Big(
	\sqrt{n}\,\overline X_n\Big) - h\Big(\sqrt{n}\big(1-2\pi f(-\pi)\big)\widehat \mu_n\Big)\bigg| &\leq H\lrCurlyBrack{\sqrt{n} A(\widehat \mu_n)} +  H\lrCurlyBrack{\sqrt{n} B_n(\widehat \mu_n)},\\
    \bigg|	h\Big(
	\sqrt{n}\lrCurlyBrack{  \overline X^*_n - \overline X_n}\Big) - h\Big(\sqrt{n}\big(1-2\pi f(-\pi)\big)\lrCurlyBrack{\widehat\mu_n^* - \widehat \mu_n}\Big)\bigg| \\
	\leq H\lrCurlyBrack{\sqrt{n} A(\widehat \mu_n^*)}+H\lrCurlyBrack{\sqrt{n} A(\widehat \mu_n)}+  H\lrCurlyBrack{\sqrt{n} B_n(\widehat \mu_n^*)}&+H\lrCurlyBrack{\sqrt{n} B_n(\widehat \mu_n)} + H\lrCurlyBrack{\sqrt{n} C_n(\widehat \mu_n^*)}.
    \end{align*} 
    Hence, we obtain by triangle inequality that 
    \begin{align*}
      & \sup_{h \in \BL}\bigg|	\EE\lrBrack{h\lrCurlyBrack{\sqrt{n}(1 - 2\pi f(-\pi)) \lrCurlyBrack{\widehat\mu_n^* - \widehat \mu_n}}\big| X_1, \dots, X_n}- \EE\lrBrack{h\lrCurlyBrack{\sqrt{n}(1 - 2\pi f(-\pi))\,\widehat \mu_n}}\bigg|\\
      = \,& \sup_{h \in \BL}\bigg|	\EE\lrBrack{h\lrCurlyBrack{\sqrt{n}(1 - 2\pi f(-\pi))\lrCurlyBrack{\widehat\mu_n^* - \widehat \mu_n}} - h\lrCurlyBrack{\sqrt{n}\lrCurlyBrack{\overline X_n^* - \overline X_n}}\big| X_1, \dots, X_n}\bigg|\\
      &+ \sup_{h \in \BL}\bigg|	\EE\lrBrack{h\lrCurlyBrack{\sqrt{n}\lrCurlyBrack{\overline X^*_n - \overline X_n}}\big| X_1, \dots, X_n}- \EE\lrBrack{h\lrCurlyBrack{
	  \sqrt{n}\,\overline X_n}}\bigg|\\
      & + \sup_{h \in \BL}\bigg|	\EE\Big[{h\lrCurlyBrack{\sqrt{n}(1 - 2\pi f(-\pi))\,\widehat \mu_n} - h\lrCurlyBrack{
	  \sqrt{n}\,\overline X_n}}\Big]\bigg|\\
	  \leq \,&\;\;\EE\lrBrack{H\big(\sqrt{n}A(\widehat \mu_n^*)\big) +H\big(\sqrt{n}A(\widehat \mu_n)\big) + H\big(\sqrt{n} B_n(\widehat \mu_n^*)\big)+H\big(\sqrt{n} B_n(\widehat \mu_n)\big) + H\big(\sqrt{n} C_n(\widehat \mu_n^*)\big) \Big| X_1, \dots, X_n}\\
	  &+  \sup_{h \in \BL}\bigg|	\EE\lrBrack{h\lrCurlyBrack{\sqrt{n}\lrCurlyBrack{\overline X^*_n - \overline X_n}}\big| X_1, \dots, X_n}- \EE\lrBrack{h\lrCurlyBrack{
	  \sqrt{n}\,\overline X_n}}\bigg|\\
	  &+\EE\lrBrack{H\big(\sqrt{n}A(\widehat \mu_n)\big)+H\big(\sqrt{n} B_n(\widehat \mu_n)\big) }.
    \end{align*}
    As for any function $h \in \BL$ it follows that $\tilde h(x) \coloneqq (1-2\pi f(-\pi))\cdot h\big(x/(1-2\pi f(-\pi)\big)$ is also bounded by one and Lipschitz with modulus one, and recalling \eqref{eq:proof-clt_bootstrap_euclidean}, it therefore suffices to show for $n \rightarrow \infty$ that 
    \begin{equation}
    \begin{aligned}
 	&\;\EE\lrBrack{H\big(\sqrt{n}A(\widehat \mu_n^*)\big) +H\big(\sqrt{n}A(\widehat \mu_n)\big) + H\big(\sqrt{n} B_n(\widehat \mu_n^*)\big)+H\big(\sqrt{n} B_n(\widehat \mu_n)\big)  + H\big(\sqrt{n} C_n(\widehat \mu_n^*)\big) \Big| X_1, \dots, X_n}\\
 	+&\; \EE\lrBrack{H\big(\sqrt{n}A(\widehat \mu_n)\big)+H\big(\sqrt{n} B_n(\widehat \mu_n)\big) } \konvPStar 0.
 	\end{aligned}
 	\label{eq:BS_Consistency_ErrorTerms1}
    \end{equation}
    The proof of the convergence in outer probability \eqref{eq:BS_Consistency_ErrorTerms1} is deferred to Proposition \ref{prop:ConvergenceToZero} in Supplement \ref{app:ProofBootstrapConsistency} and employs empirical process theory as well as the notion of Donsker classes.
\end{proof}

\begin{Rm} \label{rm:torus} These results extend at once to the $m$-torus $\TTT = \big(\SSS\big)^m$ equipped with the canonical product metric as the Fr\'echet mean on the $m$-torus is given by the vector of circular Fr\'echet means for each component of the torus. 

\end{Rm}

\subsection{Moment Convergence of Fr\'echet Means}

In order to employ the theory on the asymptotics of Fr\'echet sample means and bootstrap version for the formulation of Hotelling tests it is necessary to estimate the variance of the limit distribution. The following result guarantees that the na\"ive plug-in estimator for the covariance is indeed consistent. In fact, we prove that all moments of scaled Fr\'echet sample means and bootstrap variants converge to the corresponding moment of their limit distribution. 

\begin{Prop}\label{prop:UniformIntegrability}
  Under Assumption \ref{ass:antipodalDensityBounded} let $n\in \NN$ and consider i.i.d. random elements $X_1,\ldots,X_n\sim X$ on $\SSS$ with a measurable selection $\widehat{\mu}_n$ of Fr\'echet sample means.  Then for all $p \geq 1$ we have 
  $$\sup_{n\in \NN} \EE\left[ \left|\sqrt{n} d(\hat \mu_n, \mu)\right|^p \right] = \sup_{n\in \NN} \EE\left[ \left|\sqrt{n} \hat \mu_n\right|^p \right] < \infty. $$
  Further, for  $Z \sim \mathcal{N}\big(0,\EE[X^2]/(1-2\pi f(-\pi))^2\big)$ we have for any
  $p \geq 1 $ as $n$ tends to infinity that
  $$\EE\left[ \left( \sqrt{n} d(\hat \mu_n, \mu)\right)^p\right]=\EE\left[ \left( \sqrt{n} \hat \mu_n\right)^p\right] \rightarrow \EE[Z^p]. $$
\end{Prop} 
 
The proof is stated in Supplement \ref{app:MomentConvergence} and uses a moment convergence result for M-estimators relying on the theory by \cite{N10}. Notably, explicit bounds for $\EE\left[ \left|\sqrt{n} d(\hat \mu_n, \mu)\right|^p \right]$ and in case of more general $M$-estimators were derived by \cite{Schoetz2019}.
Moreover, similar to the consistency result on moments of the empirical Fr\'echet sample mean, the moments of bootstrap based Fr\'echet sample means are also consistent. 

\begin{Prop}\label{prop:UniformIntegrabilityBootstrap}
  Under Assumption \ref{ass:antipodalDensityBounded} let $n\in \NN$ and consider i.i.d. random elements $X_1,\ldots,X_n\sim X$ on $\SSS$ with a measurable selection $\widehat{\mu}_n$ of Fr\'echet sample means. Further, consider a bootstrap sample $X_1^*, \dots, X_n^*\iid \frac{1}{n}\sum_{i = 1}^{n}\delta_{X_i}$ with measurable selection $\widehat \mu_{n}^*$ of its Fr\'echet sample mean. Then, $$\sup_{n\in \NN} \EE\left[ \left|\sqrt{n} d(\widehat \mu_n^*, \widehat\mu_n)\right|^p \right] \leq \sup_{n\in \NN} \EE\left[ \left|\sqrt{n} (\widehat \mu_n^*- \widehat\mu_n)\right|^p \right] < \infty$$
  for all $p\geq 1$ where the expectation is taken with respect to samples $X_1, \dots, X_n\iid X$ and bootstrap based samples $X_1^*, \dots, X_n^* \iid n^{-1} \sum_{i= 1}^{n} \delta_{X_i}$. 
  Further, for  $Z \sim \mathcal{N}\big(0,\EE[X^2]/(1-2\pi f(-\pi))^2\big)$ we have for any
  $p \geq 1 $ as $n$ tends to infinity that
  $$ \EE\left[\left(\sqrt{n} (\widehat \mu_n^*- \widehat\mu_n)\right)^p \big| X_1, \dots, X_n \right] \konvPStar \EE[Z^p].$$
\end{Prop}

The proof is deferred to Supplement \ref{app:MomentConvergence} and relies on a general result for conditional moment convergence of bootstrap $M$-estimators by \cite{K11}. 
    
\section{One- and Two-Sample Tests for Fr\'echet Means under Finite Sample Smeariness}\label{scn:tests}

We begin with a brief review of the celebrated central limit theorem (CLT) by \cite{BP05} for a $m$-dimensional manifold $M$ and corresponding tests proposed therein. The CLT states that,  under suitable conditions (further clarified in \cite{BL17,EH19,EGHT19}), in a local chart $\phi : U \to \RR^m$, $U\subset M$, the fluctuation $\phi(\widehat{\mu}_n) - \phi(\mu)$ of the Fr\'echet sample mean $\widehat{\mu}_n$ about the Fr\'echet population mean $\mu$, rescaled with the square root of sample size $\sqrt{n}$, is asymptotically (for $n\to \infty$) Gaussian with zero mean and covariance given by
$$ \Sigma = H^{-1} C H^{-1}\,,$$
cf. (\ref{eq:BP-CLT}).
Here, $C$ is the population covariance of the gradient of the Fr\'echet function $F$ from (\ref{eq:population-mean}) at $\mu$ in the local chart and $H$ is twice the expected value of the Hessian of the Fr\'echet function at $\mu$ in the local chart. One of the above mentioned conditions is that $H$ be positive definite. In our language, this means that $\widehat{\mu}_n$ is nonsmeary (i.e., on the circle we have the situation of Theorem \ref{them:CLT_Circle} where $C= \EE[X^2]$ and $H=1-2\pi\, f(-\pi)$, as can be derived at once from by means of Equation \eqref{eq:Connection_SampleMeanFrechetMean}). 

For mutually independent samples $X_1,\ldots,X_n \iid X$ and $Y_1,\ldots,Y_n \iid Y$ with population Fr\'echet means $\mu^{(X)}$ and  $\mu^{(Y)}$, respectively, and $\mu_0 \in M$, consider the hypotheses
\[\begin{array}{lcr}
  H^1_0:& \mu^{(X)} = \mu_0 & \mbox{ for the one-sample test,}\\
  H^2_0:& \mu^{(X)} = \mu^{(Y)} & \mbox{ for the two-sample test.}
\end{array}\]

\paragraph{Quantile based tests.}
With a local chart $\phi :  U \to \RR^m$ where $U\subset M$ contains $\mu$ (for $H^1_0$), or $\mu^{(X)}$ and $\mu^{(Y)}$ (for $H^2_0$), respectively, \cite{BP05} thus suggest to consider the following test statistics,
\[\begin{array}{lcr}
  T^1 = &n\,\big(\phi(\widehat{\mu}_n^{(X)}) - \phi(\mu)\big)^T \widehat H_X \widehat C_X^{-1}\widehat H_X\big(\phi(\widehat{\mu}_n^{(X)} - \phi(\mu)\big)  & \mbox{ for the one-sample test,}\\
  T^2 = &(n+m)\,\big(\phi(\widehat{\mu}_n^{(X)}) - \phi(\widehat\mu_m^{(Y)})\big)^T \widehat H_{X,Y}  \widehat C_{X,Y}^{-1}\widehat H_{X,Y}\big(\phi(\widehat{\mu}_n^{(X)}) - \phi(\widehat\mu_m^{(Y)})\big) & \mbox{ for the two-sample test,}
\end{array}\]
and use $\chi^2_m$ as their asymptotic approximation under the respective null hypothesis. Here  $\widehat H_X, \widehat C_X$ or $\widehat H_{X,Y}, \widehat C_{X,Y}$ are the usual plugin estimators of $H$ and $C$ based on the first sample, or the pooled sample, respectively, cf. also \citet[Section 5.4.1]{BB12}.

\begin{Test}[\cite{BP05,BB12}]\label{tests:BP}
 For $0\leq \alpha\leq 1$ and the $1-\alpha$ quantile $\chi^2_{m,1-\alpha}$ of the $\chi^2_m$ distribution, reject at nominal level $\alpha$
 \begin{itemize}
  \item[(i)] $H_0^1$, if $T^1 \geq \chi^2_{k,1-\alpha}$,
  \item[(ii)] $H_0^2$, if $T^2 \geq \chi^2_{k,1-\alpha}$.  
 \end{itemize}

\end{Test}

Asymptotically, while both tests are independent of the chart chosen, in general, they do not keep the level promised. 

\begin{Prop}
Let $X,Y$ be random variables on $\SSS$ with unique Fr\'echet population means $\mu^{(X)},\mu^{(Y)}$, respectively, and where $X$ and $Y$ satisfy Assumption \ref{ass:antipodalDensityBounded}(ii) for respective densities $f^{(X)}, f^{(Y)}$. 
Under Type I FSS of $X$ or $Y$ the Tests \ref{tests:BP} have asymptotically a size strictly higher than the nominal size.
\end{Prop}

\begin{proof}
With the notation of Theorem \ref{them:CLT_Circle}, if $M=\SSS$ is the circle, $\mu_0 = 0$, or $\mu^{(X)} =0=\mu^{(Y)}$, respectively, and $\phi$ the identity on $(-\pi,\pi) \subset \SSS$, plugin estimators for  $H^{-1}CH^{-1}$ are $\frac{1}{n}\sum_{j=1}^nX_j^2$, or $\frac{1}{n+m}\sum_{j=1}^nX_j^2+ \frac{1}{n+m}\sum_{j=1}^mY_j^2$,  
whereas $H^{-1}CH^{-1} = \EE[X^2] \big(1-2\pi f(-\pi)\big)^{-2}$, or $H^{-1}CH^{-1}=\big(\delta \EE[X^2]\big(1-2\pi f^{(X)}(-\pi)\big)^{-2} + (1-\delta)\EE[Y^2]\big(1-2\pi f^{(Y)}(-\pi)\big)^{-2}\big)$, with $n/(n+m) \to \delta$, respectively, by Theorem \ref{them:CLT_Circle}.  Under Type I FSS Corollary \ref{cor:ConnectionFSSAndDensity} states that $f^{(X)}(-\pi)>0$ or $f^{(Y)}(-\pi)>0$. Hence, the plugin estimators are asymptotically strictly smaller than the variance of the limit law and thus the true asymptotic size of Tests \ref{tests:BP} is strictly higher than the nominal level.
\end{proof}

\begin{Rm}
On the circle, in case of Type II FSS, Tests \ref{tests:BP} also have a true size higher than their nominal size in the range where the variance modulation $\fm_n$ larger than $1$, in particular of the difference is substantial, as the following simulations in Section \ref{scn:simulations} show.
\end{Rm}

\paragraph{Bootstrap based tests.} Tests based on a bootstrap principle have also been proposed by \cite{BP05,BB12} where $\Sigma = H^{-1}CH^{-1}$ is estimated by bootstrapping from the samples.
As they have not provided details, in the following, we employ the bootstrap one- and two-sample tests by \cite{EH17}, for a given level $0 \leq \alpha \leq 1$.

Resample $B\in \NN$ times $n$-out-of-$n$ from the sample $X_1,\ldots,X_n$ and let $\mu^{(X),*}_b$ be the corresponding Fr\'echet sample means for $b=1,\ldots, B$. Mapping these sample means under $\phi$ to a Euclidean space yields the covariance estimate $\Sigma^{(X),*}_B$. From another round of resampling obtain new  $\mu^{(X),*}_b$, set 
$T^*_b = \big(\phi(\mu^{(X),*}_b) -\phi(\widehat\mu^{(X)}_n)\big)^T(\Sigma^{(X),*}_B)^{-1}  \big(\phi(\mu_b^{(X),*}) -\phi(\widehat\mu^{(X)}_n)\big)$, $b\in \{1,\ldots,B\}$ and
determine $e^*_{1-\alpha}$ such that
 \[ \frac{\sharp\{b \in \{1,\ldots,B\}: T^*_b\leq e_{1-\alpha}^*\}-1}{B}~ \leq 1-\alpha ~\leq ~\frac{\sharp\{b \in \{1,\ldots,B\}: T^*_b\leq e_{1-\alpha}^*\}}{B} \,.\]
 Further, set 
 $$ T^{1,*} = \big(\phi(\widehat\mu^{(X)}_n) -\phi(\mu_0)\big)^T(\Sigma^{(X),*}_B)^{-1}  \big(\phi(\widehat\mu^{(X)}_n) -\phi(\mu_0)\big)\,.$$

Similarly, using $m$-out-of-$m$ sampling with replacement from $Y_1,\ldots,Y_m$, obtain a bootstrap covariance estimate $\Sigma^{(Y),*}_B$. Now, set  
$A_B= \Sigma^{(X),*}_B+ \Sigma^{(Y),*}_B$. 
This choice of $A_B$, and not using the pooled variance, proves to be more robust, cf. \cite{HE_Handbook_2020}. 
Then, from another round of sampling obtain  $\mu^{(X),*}_b$ and  $\mu^{(Y),*}_b$, set $d^{(X),*}_b = \phi(\mu^{(X),*}_b) -\phi(\widehat\mu^{(X)}_n)$, $d^{(Y),*}_b = \phi(\mu^{(Y),*}_b) -\phi(\widehat\mu^{(Y)}_m)$, define 
$T^*_b = \big(d^{(X),*}_b -d^{(Y),*}_b \big)^TA^{-1}_B  \big(d^{(X),*}_b-d^{(Y),*}_b\big)$, $b\in \{1,\ldots,B\}$ and
determine $f^*_{1-\alpha}$ such that
\[ \frac{\sharp\{b \in \{1,\ldots,B\}: T^*_b\leq f_{1-\alpha}^*\}-1}{B}~ \leq 1-\alpha ~\leq ~\frac{\sharp\{b \in \{1,\ldots,B\}: T^*_b\leq f_{1-\alpha}^*\}}{B} \,.\]
Further, set 
$$ T^{2,*} = \big(\phi(\widehat\mu^{(X)}_n) -\phi(\widehat \mu^{(Y)}_m)\big)^T A^{-1}_B  \big(\phi(\widehat\mu^{(X)}_n) -\phi(\widehat \mu^{(Y)}_m)\big)\,.$$

\begin{Tests}[Bootstrap Based]\label{tests:bootstrap}
 With the notation above, for $0\leq \alpha\leq 1$, reject at level $\alpha$
 \begin{itemize}
  \item[(i)] $H_0^1$, if $T^{1,*} \geq e^*_{1-\alpha}$,
  \item[(ii)] $H_0^2$, if $T^{2,*} \geq f^*_{1-\alpha}$.  
 \end{itemize}
\end{Tests}

Again, asymptotically, both tests are independent of the chart chosen. 

\begin{Prop}
    Consider mutually independent samples $X_1, \dots, X_n \iid X$ and $Y_1, \dots, Y_n \iid Y$ on $\SSS$, where  $X$ and $Y$ have unique Fr\'echet population means $\mu^{(X)}=0$ and $\mu^{(Y)}$. Further, assume that $X$ and $Y$ are not concentrated at $\mu^{(X)}$ and $\mu^{(Y)}$, respectively, and fulfill Assumption \ref{ass:antipodalDensityBounded}(ii) for respective densities $f^{(X)}, f^{(Y)}$. Then it follows for $0< \alpha< 1$ as $n,B\rightarrow \infty$
    \begin{enumerate}
        \item[(i)]  under $\mu^{(X)} = \mu_0$ that $\mathbb{P}\left( T^{1,*} \geq e^*_{1-\alpha}\Big| X_1, \dots, X_n\right)\konvPStar \alpha,$ \\
        \item[(ii)] and under  $\mu^{(X)} \neq \mu_0$ that $\mathbb{P}\left( T^{1,*} \geq e^*_{1-\alpha}\Big| X_1, \dots, X_n\right)\konvPStar 1.$ 
        \item[(iii)]  Moreover,  under $\mu^{(X)} = \mu^{(Y)}$ that $\mathbb{P}\left( T^{2,*} \geq  f^*_{1-\alpha}\Big| X_1, \dots, X_n\right)\konvPStar \alpha, $\\
        \item[(iv)] whereas under $\mu^{(X)} \neq \mu^{(Y)}$ it holds $\mathbb{P}\left( T^{2,*}\geq  f^*_{1-\alpha}\Big| X_1, \dots, X_n\right)\konvPStar 1. $
    \end{enumerate}
\end{Prop}

This shows that the bootstrap based test is indeed consistent under FSS of Type I and II for circular distributions which have a density near the antipode of the Fr\'echet population mean.

\begin{proof}
  We only show Assertions (i) and (ii), the proof for the remaining assertions is analogous. Consider $\phi$ the identity on $(-\pi, \pi]$ and define for $t>0$ the Lipschitz function $$\Psi^{(t)}\colon \RR\rightarrow [-t,t], \Psi^{(t)}(x) \coloneqq \sign(x) \min(|x|, t).$$ Suppose $\mu^{(X)} = \mu_0=0$, then it follows for $n \rightarrow \infty$ that $$\frac{1}{ \EE[n(\widehat\mu_n^{(X)})^2]} \rightarrow \frac{(1-2\pi f(\overline{\mu}^{(X)}))^2}{\EE[X^2]},$$ cf. Proposition \ref{prop:UniformIntegrability}. 
  This asserts that the Lipschitz function $g^{(t)}_{n}(x) \coloneqq \Psi^{(t)}(x^2(\EE[n(\widehat\mu_n^{(X)})^2])^{-1})$ converges uniformly for $n\rightarrow\infty$ to $g^{(t)}(x) \coloneqq \Psi^{(t)}(x^2\frac{(1-2\pi f(\overline{\mu}^{(X)}))^2}{\EE[X^2]} )$. Likewise, it holds for 
  $n\rightarrow \infty$ that $$\frac{1}{\EE[n\Sigma^{(X),*}_B|X_1, \dots, X_n]}\konvPStar \frac{(1-2\pi f(\overline{\mu}^{(X)}))^2}{\EE[X^2]},$$cf. second assertion from Proposition \ref{prop:UniformIntegrabilityBootstrap}.
  Moreover, as a consequence of the first part of Proposition \ref{prop:UniformIntegrabilityBootstrap} it follows using Markov's inequality that  $\var[\sqrt{n}(\widehat \mu_n^{(X),*}-\widehat \mu_n^{(X)})|X_1, \dots, X_n]$ is stochastically bounded (see Relation \eqref{eq:OPConditionalExp} in Supplement C), therefore asserting that  $$\var[n\Sigma^{(X),*}_B|X_1, \dots, X_n]=O_P(1/B).$$
  Hence,  for $t>0$ the Lipschitz function $g^{*(t)}_{n,B}(x) \coloneqq \Psi^{(t)}(x^2 (n\Sigma^{(X),*}_B)^{-1}))$  converges uniformly for $n, B\rightarrow \infty$ conditional on $X_1, \dots, X_n$ in outer probability to $g^{(t)}$. This yields by Theorem \ref{them:BS_CLT_Circle} for $n, B \rightarrow \infty$ that $$  \sup_{h\in \BL}\bigg|	\EE\lrBrack{h\circ g_{n,B}^{*(t)} \lrCurlyBrack{\sqrt{n}\lrCurlyBrack{\widehat\mu_n^* - \widehat \mu_n}}\big| X_1, \dots, X_n}- \EE\lrBrack{h\circ g_n^{(t)}\lrCurlyBrack{\sqrt{n}\,\widehat \mu_n}}\bigg|\konvPStar 0. $$
  Notably, it holds that $g_{n,B}^{*(t)}\lrCurlyBrack{\sqrt{n}\lrCurlyBrack{\widehat\mu_n^* - \widehat \mu_n}} = \Psi^{(t)}(T^{1,*})$ and $g_n^{(t)}\lrCurlyBrack{\sqrt{n}\,\widehat \mu_n} = \Psi^{(t)}((\widehat \mu_n)^2/\EE[(\widehat\mu_n^{(X)})^2])$.
  Choosing $t$ sufficiently large depending on $\alpha\in (0,1)$ concludes for $n, B \rightarrow \infty$ under $\mu^{(X)} = \mu_0$ that the bootstrap-based quantity $e^*_{1-\alpha}$ is an asymptotically consistent estimator for the $(1-\alpha)$-quantile of $(\widehat \mu_n)^2/\EE[(\widehat\mu_n^{(X)})^2]$ and thus Assertion (i) holds. More precisely, since $(\widehat \mu_n)^2/\EE[(\widehat\mu_n^{(X)})^2]$ asymptotically follows a $\chi^2_1$-distribution (Theorem \ref{them:CLT_Circle}) the quantity $e^*_{1-\alpha}$ given $X_1, \dots, X_n$ tends in outer probability for $n,B\rightarrow\infty$ to the respective $(1-\alpha)$ quantile of the $\chi^2_1$ distribution. 
 
  In particular, for Assertion (ii) we note that $\Sigma^{(X),*}_B$ converges in outer probability conditioned on $X_1, \dots, X_n$ to zero, whereas by assumption $$\widehat\mu^{(X)}_n -\mu_0 \xrightarrow{a.s.} \mu^{(X)} - \mu_0 \neq 0,$$  cf. \cite{Z77}. Hence, conditioned on $X_1, \dots, X_n$ the test statistic  $T^{1,*}$ diverges in outer probability to infinity and the assertion follows.
\end{proof}

\begin{Rm}\label{rmk:bootstrap-test} Simulations in Section \ref{scn:simulations} depicted in Figures \ref{fig:OneSampleTestSimulation} and \ref{fig:TwoSampleTestSimulation} show that the Tests \ref{tests:bootstrap} keep the nominal level $\alpha = 0.05$ fairly well, in particular for Type I FSS. Upon (very) close inspection, for Type II FSS, the Tests  \ref{tests:bootstrap} may be slightly too conservative, for Type I FSS too liberal. Investigating this effect and correcting for it is left for future research and beyond the scope of this paper.
\end{Rm}

\begin{Rm}
Under uniqueness of means, the bootstrap based test is also asymptotically consistent on tori if the covariance of $X$ (one-sample) or the sum of covariances of $X$ and $Y$ (two-samples), respectively, is non-singular and each marginal distribution has a density near the antipode fulfilling Assumption \ref{ass:antipodalDensityBounded}(ii). This follows from the bootstrap consistency of the Fr\'echet sample mean on tori (Theorem \ref{them:BS_CLT_Circle} and Remark \ref{rm:torus}) and conditional convergence in outer probability of the covariance estimators $\Sigma_B^{(X),*}$ and $A_B$ for $\phi(\widehat\mu_n^{(X)})$ and $\phi(\widehat\mu_n^{(X)}) - \phi(\widehat\mu_m^{(Y)})$, respectively, as $n, B \rightarrow \infty$ to the corresponding population covariance quantities.
\end{Rm}

\section{Simulations}\label{scn:simulations}

To assess the performance of the quantile-based and bootstrap-based tests under the presence of FSS we consider i.i.d. samples of  some nonaltered (Type I FSS) von Mises mixtures, introduced in (\ref{eq:von-Mises}) and denoted by $\Prb^{\kappa, \beta, \lambda}_{vMm}$, and altered (Type II FSS) von Mises mixtures, introduced in (\ref{eq:FvM-modified}) and denoted by  $\Prb^{\kappa, \beta, \lambda,r}_{vMm}$. By Theorem \ref{thm:CirclS-FSS} all of the $\Prb_{vMm}^{\kappa, \beta, \lambda, r}$ with unique mean $\mu=0$ and $r < \pi/2 - \eps$ for some $\eps>0$ are FSS if they are not smeary themselves. 
For the following simulations we considered parameters as described in Table \ref{tab:ParametersVonMises}. All of them give unique means at $\mu=0$, none of which is smeary. Only the nearby $\Prb_{vMm}^{3, 0.5, \lambda, 0}$ is smeary for  $\lambda\approx 0.8683$, (following the notation of \cite{EH19}, the order of smeariness is equal to $2$, according to Theorem 3(ii) in \cite{HH15}).

\begin{figure}[H]\label{OneSampleTest:fig}
  \includegraphics[width=0.95\textwidth]{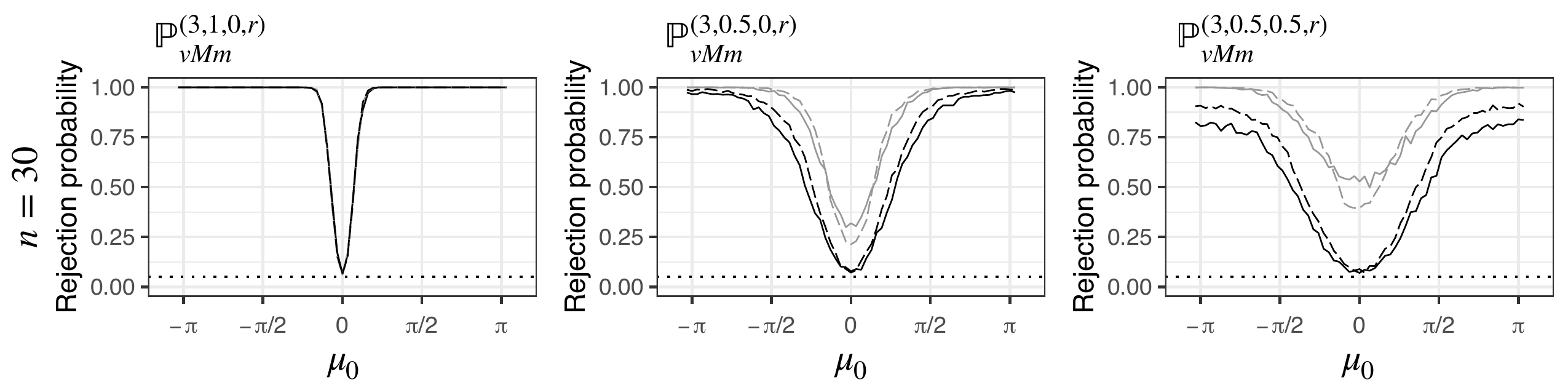}
  \includegraphics[width=0.95\textwidth]{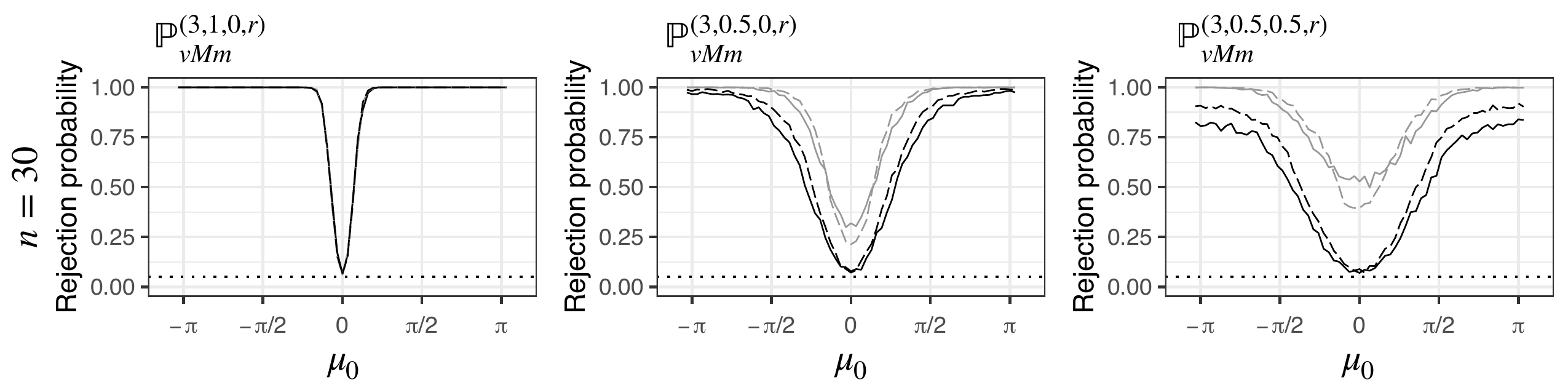}
  \includegraphics[width=0.95\textwidth]{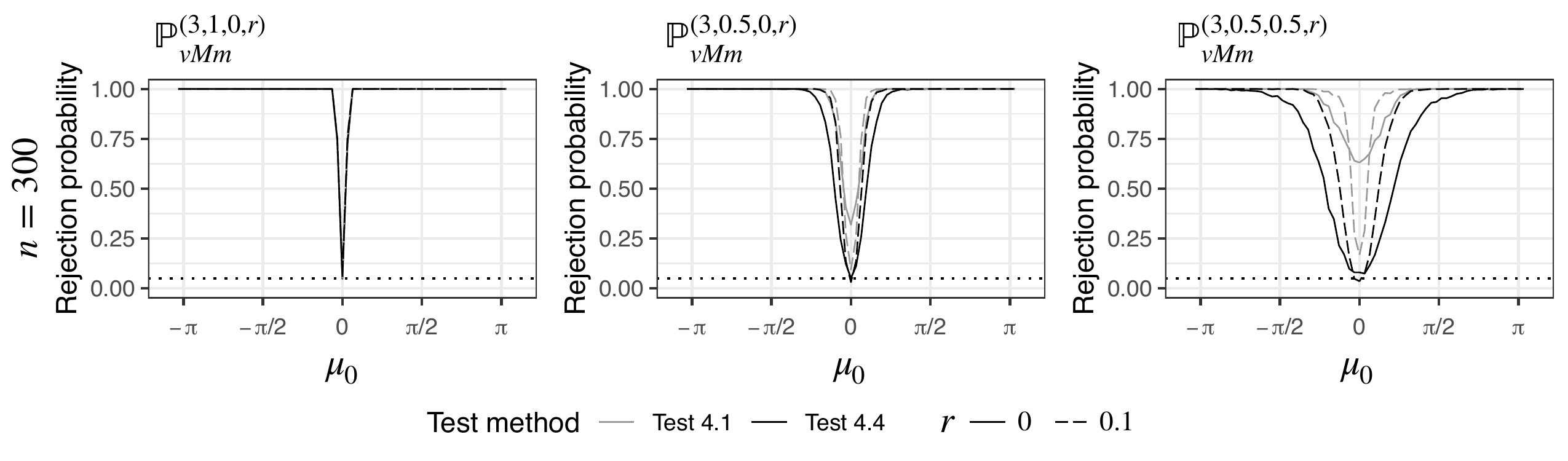}
  \caption{\it Rejection probabilities for $H_0^1$: $\mu=\mu_0$ for varying $\mu_0\in [-\pi,\pi)$ under quantile based tests (Test \ref{tests:BP}, gray) and bootstrap based tests for $B = 1000$ (Test \ref{tests:bootstrap}, black) at nominal level 5\% (dotted horizontal) based on 1000 simulation runs with one sample of size $n=30$ (top row), $n=100$ (middle row), and $n=300$ (bottom row). 
  The solid lines represent samples which were generated from mixed von Mises distributions $\Prb^{\kappa, \beta, \lambda,r}_{vMm}$, i.e., $r=0$. The dashed lines correspond to samples from $\Prb^{\kappa, \beta, \lambda,r}_{vMm}$ where all elements closer to $-\pi$ than $r=0.1$ were mirrored.  Table \ref{tab:ParametersVonMises} gives an overview of  parameters.
  }\label{fig:OneSampleTestSimulation}
\end{figure}

\begin{figure}[H]\label{TwoSample_SameDistr:fig}
  \centering
  \includegraphics[width=0.9\textwidth]{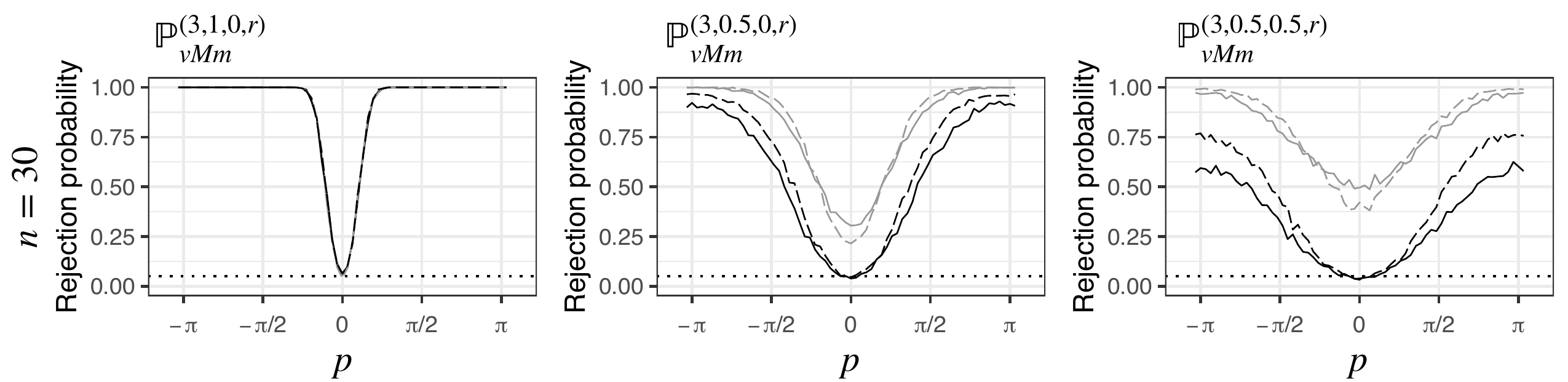}
    \includegraphics[width=0.9\textwidth]{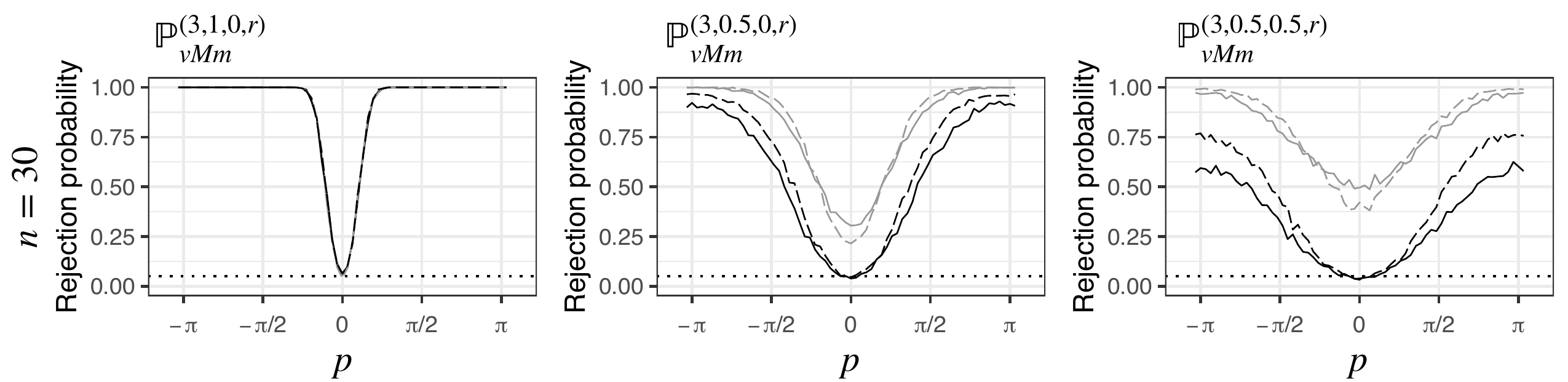}
  \includegraphics[width=0.9\textwidth]{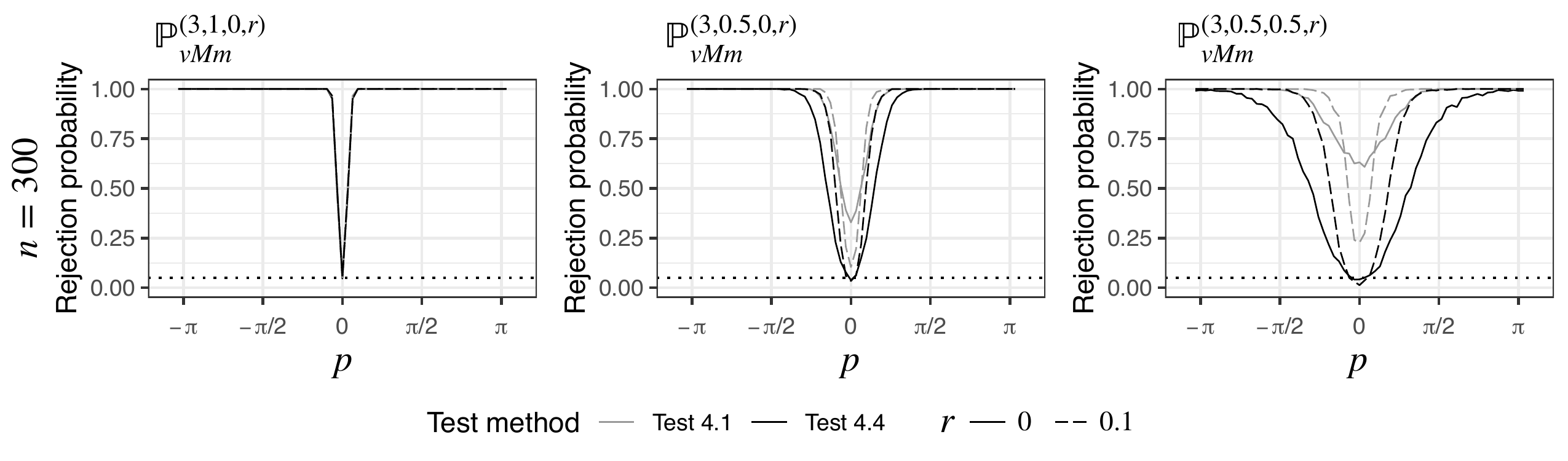}
  \caption{\it Rejection probabilities for $H_0^2$ over angle $p\in [-\pi,\pi)$ between $\mu^{(X)}$ and $\mu^{(Y)}$ under quantile based test (Test \ref{tests:BP}, gray) and bootstrap based test for $B = 1000$ (Test \ref{tests:bootstrap}, black) at nominal level 5\% (dotted horizontal) based on 1000 simulation runs  with two samples from the same distribution of size $n=30$ (top row), $n=100$ (middle row), and $n=300$ (bottom row) but where the latter sample is rotated by $p$. 
 The solid lines represent samples which were generated from mixed von Mises distributions $\Prb^{\kappa, \beta, \lambda,r}_{vMm}$, i.e., $r=0$. The dashed lines correspond to samples from $\Prb^{\kappa, \beta, \lambda,r}_{vMm}$ where all elements closer to $-\pi$ than $r=0.1$ were mirrored. 
  }
  \label{fig:TwoSampleTestSimulation}
\end{figure}

In Figures \ref{fig:OneSampleTestSimulation} and \ref{fig:TwoSampleTestSimulation} we compare the power functions at nominal level $\alpha = 0.05$ of the quantile tests (gray lines, Test \ref{tests:BP}) to the power functions of the bootstrap tests (black lines, Test \ref{tests:bootstrap}). For the one-sample tests (Figure \ref{fig:OneSampleTestSimulation}) we consider simulations of $\Prb_{vMm}^{\kappa, \beta, \lambda, r}$ and rotated null hypotheses $\mu_0 \in [-\pi,\pi)$, for the two sample-tests (Figure \ref{fig:TwoSampleTestSimulation}) we consider two simulations of $\Prb_{vMm}^{\kappa, \beta, \lambda, r}$ that are rotated with respect to one another by the angle $p \in [-\pi,\pi)$. Random variables from von Mises distributions were generated using the \textsf{R}-package \emph{circular} \citep{L17R_Pack}.

\begin{table}[H]
\centering
\begin{tabular}{|r|c|c|c|}
\hline
$\kappa$ & 3 & 3 & 3 \\
$\beta$ & 1 & 1/2 & 1/2 \\
$\lambda$ & 0 & 0 & 1/2 \\
\hline
$r$ & \textcolor{white}{0.}0 \;\vline \; 0.1 &  \textcolor{white}{0.}0\; \vline \; 0.1 & \textcolor{white}{0.0}0\; \vline \; 0.1  \\
\hline
$\fm_{30}$\; & 1.0 \;\vline \; 1.0 & 3.7 \;\vline\; 3.0 & \textcolor{white}{0}7.2 \;\vline \; 5.4\\
$\fm_{100}$ & 1.0 \;\vline\; 1.0 & 4.2 \;\vline\; 2.7 & 11.6 \;\vline\; 6.6 \\
$\fm_{300}$ & 1.0 \;\vline\; 1.0 & 4.4 \;\vline\; 2.0 & 15.9 \;\vline\; 5.6 \\ \hline
\end{tabular}
\caption{\it Parameters of $\Prb_{vMm}^{\kappa, \beta, \lambda, r}$ with variance modulation $\fm_n$ of distributions considered for simulations on performance of Tests \ref{tests:BP} and \ref{tests:bootstrap} and sample sizes $n \in \{30, 100,300\}$.}
\label{tab:ParametersVonMises}
\end{table}

With increasing scale of FSS (see Table \ref{tab:ParametersVonMises}) we see that the quantile tests (gray) become more and more liberal while the bootstrap tests (black) maintain the correct level. 
In particular, the quantile based tests perform poorly in the presence of considerable Type I FSS, in consequence of the scaled variance of intrinsic sample means $n \EE[d(\widehat{\mu}_n, \mu)^2]$ being larger than the Euclidean variance $\sigma^2$. 
Upon very close inspection, in case of Type II FSS (dashed lines, $r>0$) we see that the bootstrap tests may be slightly too conservative. Conversely, in case of Type I FSS (solid lines, $r=0$) the bootstrap tests may be slightly to liberal. Both effects may be due to a systematic bias, cf. Remark \ref{rmk:bootstrap-test}.

\begin{figure}[H]\label{Winddata_FrechetMeans:fig}
\includegraphics[width = 0.95\textwidth]{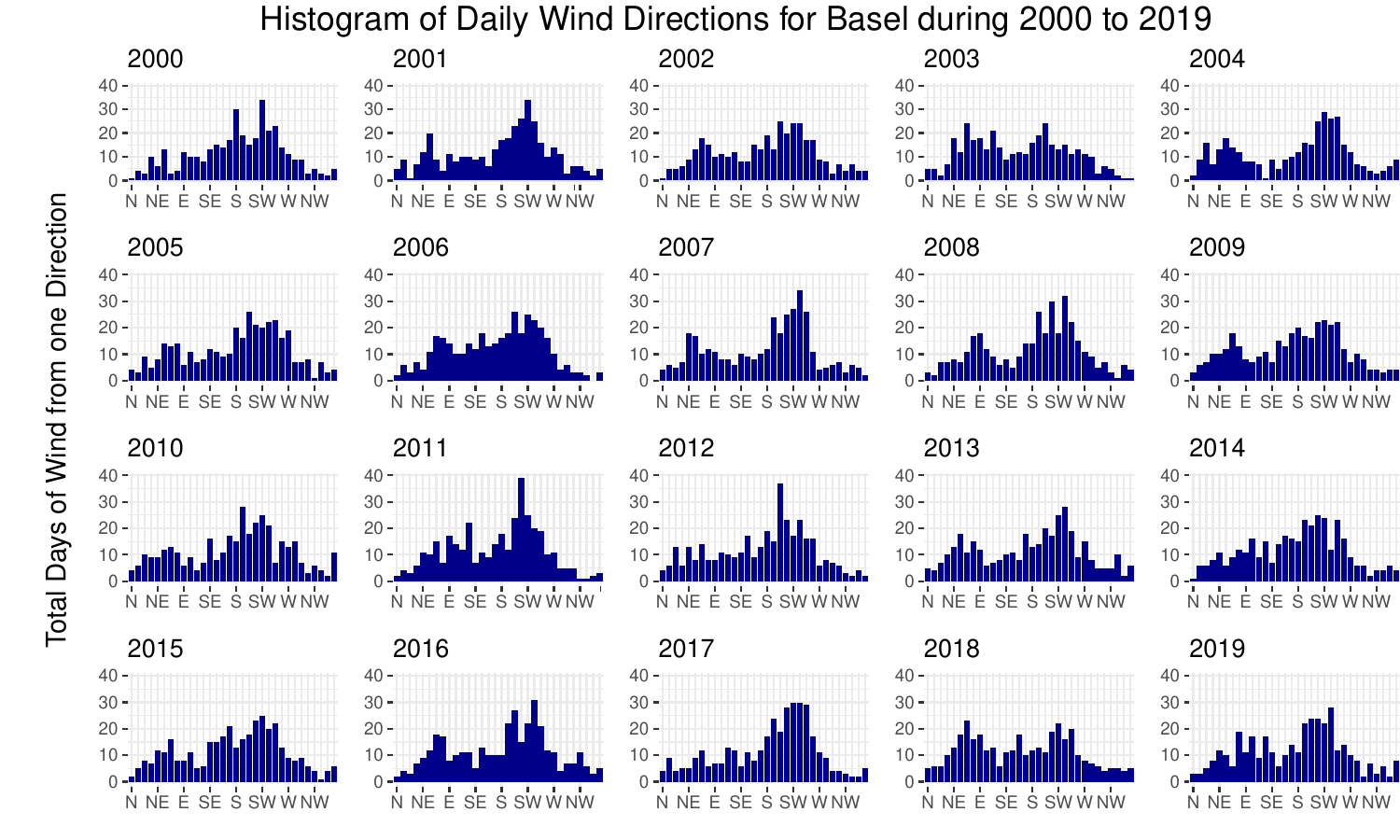}
\includegraphics[width = 0.95\textwidth]{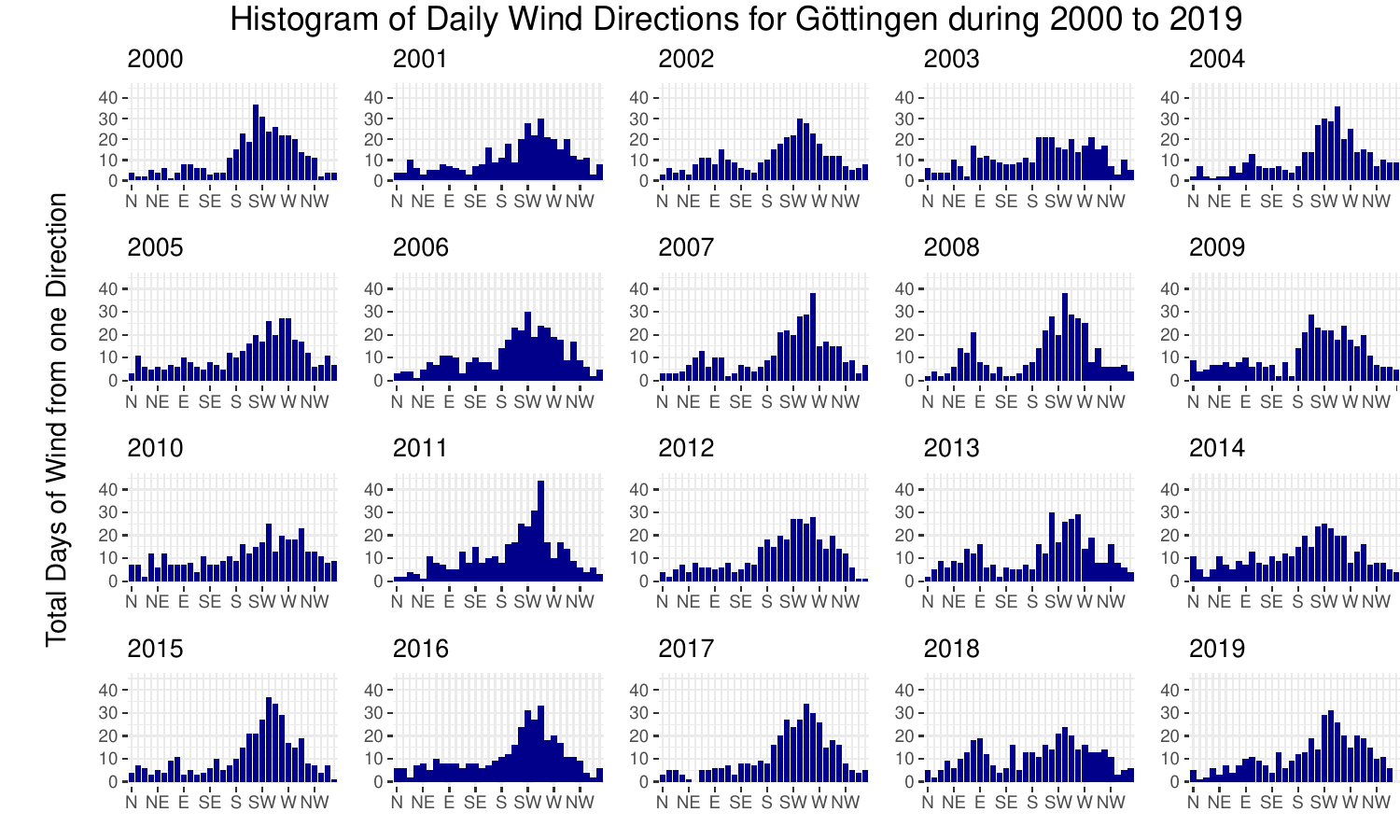}
  \caption{\it Histograms of Fr\'echet means of daily wind direction data by \cite{meteoblue} for Basel (top) and G\"ottingen (bottom) for the years $2000$ to $2019$. The $x$-axis is divided into 32 segments with labels  $N:$ ``North'', $E:$ ``East'', $S:$ ``South'', and $W:$ ``West'', indicating the average direction of wind origin.  \label{fig:wind-data}
  }
\end{figure}

\section{Assessing Significant Change of Wind Direction}\label{scn:wind}

In application of our methods dealing with FSS, we analyze wind data from Basel and G\"ottingen (the city of the authors' institution) provided  
by \cite{meteoblue}. For our purpose, we consider daily Fr\'echet mean wind directions for the years 2000 to 2019 giving for each city 20 samples of two-dimensional circular data of size $n= 365$. The respective daily wind directions are illustrated in Figure~\ref{fig:wind-data}.  To assess a possible effect of climate change, we test for a significant change in wind direction.

For each of these 40 samples we computed the estimated variance modulation $\fmnhat$ from \eqref{eq:ScaleFSS} with $B= 1000$, cf. Table \ref{tab:ScaleFSSWinddata}. Remarkably, for both cities all of the scales of FSS statistically indicate presence of FSS (Test \ref{tests:bootstrap} rejects the absence of FSS with the lowest $p = B^{-1}$ possible) except for G\"ottingen in the year 2017.

\begin{table}[H]
\centering
\begin{tabular}{|l||r|r|r|r|r|r|r|r|r|r|}
\hline
Year           & 2000  & 2001  & 2002  & 2003   & 2004   & 2005  & 2006  & 2007   & 2008  & 2009  \\ \hline
Basel      & 1.601 & 4.067 & 1.542 & 1.197  & 32.695 & 2.683 & 1.845 & 2.096  & 1.676 & 2.344 \\
G\"ottingen & 3.317 & 2.734 & 2.760 & 3.065  & 1.492  & 4.171 & 1.605 & 15.493 & 3.861 & 8.796 \\ \hline \hline
 Year          & 2010  & 2011  & 2012  & 2013   & 2014   & 2015  & 2016  & 2017   & 2018  & 2019  \\ \hline
Basel      & 5.010 & 1.775 & 1.466 & 3.211  & 1.491  & 1.741 & 1.705 & 5.229  & 2.694 & 2.481 \\
G\"ottingen & 8.003 & 1.685 & 4.457 & 36.365 & 3.446  & 3.305 & 5.636 & 1.037  & 1.804 & 2.592 \\ \hline
 
\end{tabular}
\caption{\it Empirical variance modulation $\fmnhat$ for $n = 365$ from \eqref{eq:ScaleFSS} for daily Fr\'echet mean wind directions in Basel and G\"ottingen for the years 2000 to 2019. All of these values indicate presence of FSS (Test \ref{tests:bootstrap} rejects absence of FSS with the lowest p-value possible: $p = 10^{-3}$) except for G\"ottingen in the year 2017.}
\label{tab:ScaleFSSWinddata}
\end{table}

\begin{figure}[H]
  \centering
  \includegraphics[width = 0.9\textwidth , trim = 0 0 0 300, clip]{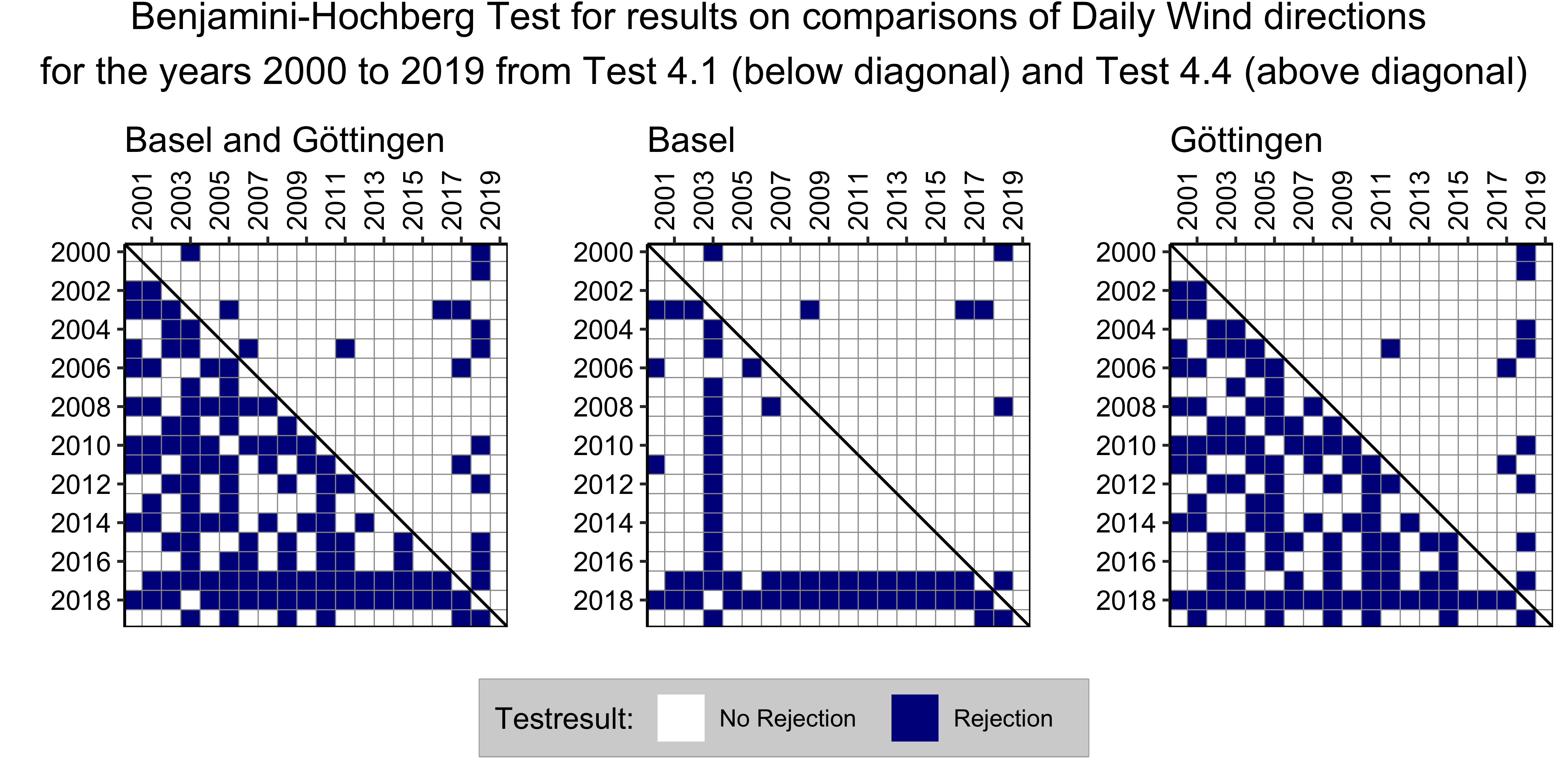}
  \vspace{-0.1cm}
  \caption{\it Benjamini-Hochberg corrected test results (above diagonals: Test \ref{tests:bootstrap} with $B = 1000$, below diagonals: Test \ref{tests:BP}) of comparisons of daily Fr\'echet means of wind direction data for the years 2000 to 2019 for Basel and G\"ottingen on $\mathbb T^2$ (left), Basel on $\SSS$ (middle), and G\"ottingen on $\SSS$ (right) for a significance level of 5\%.
   \label{fig:wind-BH-tests}}
\end{figure}

In consequence, we expect that the quantile based two-sample Test \ref{tests:BP} will feature a considerably high error of the first kind as compared to the bootstrap based Test \ref{tests:bootstrap}. For each series of $20\cdot 19 /2 = 190$ tests at nominal level $\alpha=0.05$ we performed a Benjamini-Hochberg correction, cf. Figure \ref{fig:wind-BH-tests}. 

The first series of tests is performed on the two-torus $\mathbb T^2 = \SSS\times \SSS$ for Basel and G\"ottingen jointly (left panel of Figure \ref{fig:wind-BH-tests}). 
According to the quantile Test \ref{tests:BP}, a majority of years seem to be significantly different from a great number of others.  
Applying the bootstrap Test \ref{tests:bootstrap} for $B = 1000$ we see this ``noise'' disappearing, leaving only the years 2003, 2005, 2017 and 2018 as moderately exceptional, where 2018 appears to be more exceptional  than the other three. 

Looking only at Basel (now testing only on $\SSS$, middle panel of  Figure \ref{fig:wind-BH-tests}), Test \ref{tests:BP} seems to suggest very clearly that the years 2003, 2017 and 2018 are exceptional. Test \ref{tests:bootstrap}, however, clarifies the picture: 2017 is not exceptional and 2003 and 2018 are only moderately exceptional. 

Comparison with G\"ottingen only  (again testing on $\SSS$, right panel of  Figure \ref{fig:wind-BH-tests}) shows the ``noise'' from Test \ref{tests:BP} is rooted in G\"ottingen. Again, Test \ref{tests:bootstrap} clarifies the picture: only the years 2005, 2017 are mildly exceptional. The year 2018 remains more prominently exceptional. The year 2010, which appeared almost as strongly exceptional as year 2018 for Test \ref{tests:BP}, is no longer exceptional for Test \ref{tests:bootstrap}. Notably, 2017 is also exceptional through absence of FSS, cf. Table \ref{tab:ScaleFSSWinddata}. 

In conclusion, using the bootstrap test which preserves the nominal level, the years 2003 and 2018 appear exceptional along with 2005 and 2017. 
These findings fall well into the climatological context of central European heat waves linked to exceptional wind constellations (\cite{kornhuber2019extreme} identify a \emph{recurrent wave-7} wind pattern for the years 2003, 2006, 2015 and 2018). While the 2003 heat wave occurred in most of Europe, the 2018 heat wave manifested in a climatic dipole: hot and dry north of the Alps, comparably cool and moist across large parts of Mediterranean, cf. \cite{buras2019quantifying}. This is in agreement with our finding that the wind anomaly of 2018 was more prominent for G\"ottingen (in the northern part of Germany), rather than for Basel (edging the southern border of Germany) which is barely north of the Alps. 

In a debate quantifying climate change, its  anthropogenic component and future costs, linking to changes of wind patterns (e.g. \cite{mcinnes2011global}), our new inferential tools for cyclic data presented here, warrant a more detailed application in future work.

\section{Discussion and Outlook}\label{scn:outlook}

In this contribution, we have investigated two manifolds with codimension one cut loci, namely circles and tori and found FSS manifesting in two different types. We expect similar findings for other manifolds with codimension one cut loci, such as real projective spaces, modeling projective shapes, say, as in \cite{MP05,HotzKelmaKent2016}. We conjecture that this is, using the language of \cite{E19}, a consequence of \emph{topological smeariness}. On manifolds with higher codimension cut loci, in the language of \cite{E19}, for instance on arbitrary spheres, there is the different phenomenon of \emph{geometrical smeariness}. We conjecture that this leads to Type I FSS only.

Moreover, we have shown that the proposed bootstrap tests are asymptotically consistent and demonstrated in simulations that they preserve the level fairly well for reasonable sample sizes. Inspired by recent findings by \cite{ZH20} on the non-asymptotic accuracy of the bootstrap for Euclidean sample means for the finite sample regime, we deem it possible to derive similar non-asymptotic results for the bootstrap Fr\'echet sample mean. This issue is beyond the scope of this work and left for future work.

\section*{Acknowledgement}
    S. Hundrieser acknowledges funding by the Deutsche Forschungsgemeinschaft (DFG, German Research Foundation) under Germany's Excellence Strategy - EXC 2067/1- 390729940. 
    The authors B. Eltzner and S. Huckemann gratefully acknowledge support by the DFG HU 1575/7, the DFG CRC 1456, the Niedersachsen Vorab of the Volkswagen Foundation and the Felix-Bernstein Insitute for Mathematical Stochastics in the Biosciences. The authors also thank meteoblue AG for providing the wind data sets and Prof. Nishiyama for providing details to his work on moment convergence of $M$-estimators.

\bibliographystyle{chicago.bst}
\bibliography{bibliography}

\begin{thebibliography}{}

\bibitem[\protect\citeauthoryear{Afsari}{Afsari}{2009}]{Afsari09}
Afsari, B. (2009).
\newblock {\em Means and averaging on Riemannian manifolds}.
\newblock University of Maryland.

\bibitem[\protect\citeauthoryear{Barden, Le, and Owen}{Barden
  et~al.}{2013}]{BLO13}
Barden, D., H.~Le, and M.~Owen (2013).
\newblock Central limit theorems for {F}r{\'e}chet means in the space of
  phylogenetic trees.
\newblock {\em Electron. J. Probab\/}~{\em 18\/}(25), 1--25.

\bibitem[\protect\citeauthoryear{Barden, Le, and Owen}{Barden
  et~al.}{2018}]{BLO18}
Barden, D., H.~Le, and M.~Owen (2018).
\newblock Limiting behaviour of fr{\'e}chet means in the space of phylogenetic
  trees.
\newblock {\em Annals of the Institute of Statistical Mathematics\/}~{\em
  70\/}(1), 99--129.

\bibitem[\protect\citeauthoryear{Bhattacharya and Bhattacharya}{Bhattacharya
  and Bhattacharya}{2012}]{BB12}
Bhattacharya, A. and R.~Bhattacharya (2012).
\newblock {\em Nonparametric Inference on Manifolds}.
\newblock Cambridge University Press.

\bibitem[\protect\citeauthoryear{Bhattacharya and Lin}{Bhattacharya and
  Lin}{2017}]{BL17}
Bhattacharya, R. and L.~Lin (2017).
\newblock Omnibus {CLT} for {F}r\'echet means and nonparametric inference on
  non-{Euclidean} spaces.
\newblock {\em Proceedings of the American Mathematical Society\/}~{\em
  145\/}(1), 413--428.

\bibitem[\protect\citeauthoryear{Bhattacharya and Patrangenaru}{Bhattacharya
  and Patrangenaru}{2005}]{BP05}
Bhattacharya, R.~N. and V.~Patrangenaru (2005).
\newblock Large sample theory of intrinsic and extrinsic sample means on
  manifolds {II}.
\newblock {\em The Annals of Statistics\/}~{\em 33\/}(3), 1225--1259.

\bibitem[\protect\citeauthoryear{Billera, Holmes, and Vogtmann}{Billera
  et~al.}{2001}]{BHV01}
Billera, L., S.~Holmes, and K.~Vogtmann (2001).
\newblock Geometry of the space of phylogenetic trees.
\newblock {\em Advances in Applied Mathematics\/}~{\em 27\/}(4), 733--767.

\bibitem[\protect\citeauthoryear{Billingsley}{Billingsley}{1995}]{Billingsley1995}
Billingsley, P. (1995).
\newblock {\em Probability and measure}, Volume 939.
\newblock John Wiley \& Sons.

\bibitem[\protect\citeauthoryear{Buras, Rammig, and Zang}{Buras
  et~al.}{2020}]{buras2019quantifying}
Buras, A., A.~Rammig, and C.~S. Zang (2020).
\newblock Quantifying impacts of the drought 2018 on {E}uropean ecosystems in
  comparison to 2003.
\newblock {\em Biogeosciences\/}~{\em 17\/}(6), 1655--1672.

\bibitem[\protect\citeauthoryear{Dubey and M{\"u}ller}{Dubey and
  M{\"u}ller}{2019}]{dubey2019frechet}
Dubey, P. and H.-G. M{\"u}ller (2019).
\newblock Fr{\'e}chet analysis of variance for random objects.
\newblock {\em Biometrika\/}~{\em 106\/}(4), 803--821.

\bibitem[\protect\citeauthoryear{{Eltzner}}{{Eltzner}}{2019}]{E19}
{Eltzner}, B. (2019).
\newblock {Geometrical Smeariness -- A new Phenomenon of {Fr\'echet} Means}.
\newblock {\em arXiv e-prints 1908.04233; to appear in Bernoulli\/}.

\bibitem[\protect\citeauthoryear{Eltzner, Galaz-Garcia, Huckemann, and
  Tuschmann}{Eltzner et~al.}{2021}]{EGHT19}
Eltzner, B., F.~Galaz-Garcia, S.~F. Huckemann, and W.~Tuschmann (2021).
\newblock Stability of the cut locus and a central limit theorem for
  {F}r{\'e}chet means of {R}iemannian manifolds.
\newblock {\em Proceedings of the American Mathematical Society\/}~{\em
  149\/}(9), 3947–3963.

\bibitem[\protect\citeauthoryear{Eltzner and Huckemann}{Eltzner and
  Huckemann}{2017}]{EH17}
Eltzner, B. and S.~Huckemann (2017).
\newblock Bootstrapping descriptors for non-{Euclidean} data.
\newblock In F.~Nielsen and F.~Barbaresco (Eds.), {\em Geometric Science of
  Information}, pp.\  12--19. Springer International Publishing.

\bibitem[\protect\citeauthoryear{Eltzner and Huckemann}{Eltzner and
  Huckemann}{2019}]{EH19}
Eltzner, B. and S.~F. Huckemann (2019).
\newblock A smeary central limit theorem for manifolds with application to high
  dimensional spheres.
\newblock {\em The Annals of Statistics\/}~{\em 47\/}(6), 3360--3381.

\bibitem[\protect\citeauthoryear{Eltzner, Hundrieser, and Huckemann}{Eltzner
  et~al.}{2021}]{EHH-GSI21}
Eltzner, B., S.~Hundrieser, and S.~F. Huckemann (2021).
\newblock Finite sample smeariness on spheres.
\newblock In F.~Nielsen and F.~Barbaresco (Eds.), {\em Geometric Science of
  Information}. Springer International Publishing.
\newblock accepted.

\bibitem[\protect\citeauthoryear{Fr\'echet}{Fr\'echet}{1948}]{F48}
Fr\'echet, M. (1948).
\newblock Les \'el\'ements al\'eatoires de nature quelconque dans un espace
  distanci\'e.
\newblock {\em Annales de l'Institut de Henri Poincar\'e\/}~{\em 10\/}(4),
  215--310.

\bibitem[\protect\citeauthoryear{Hotz and Huckemann}{Hotz and
  Huckemann}{2015}]{HH15}
Hotz, T. and S.~Huckemann (2015).
\newblock Intrinsic means on the circle: Uniqueness, locus and asymptotics.
\newblock {\em Annals of the Institute of Statistical Mathematics\/}~{\em
  67\/}(1), 177--193.

\bibitem[\protect\citeauthoryear{Hotz, Huckemann, Le, Marron, Mattingly,
  Miller, Nolen, Owen, Patrangenaru, and Skwerer}{Hotz et~al.}{2013}]{HHMMN13}
Hotz, T., S.~Huckemann, H.~Le, J.~S. Marron, J.~Mattingly, E.~Miller, J.~Nolen,
  M.~Owen, V.~Patrangenaru, and S.~Skwerer (2013).
\newblock Sticky central limit theorems on open books.
\newblock {\em Annals of Applied Probability\/}~{\em 23\/}(6), 2238--2258.

\bibitem[\protect\citeauthoryear{Hotz, Kelma, and Kent}{Hotz
  et~al.}{2016}]{HotzKelmaKent2016}
Hotz, T., F.~Kelma, and J.~T. Kent (2016).
\newblock Manifolds of projective shapes.
\newblock {\em arXiv e-prints 1602.04330\/}.

\bibitem[\protect\citeauthoryear{Huckemann, Mattingly, Miller, and
  Nolen}{Huckemann et~al.}{2015}]{H_Mattingly_Miller_Nolen2015}
Huckemann, S., J.~Mattingly, E.~Miller, and J.~Nolen (2015).
\newblock Sticky central limit theorems at isolated hyperbolic planar
  singularities.
\newblock {\em Electronic Journal of Probability\/}~{\em 20}, 1--34.

\bibitem[\protect\citeauthoryear{Huckemann and Eltzner}{Huckemann and
  Eltzner}{2020}]{HE_Handbook_2020}
Huckemann, S.~F. and B.~Eltzner (2020).
\newblock Statistical methods generalizing principal component analysis to
  non-{E}uclidean spaces.
\newblock In {\em Handbook of Variational Methods for Nonlinear Geometric
  Data}, pp.\  317--388. Springer.

\bibitem[\protect\citeauthoryear{Kato}{Kato}{2011}]{K11}
Kato, K. (2011).
\newblock A note on moment convergence of bootstrap {M}-estimators.
\newblock {\em Statistics \& Decisions\/}~{\em 28\/}(1), 51--61.

\bibitem[\protect\citeauthoryear{Kornhuber, Osprey, Coumou, Petri, Petoukhov,
  Rahmstorf, and Gray}{Kornhuber et~al.}{2019}]{kornhuber2019extreme}
Kornhuber, K., S.~Osprey, D.~Coumou, S.~Petri, V.~Petoukhov, S.~Rahmstorf, and
  L.~Gray (2019).
\newblock Extreme weather events in early summer 2018 connected by a recurrent
  hemispheric wave-7 pattern.
\newblock {\em Environmental Research Letters\/}~{\em 14\/}(5), 054002.

\bibitem[\protect\citeauthoryear{Le~Gall}{Le~Gall}{2016}]{G16}
Le~Gall, J. (2016).
\newblock {\em Brownian Motion, Martingales, and Stochastic Calculus}.
\newblock Graduate Texts in Mathematics. Springer International Publishing.

\bibitem[\protect\citeauthoryear{Lund, Agostinelli, et~al.}{Lund
  et~al.}{2017}]{L17R_Pack}
Lund, U., C.~Agostinelli, et~al. (2017).
\newblock Circular.
\newblock \url{https://cran.r-project.org/web/packages/circular/index.html}.
\newblock R package version 0.4-93.

\bibitem[\protect\citeauthoryear{Mardia and Patrangenaru}{Mardia and
  Patrangenaru}{2005}]{MP05}
Mardia, K. and V.~Patrangenaru (2005).
\newblock Directions and projective shapes.
\newblock {\em The Annals of Statistics\/}~{\em 33}, 1666--1699.

\bibitem[\protect\citeauthoryear{Mardia and Jupp}{Mardia and Jupp}{2000}]{MJ00}
Mardia, K.~V. and P.~E. Jupp (2000).
\newblock {\em Directional Statistics}.
\newblock New York: Wiley.

\bibitem[\protect\citeauthoryear{McInnes, Erwin, and Bathols}{McInnes
  et~al.}{2011}]{mcinnes2011global}
McInnes, K.~L., T.~A. Erwin, and J.~M. Bathols (2011).
\newblock Global climate model projected changes in 10 m wind speed and
  direction due to anthropogenic climate change.
\newblock {\em Atmospheric Science Letters\/}~{\em 12\/}(4), 325--333.

\bibitem[\protect\citeauthoryear{McKilliam, Quinn, and Clarkson}{McKilliam
  et~al.}{2012}]{MQC12}
McKilliam, R.~G., B.~G. Quinn, and I.~V.~L. Clarkson (2012).
\newblock Direction estimation by minimum squared arc length.
\newblock {\em IEEE Transactions on Signal Processing\/}~{\em 60\/}(5),
  2115--2124.

\bibitem[\protect\citeauthoryear{meteoblue AG}{meteoblue AG}{2020}]{meteoblue}
meteoblue AG (2020).
\newblock history+ platform.
\newblock
  \url{www.meteoblue.com/en/weather/archive/export/basel_switzerland_2661604},
  \url{www.meteoblue.com/en/weather/archive/export/goettingen_germany_2918632}.
\newblock Last checked on 06/04/2020.

\bibitem[\protect\citeauthoryear{Nishiyama}{Nishiyama}{2010}]{N10}
Nishiyama, Y. (2010).
\newblock Moment convergence of {M}-estimators.
\newblock {\em Statistica Neerlandica\/}~{\em 64\/}(4), 505--507.

\bibitem[\protect\citeauthoryear{Pennec}{Pennec}{2019}]{Pennec19}
Pennec, X. (2019).
\newblock Curvature effects on the empirical mean in {R}iemannian and affine
  manifolds: a non-asymptotic high concentration expansion in the small-sample
  regime.
\newblock {\em arXiv e-prints 1906.07418\/}.

\bibitem[\protect\citeauthoryear{Sch{\"o}tz}{Sch{\"o}tz}{2019}]{schotz2019arbitrary}
Sch{\"o}tz, C. (2019).
\newblock Arbitrary rates of convergence for projected and extrinsic means.
\newblock {\em arXiv e-prints arXiv:1910.11223\/}.

\bibitem[\protect\citeauthoryear{Sch\"otz}{Sch\"otz}{2019}]{Schoetz2019}
Sch\"otz, C. (2019).
\newblock Convergence rates for the generalized {Fr\'echet} mean via the
  quadruple inequality.
\newblock {\em Electron. J. Statist.\/}~{\em 13\/}(2), 4280--4345.

\bibitem[\protect\citeauthoryear{Tran, Eltzner, and Huckemann}{Tran
  et~al.}{2021}]{DHH-GSI21}
Tran, D., B.~Eltzner, and S.~F. Huckemann (2021).
\newblock Smeariness begets finite sample smeariness.
\newblock In F.~Nielsen and F.~Barbaresco (Eds.), {\em Geometric Science of
  Information}. Springer International Publishing.
\newblock accepted.

\bibitem[\protect\citeauthoryear{van~der Vaart}{van~der Vaart}{2000}]{vdV00}
van~der Vaart, A. (2000).
\newblock {\em {Asymptotic statistics}}.
\newblock Cambridge Univ. Press.

\bibitem[\protect\citeauthoryear{van~der Vaart and Wellner}{van~der Vaart and
  Wellner}{1996}]{vdVW96}
van~der Vaart, A. and J.~Wellner (1996).
\newblock {\em {Weak Convergence and Empirical Processes}}.
\newblock Springer.

\bibitem[\protect\citeauthoryear{Zhilova}{Zhilova}{2020}]{ZH20}
Zhilova, M. (2020).
\newblock {Nonclassical Berry–Esseen inequalities and accuracy of the
  bootstrap}.
\newblock {\em The Annals of Statistics\/}~{\em 48\/}(4), 1922 -- 1939.

\bibitem[\protect\citeauthoryear{Ziezold}{Ziezold}{1977}]{Z77}
Ziezold, H. (1977).
\newblock Expected figures and a strong law of large numbers for random
  elements in quasi-metric spaces.
\newblock {\em Transaction of the 7th Prague Conference on Information Theory,
  Statistical Decision Function and Random Processes\/}~{\em A}, 591--602.

\end{thebibliography}
    
\appendix
\section*{Appendix}
\section{Lemmata for the Proof of Theorem \ref{thm:CirclS-FSS}.}    \label{app:ProofModulationCurve}

\begin{Lem}\label{lem:FSS-foundation1}
 Consider $X_1,\ldots,X_n\iid X$ on $\SSS = [-\pi,\pi)/\sim$ with Euclidean sample and Fr\'echet sample mean $\overline X_n$ and $\widehat\mu_n$, respectively.
 Then 
 \begin{itemize}
  \item[(i)] there is $j \in  \{-n,\ldots,n\}$ such that $\widehat\mu_n = \overline X_n + \frac{2\pi\,j}{n}$, and 
  \item[(ii)] $|\widehat\mu_n|  \geq |\overline X_n|$.
 \end{itemize}
 Furthermore, assume that $X$ has a unique propulation mean $\mu=0$ and that $\widehat\mu_n$ is a measurable selection of Fr\'echet samples means. Then
  \begin{itemize}
  \item[(iii)] if $ \Prb\{|\widehat\mu_n| \neq |\overline X_n|\}>0$ then $\fm_n> 1$.
  \end{itemize}
\end{Lem}

\begin{proof} The first assertion can be found in \citet[proof of Corollary 4]{HH15}. For convenience we reproduce the short argument.
 For arbitrary fixed $\mu \in \SSS$ consider the index sets $I_0 = \{j: -\pi \leq X_j - \mu\leq \pi\}$, $I_1 = \{j: X_j - \mu > \pi\}$ and $I_2 = \{j: X_j - \mu < - \pi\}$. Then
 \begin{eqnarray}\nonumber 
    \sum_{j=1} d(X_j,\mu)^2 &=& \sum_{j\in I_0} (X_j-\mu)^2 + \sum_{j\in I_1} (X_j-2\pi -\mu)^2 + \sum_{j\in I_2} (X_j+2\pi -\mu)^2\\
    \label{eq:circular-Frechet-fcn}
    &=& \sum_{j=1}^n (X_j-\mu)^2 - 4\pi \sum_{j\in I_1} (X_j-\mu- \pi) +  4\pi \sum_{j\in I_2} (X_j-\mu+\pi) 
 \end{eqnarray}
 This is minimized by 
 \begin{eqnarray}\label{eq:circular-sample-mean} 
  \widehat\mu_n &=& \overline X_n + 2\pi\,\frac{|I_2| - |I_1|}{n}\,,
  \end{eqnarray}
 yielding the first assertion.
 
 For the second assertion, we consider  the above introduced $I_0,I_1,I_2$, now with $\widehat{\mu}_n$ instead of $\mu$. 
 
 Assume first that $\widehat\mu_n < 0 \leq \overline X_n$. Then $I_2=\emptyset$. Setting  $m=|I_1|$  we have thus by (\ref{eq:circular-sample-mean}) that $0 < \overline X_n - \widehat\mu_n = \frac{2\pi\,m}{n}$, whence by (\ref{eq:circular-Frechet-fcn}),
  \begin{eqnarray*} 
    \sum_{j=1} d(X_j,\widehat\mu_n)^2 &=&  \sum_{j=1}^n (X_j-\overline X_n + \overline X_n - \widehat\mu_n)^2  -  4\pi \sum_{j\in I_1} (X_j - \pi-\overline X_n + \overline X_n - \widehat\mu_n )\\
    &=& \sum_{j=1}^n (X_j-\overline X_n)^2 + n\left(\frac{2\pi\,m}{n}\right)^2 + 4\pi \sum_{j\in I_1} (\overline X_n + \pi - X_j) - 4\pi\,m \,\frac{2\pi\, m}{n}\\
    &=& \sum_{j=1}^n (X_j-\overline X_n)^2  + 4\pi \sum_{j\in I_1} (\overline X_n + \pi - X_j) - 4\pi\,m\, \frac{\overline X_n - \widehat\mu_n}{2}\\
    &=& \sum_{j=1}^n (X_j-\overline X_n)^2  + 4\pi \sum_{j\in I_1} \left(\frac{\overline X_n +\widehat\mu_n}{2} + \pi - X_j\right)\,.
\end{eqnarray*} 
Note that $\pi-X_j \geq 0$ always. Hence if $- \overline X_n < \widehat\mu_n$ then $\sum_{j=1} d(X_j,\widehat\mu_n)^2 > \sum_{j=1}^n (X_j-\overline X_n)^2 \geq \sum_{j=1} d(X_j,\overline X_n)^2 $, so that $\widehat\mu_n$ cannot be a Fr\'echet mean. In consequence we have shown that $\widehat\mu_n < 0 \leq \overline X_n$ implies $|\widehat\mu_n| \geq |\overline X_n|$.

Next assume that $0 \leq \widehat\mu_n  \leq \overline X_n$. Then $I_1=\emptyset$ and setting  $m=|I_2|\geq 0$ we infer from (\ref{eq:circular-sample-mean}) that $0\leq \overline X_n - \widehat\mu_n = -\frac{2\pi\,m}{n}$ and thus $m=0$ implying $\widehat\mu_n = \overline X_n$. 

With the above, this yields the second assertion.

Since $\fm_n = n \EE[d(\widehat\mu_n,0)^2]/\EE[d(X,0)^2]$, the last assertion follows at once from
  \begin{eqnarray*} 
    \EE[d(X,0)^2]&=& n\EE[|\overline X_n|^2] ~<~ n \EE[|\widehat\mu_n|^2] = n \EE[d(\widehat\mu_n,0)^2]\,,
    \end{eqnarray*}
    where the inequality is due to $1 = \Prb\{|\widehat\mu_n| = |\overline X_n]\} + \Prb\{|\widehat\mu_n| >|\overline X_n]\}$ by (ii), since the second probability is positive by hypothesis. 
\end{proof}

\begin{Lem}\label{lem:FSS-foundation2}
 Let $n\geq 2$, $a\in (0,\pi/2]$ and consider $X_1,\ldots,X_n\iid X$ on $\SSS$ with unique population Fr\'echet mean $\mu=0$.
 \begin{itemize}
  \item[(i)] If $\Prb\{X_1,\ldots,X_m \in [a,a+\eps) \mbox{ and } X_{m+1},\ldots, X_n \in (a-\pi-\eps, a-\pi]\} >0$ for some $0\leq \eps < \min\left\{a,\pi - a,\frac{\pi}{n-1}\right\}$ and $m\in \NN$ with $1\leq m \leq n-1$, then $\fm_n > 1$.
  \item[(ii)] If $\Prb\{a- \pi < X \leq a\} = 1$
or
  $\Prb\{a- \pi \leq X < a\} = 1$,
 then $\fm_n = 1$.
 \end{itemize}
\end{Lem}

\begin{proof}
For (i) we show that under the assumption, 
\begin{eqnarray}\label{proof1:lem:FSS-foundation2}\Prb\{|\overline X_n|   < |\widehat\mu_n|\} &>& 0\end{eqnarray}
whence $\Prb\{|\widehat\mu_n| \neq |\overline X_n|\} >0$ and in consequence of Lemma \ref{lem:FSS-foundation1} (iii), $\fm_n > 1$.

Let  $A = \{X_1,\ldots,X_m \in [a,a+\eps) \mbox{ and } X_{m+1},\ldots, X_n \in (a-\pi-\eps, a-\pi]\}$ with $\Prb(A) >0$. By hypothesis on $\eps$ we have $(a-\pi -\eps,a+\eps) \subset (-\pi,\pi)$ yielding that for all $\omega \in A$,
\begin{eqnarray*}\overline X_n &\in& \left(\frac{n-m}{n}(a- \pi - \eps) + \frac{m}{n}\,a,~\frac{n-m}{n}(a- \pi) + \frac{m}{n}(a +\eps)\right)\\
&=& \left(a - \frac{n-m}{n}(\pi+\eps) , a - \frac{n-m}{n}\,\pi + \frac{m}{n}\,\eps\right)\,,
\end{eqnarray*}
i.e.
\begin{align}\label{proof2:lem:FSS-foundation2}
|\overline X_n| &< \max\left(  a - \frac{n-m}{n}\,\pi + \frac{m}{n}\,\eps ,  \frac{n-m}{n}(\pi+\eps) -a \right) \,.
\end{align}
Note that $\eps < \frac{1}{n-1}\,\pi = \frac{n- (n-1)}{n-1}\,\pi \leq \frac{n-m}{m}\,\pi$ for all $0<m<n$ and $\eps < \frac{1}{n-1}\,\pi \leq \frac{m}{n-m}\,\pi$ for all $1\leq m < n$
ensure that 
\begin{align}\label{proof3:lem:FSS-foundation2}
a - \frac{n-m}{n}\,\pi + \frac{m}{n} < a \leq \frac{\pi}{2} \quad \text{ and } \quad \frac{n-m}{n}(\pi+\eps) -a <\pi-a
\end{align}
for all $1 \leq m \leq n-1$.
On the other hand, we have a measurable subset $B$ of $A$ with $\Prb(B)>0$ such that $\widehat\mu_n\in [-\pi, a-\pi) \cup  (a,\pi)$ for all $\omega \in B$ (we have $B=A$ unless $A=\{X_1,\ldots,X_n = a$ and $X_{m+1},\ldots X_n =\pi-a\}$ , then $0<\Prb(B) = \Prb(A)/2$ according to our agreement how to draw measurable selections from nonunique means).

In case of $\widehat\mu_n \in [-\pi,a-\pi)$ this means $|\widehat\mu_n| > \pi - a \geq  \frac{\pi}{2}$, which, for all $\omega \in B$, is greater than any of the bounds for $|\overline X_n|$ given by (\ref{proof2:lem:FSS-foundation2}) and (\ref{proof3:lem:FSS-foundation2}), yielding (\ref{proof1:lem:FSS-foundation2}). 

In case of $\widehat\mu_n \in (a,\pi)$, we have for all $\omega\in B$ that (the first inequality is due to the fact that unwrapping to the tangent line at $\widehat\mu_n$, its origin is the Euclidean mean)
\begin{eqnarray*}\widehat\mu_n &>& \frac{n-m}{n}(a+\pi - \eps) + \frac{m}{n}\,a ~=~a + \frac{n-m}{n}(\pi-\eps)\,,
\end{eqnarray*}
which is larger than the upper line of (\ref{proof2:lem:FSS-foundation2}) due to the upper line of (\ref{proof3:lem:FSS-foundation2}). It is also larger than the lower line of  (\ref{proof2:lem:FSS-foundation2}) since $\eps < a < \frac{n}{n-1}\,a \leq \frac{n}{n-m}\,a$ for all $1\leq m \leq n-1$  yields
$$ a + \frac{n-m}{n}(\pi-\eps) - \left(\frac{n-m}{n}(\pi+\eps) -a\right)  = 2 a - 2\,\frac{n-m}{n}\,\eps > 0\,.$$
In conclusion, $|\widehat\mu_n| > |\overline X_n|$ for all $\omega \in B$ which in conjunction with $\Prb(B)>0$ yields again (\ref{proof1:lem:FSS-foundation2}).
 
In case (ii) we infer $\widehat\mu_n = \overline X_n$ from (\ref{eq:circular-sample-mean}) since with the notation there, $I_1 = \emptyset = I_n$. Hence, taking into account $n \EE(d(0,\overline X_n)^2] = \EE[d(0,X)^2]$, $\fm_n = 1$.
\end{proof}

\section{Lemmata and Propositions for Proof of Theorem \ref{them:BS_CLT_Circle}}
\label{app:ProofBootstrapConsistency}

In this section, we prove that the terms from \eqref{eq:BS_Consistency_ErrorTerms1}
converge in outer probability to zero and thus complete the proof on consistency of the bootstrap for circular Fr\'echet means. 
  
Throughout this section,  under Assumption \ref{ass:antipodalDensityBounded}, we consider a sample $X_1, \dots, X_n\iid X$ with a measurable selection $\widehat{\mu}_n$ of its Fr\'echet sample means, as well as a bootstrap sample $X_1^*, \dots, X_n^*\iid \frac{1}{n} \sum_{i = 1}^{n} \delta_{X_i}$ with a measurable selection $\widehat{\mu}^*_n$ of its Fr\'echet sample means.
For convention we denote the underlying probability measure of $X$ by $\mathbb{P}$, the empirical measure of $X_1, \dots, X_n$ by $\mathbb{P}_n$ and the bootstrap empirical measure of $X_1^*, \dots, X_n^*$ by $\mathbb{P}_n^*$. 
Further, recall that we define for $t \in \RR$ the function $H(t) \coloneqq \min(2,|t|)$ which satisfies for any $h \in \BL$ and $t' \in \mathbb{R}$ the inequality $|h(t+t') - h(t)| \leq H(t')$. Further, let us recall the c\`adl\`ag functions $A, B_n, C_n \colon [-\pi, \pi)\rightarrow \RR$ which are defined for $x\in [-\pi,\pi)$ by \begin{align}
	A(x) &\coloneqq \begin{cases}
		 f(-\pi)x - \mathbb{P}(X \leq x-\pi)   & \text{ if } x \geq 0,\\
		 f(-\pi)x + \mathbb{P}(X > x+\pi)   & \text{ if } x <  0,
	\end{cases} \label{eq:DefFunctionA}\\
	B_n(x) &\coloneqq \begin{cases}
		 \mathbb{P}(X \leq x-\pi)  -\frac{1}{n} \lrCurlyBrack{\sum_{X_j\leq  x-\pi} 1} &  \text{ if } x \geq 0,\\
		- \mathbb{P}(X >x+\pi) +\frac{1}{n} \lrCurlyBrack{\sum_{X_j > x+\pi} 1} &  \text{ if } x < 0,
	\end{cases}\label{eq:DefFunctionBn}\\
	C_n(x) & \coloneqq \begin{cases}
		 \frac{1}{n}\lrCurlyBrack{\sum_{X_j \leq x-\pi} 1} - \frac{1}{n} \lrCurlyBrack{ \sum_{X_j^* \leq x-\pi}1} & \text{ if } x \geq 0,\\
		  -\frac{1}{n}\lrCurlyBrack{\sum_{X_j > x-\pi} 1} + \frac{1}{n} \lrCurlyBrack{ \sum_{X_j^* > x-\pi}1} & \text{ if } x < 0.
	\end{cases}\label{eq:DefFunctionCn}
\end{align}

\begin{Prop}\label{prop:ConvergenceToZero}
For $n \rightarrow \infty$, we have
\begin{equation}
 \begin{aligned}
 	&\;\EE\lrBrack{H\big(\sqrt{n}A(\widehat \mu_n^*)\big) +H\big(\sqrt{n}A(\widehat \mu_n)\big) + H\big(\sqrt{n} B_n(\widehat \mu_n^*)\big)+H\big(\sqrt{n} B_n(\widehat \mu_n)\big)  + H\big(\sqrt{n} C_n(\widehat \mu_n^*)\big) \Big| X_1, \dots, X_n}\\
 	+&\; \EE\lrBrack{H\big(\sqrt{n}A(\widehat \mu_n)\big)+H\big(\sqrt{n} B_n(\widehat \mu_n)\big) } \konvPStar 0\,.
 	\end{aligned}
 	\label{eq:BS_Consistency_ErrorTermsNEW}
\end{equation}
\end{Prop}
 
\begin{proof}
We prove in Proposition \ref{lem:Convergence_B_C_Functions} for $n \rightarrow \infty$ that the terms \eqref{eq:BS_Consistency_ErrorTermsNEW} involving $B_n$ and $C_n$ (conditionally) converge (in outer probability) to zero.
For the terms involving the function $A$ we show (conditional) convergence (in outer probability) to zero in Proposition \ref{prop:conditonalTightnessOfMuNSetting1}.
\end{proof}
 
\paragraph{Outline of proof.}
We begin in Lemma \ref{lem:ConditionalConvergenceInProbability} with a characterization of conditional convergence in outer probability which will be used throughout the rest of this section. Afterwards, in Lemma \ref{lem:BootstrapConsistencyFrechetMean} we verify that the bootstrap Fr\'echet sample mean conditionally converges to the Fr\'echet population mean as the sample size $n$ tends to infinity. In Proposition \ref{lem:Convergence_B_C_Functions} we show that the terms in \eqref{eq:BS_Consistency_ErrorTermsNEW} with $B_n$ and $C_n$ (conditionally) converge to zero. For this purpose, we show in Proposition \ref{lem:Convergence_B_C_Functions} that the (random) functions $\sqrt{n}B_n, \sqrt{n}C_n$ converge weakly in the space of c\` adl\` ag functions on $[-\pi, \pi)$ to a tight limit process with small fluctuation close to $0$. In conjunction with (conditional) convergence in probability of $\widehat \mu_n$ \citep{Z77} and $\widehat \mu_n^*$ (Lemma \ref{lem:BootstrapConsistencyFrechetMean}) to $\mu = 0$ as well as our characterization of conditional convergence (Lemma \ref{lem:ConditionalConvergenceInProbability}) the claim follows. 
To show (conditional) convergence of the terms in \eqref{eq:BS_Consistency_ErrorTermsNEW} involving $A$ we notice by Assumption \ref{ass:antipodalDensityBounded} that $A(x) = o(x)$ for $x \rightarrow 0$. To make use of this fact for our convergence analysis, we show a (conditional) tightness property for $\sqrt{n} \hat \mu_n$ and $\sqrt{n} \hat \mu_n^*$ given $X_1, \dots, X_n$.
For this purpose, we first show in Lemma \ref{lem:ConditionalTightnessBootstrapEuclideanMean} that the Euclidean bootstrap sample mean is conditionally tight. Afterwards, we the prove in Proposition \ref{prop:conditonalTightnessOfMuNSetting1} the required (conditional) tightness property for the Fr\'echet means and verify the convergence of the terms in \eqref{eq:BS_Consistency_ErrorTermsNEW} involving $A$.
 
\begin{Lem}\label{lem:ConditionalConvergenceInProbability}
 	For a random variable $Y_n$ on $\RR$ that is measurable w.r.t.  $X_1, \dots, X_n$ we have
 	$$\EE\lrBrack{H( Y_n )| X_1, \dots, X_n}\konvPStar 0\mbox{ for }n\to\infty $$
 	if and only if for all $\epsilon>0$,
 	$$\mathbb{P}\big(|Y_n| \geq \eps \big| X_1, \dots, X_n \big) \konvPStar 0\mbox{ for }n\to \infty\,.  $$
\end{Lem}
\begin{proof}
	Suppose for all $\eps>0$ holds $\mathbb{P}\big(|Y_n| \geq \eps \big| X_1, \dots, X_n \big) \konvPStar 0$. Fix $\eps>0$. Define $$K_n(\eps) \coloneqq \{\mathbb{P}\big(|Y_n| \geq \eps \big| X_1, \dots, X_n \big) \geq \eps\}$$ and set $N(\eps)\in \NN$ such that for all $n \geq N(\eps)$ it holds that $\mathbb{P}^*(K_n(\eps))\leq \eps$. Conditioned on $K_n(\eps/4)^c$ it follows for $n \geq N(\eps/4)$ that $\mathbb{P}(|Y_n| \geq \eps/4 | X_1, \dots, X_n ) \leq \eps/4$, thus by definition of $H(z)= \min(2,|z|)$ \begin{align*}
	\EE\lrBrack{H( Y_n )\Big| X_1, \dots, X_n} &= \EE\lrCurlyBrack{H( Y_n )1\lrCurlyBrack{|Y_n| < \frac{\eps}{4}} \Big| X_1, \dots, X_n}+ \EE\lrBrack{H( Y_n )1\lrCurlyBrack{|Y_n| \geq \frac{\eps}{4}} \Big| X_1, \dots, X_n}\\
	& \leq  \frac{\eps}{4} + 2 \mathbb{P}\lrCurlyBrack{|Y_n| \geq \frac{\eps}{4} \Big| X_1, \dots, X_n} \leq \frac{3}{4} \eps < \eps.
	\end{align*}
    Consequently, for $n \geq N(\eps/4)$ we assert that
    $$ \begin{aligned}	& \mathbb{P}^*\Big(	\EE\lrBrack{H( Y_n )\Big| X_1, \dots, X_n} \geq \eps \Big)  \\
    \leq \;& \mathbb{P}^*\Big(K_n(\eps/4) \Big) + \mathbb{P}^*\Big(K_n(\eps/4)^c	\cap \left\{\EE\lrBrack{H( Y_n )\Big| X_1, \dots, X_n} \geq \eps \right\}\Big) \leq \frac{\eps}{4} \leq \eps,	
    \end{aligned}$$
     which proves $\EE\lrBrack{H( Y_n )| X_1, \dots, X_n}\konvPStar 0$. 
     For the converse, assume there exists $\eps>0$ such that $\mathbb{P}^*\big(K_n(\eps)\big)> \eps$ for all $n \in \NN$. Then it follows conditioned on $K_n(\eps)$ for each $n \in \NN$ that 
     \begin{align*}
    	\EE\lrBrack{H( Y_n )\Big| X_1, \dots, X_n} &\geq  \EE\lrBrack{H( Y_n )1(|Y_n| \geq  \eps)\Big| X_1, \dots, X_n} \geq  \eps \,\mathbb{P}(|Y_n| \geq \eps | X_1, \dots, X_n ) > \eps^2.
    	\end{align*}
     Hence, we have for all $n \in \NN$ that  
     $$ \begin{aligned}	\mathbb{P}^*\Big(	\EE\lrBrack{H( Y_n )\Big| X_1, \dots, X_n} \geq \eps^2 \Big)  &\geq \mathbb{P}^*\Big(K_n(\eps) \Big) > \eps.
     \end{aligned}$$
     which shows that $\EE\lrBrack{H( Y_n )| X_1, \dots, X_n}$ does not converge in outer probability to zero.
 \end{proof}

For the next results we use concepts of empirical process theory. In particular, we employ the notion of general weak convergence on Banach spaces, bracketing numbers, and Donsker function classes. For an introduction to these concepts we refer to Van der Vaart \& Wellner (1996) and Van der Vaart (1998). 

First, we establish consistency of the bootstrap sample mean $\widehat \mu_n^*$ to the population mean $\mu$. 
  
\begin{Lem}\label{lem:BootstrapConsistencyFrechetMean}

There exists for each $\eps>0$ an $N \in \NN$ such that $\mathbb{P}^*\lrCurlyBrack{\mathbb{P}\lrCurlyBrack{|\widehat \mu_n^*| \geq \eps| X_1, \dots, X_n} \geq \eps} \leq \eps$ for all $n \geq N$.
\end{Lem}
\begin{proof}
    By uniqueness of the Fr\'echet population mean $\mu=0$ in conjunction with continuity of the Fr\'echet function $F$ it follows for any $\eps>0$ that there exists $\delta>0$ such that $F^{-1}( [F(\mu), F(\mu) + \delta)] ) \subseteq [-\eps, \eps]$. Hence, for any continuous function $G\colon \SSS \rightarrow \RR$ such that $\sup_{x \in \SSS}|F(x)-G(x)| \eqqcolon \|F - G\|_\infty< \delta/3$ we see for $x \notin[-\eps, \eps]$ that $$G(x) \geq F(x) - \frac{\delta}{3} \geq F(\mu) + \frac{2\delta}{3} \geq G(\mu) + \frac{\delta}{3},$$
    which yields that $\argmin_{x\in \SSS}G(x) \in [-\eps, \eps]$.
    
    Next, we show for sufficiently large $n$ that the bootstrap empirical Fr\'echet function $F_n^*$ is close to $F$ in supremum norm with high probability and thus the bootstrap empirical Fr\'echet mean $\widehat \mu_n^*$ is  close to $\mu =0$.
    For this purpose, consider the function class $\mathcal{F} \coloneqq \{ d^2(\cdot, x) \colon x \in \SSS \}$ and denote by $\ell^\infty(\mathcal{F})$ the Banach space of bounded linear functionals on $\mathcal{F}$ equipped with supremum norm. Note that the probability measure $\mathbb{P}$ associated to $X$, the empirical measure  $\mathbb{P}_n$ of $X_1, \dots, X_n$ and the bootstrap empirical measure $\mathbb{P}_n^*$ of $X_1^*, \dots, X_n^*$ are all contained in $\ell^\infty(\mathcal{F})$ since $\mathcal{F}$ is uniformly bounded. In particular, each probability measure is assigned to its corresponding Fr\'echet function. 
    
    Further, note that there exists $M>0$ such that for all $x,y,z\in [-\pi, \pi)$ it holds that $$|d^2(y,x) - d^2(z,x)| \leq M |y-z|.$$ Hence, by \cite[Theorem 2.7.11]{vdVW96} in conjunction with boundedness of $[-\pi, \pi)$ it follows that the bracketing number $$N_{[]}(\eps, \mathcal{F}, L^2(\mathbb{P}))\leq C \eps^{-1}$$ for all $0<\epsilon \leq 1$ and some constant $C>0$ that only depends on $M$. Thus, by \cite[Theorem 19.5]{vdV00} the function class $\mathcal{F}$ is $\mathbb{P}$-Donsker. This implies for the Fr\'echet population function $F$ and the empirical Fr\'echet function $F_n$ that $\sqrt{n}(F_n- F)$ converges weakly in the space of continuous functions on $\SSS$, denoted by $C(\SSS)$, for $n\rightarrow \infty$, to a tight limit. Additionally, since $\mathcal{F}$ is $\mathbb{P}$-Donsker, it follows by \cite[Theorem 3.6.1]{vdVW96} for the bootstrap empirical Fr\'echet function that $$ \sup_{h \in \text{BL}_1(C(\SSS)} \left\| \EE\lrBrack{h\Big(\sqrt{n}(F_n^* - F_n)\Big)\Big| X_1, \dots, X_n} - \EE\lrBrack{h\Big(\sqrt{n}(F_n - F)\Big)} \right\| \konvPStar 0, $$
    where $\text{BL}_1(C(\SSS))$ denotes the space of 1-Lipschitz functions mapping from $C(\SSS)$ to $\RR$ which are bounded by one.
    
    To prove the assertion consider $\eps>0$ and let $\delta>0$ such that $F^{-1}( [F(\mu), F(\mu) + \delta)] ) \in [-\eps, \eps]$. 	
    By weak convergence of $\sqrt{n}(F_n- F)$ there exists $M>0$ and  $N_1\in \NN$ such that for $n\geq N_1$ holds $\mathbb{P}(\sqrt{n}\|F_n- F\|_{\infty}\geq M) \leq \eps/3$. Define the Lipschitz function $h(f) = \min(\max(\| f\|_\infty - M ,0),1)$ for $f \in C([-\pi, \pi))$ which satisfies \begin{equation}\label{eq:BoundsLipschitzFunction}
    1(\| f\|_\infty  \geq M+1) \leq h(f)   \leq 1(\| f\|_\infty  \geq M).
    \end{equation}
    Then it holds that $\EE[h(\sqrt{n}(F_n- F))]  \leq \eps/3$ and there exists $N_2 \in \NN$ such that for $n \geq N_2$ follows
    $$\begin{aligned}
    \mathbb{P}^*\bigg(\underbrace{\left|\EE\lrBrack{h\Big(\sqrt{n}(F_n^* - F_n)\Big)\Big| X_1, \dots, X_n} - \EE\lrBrack{h\Big(\sqrt{n}(F_n- F)\Big)}\right| \geq \frac{\eps}{3}}_{\eqqcolon K_n^{(1)}}\bigg) \leq \frac{\eps}{3}.
    \end{aligned} $$
    Define $N_3\coloneqq \lceil 9(M+1)^2/\delta^2 \rceil$, then Inequality  \eqref{eq:BoundsLipschitzFunction} asserts  for $n\geq \max( N_1, N_2, N_3)$ on the complement $(K_n^{(1)})^c$ that
    $$ \begin{aligned}
    & \mathbb{P} \lrCurlyBrack{\|F_n^* - F_n\|_\infty \geq \frac{\delta}{3} \Big| X_1, \dots, X_n} = \mathbb{P} \lrCurlyBrack{\sqrt{n} \|F_n^* - F_n\|_\infty \geq \frac{\delta}{3}\sqrt{n} \Big| X_1, \dots, X_n} \\
    \leq \;&  \mathbb{P} \big(\sqrt{n}\|F_n^* - F_n\|_\infty \geq M+1 \Big| X_1, \dots, X_n\big)\leq\EE\lrBrack{h\Big(\sqrt{n}(F_n^* - F_n)\Big)\Big| X_1, \dots, X_n} \\
    \leq \;& \EE\lrBrack{h\Big(\sqrt{n}(F_n- F)\Big)} + \frac{\eps}{3} \leq \frac{2}{3} \eps
    \end{aligned} $$
    In particular, conditioned on $\{\sqrt{n}\|F_n- F\|_{\infty}<M\} \cap (K_n^{(1)})^c$ it holds for $n \geq \max(N_1, N_2, N_3)$ that
    $$ \begin{aligned}
    & \mathbb{P} \lrCurlyBrack{\|F_n^* - F\|_\infty \geq \frac{2 \delta}{3} \Big| X_1, \dots, X_n} \\
    \leq \;& \mathbb{P} \lrCurlyBrack{\|F_n^* - F_n\|_\infty \geq \frac{\delta}{3} \Big| X_1, \dots, X_n} + \mathbb{P} \lrCurlyBrack{\|F_n - F\|_\infty \geq \frac{\delta}{3} \Big| X_1, \dots, X_n}\leq \frac{2}{3}\eps  + 0.
    \end{aligned}  $$
    Recall by our first part of the proof that $\|F_n^* - F\|_\infty \leq \delta$ implies by continuity of $F_n^*$ that $\widehat\mu_n^* \in [-\eps, \eps]$. Thus, we see  conditioned on $\{\sqrt{n}\|F_n- F\|_{\infty}<M\} \cap (K_n^{(1)})^c$ for $n\geq \max(N_1, N_2, N_3)$ that
    $$ \mathbb{P} \lrCurlyBrack{|\widehat \mu_n^*| \geq \eps \Big| X_1, \dots, X_n} \leq \frac{2}{3}\eps < \eps.$$
    Concluding, the assertion follows since for all $n\geq \max(N_1, N_2, N_3)$ holds 
    $$\begin{aligned}
    & \mathbb{P}^*\lrCurlyBrack{ \mathbb{P}\Big(|\widehat \mu_n^*| \geq \eps \Big| X_1, \dots, X_n\Big)\geq \eps } \leq  \mathbb{P}\lrCurlyBrack{\sqrt{n}\|F_n- F\|_{\infty}<M} + \mathbb{P}^*(K_n^{(1)})\\
    &  + \mathbb{P}^*\lrCurlyBrack{\{\sqrt{n}\|F_n- F\|_{\infty}<M\} \cap (K_n^{(1)})^c \cap \left\{ \mathbb{P}\Big(|\widehat \mu_n^*| \geq \eps \Big| X_1, \dots, X_n\Big)\geq \eps \right\}}  \leq \frac{2}{3}\eps + 0 \leq \eps. \qedhere
    \end{aligned} $$
\end{proof}

This result in conjunction with Ziezold's strong law, stating that $\widehat \mu_n \xrightarrow{a.s.} 0 $ \citep{Z77}, is crucial to proving that the unconditional expectation of $H\big(\sqrt{n} B_n(\widehat \mu_n)\big)$  as well as the conditional expectations of $H\big(\sqrt{n} B_n(\widehat \mu_n^*)\big)$, $H\big(\sqrt{n} C_n(\widehat \mu_n^*)\big)$, and $H\big(\sqrt{n} B_n(\widehat \mu_n)\big)$ given $X_1, \dots, X_n$  converge in outer probability to zero.

\begin{Prop}\label{lem:Convergence_B_C_Functions}
	With the functions $B_n$ and $C_n$ defined in \eqref{eq:DefFunctionBn} and \eqref{eq:DefFunctionCn}, respectively we have for $n \rightarrow \infty$ that $$ \EE\lrBrack{H\big(\sqrt{n}B_n(\widehat \mu_n)\big)} +  \EE\lrBrack{H\big(\sqrt{n} B_n(\widehat \mu_n)\big) +  H\big(\sqrt{n} B_n(\widehat \mu_n^*)\big)+H\big(\sqrt{n} C_n(\widehat \mu_n^*)\big)\Big| X_1, \dots, X_n}  \konvPStar 0.$$
\end{Prop}

\begin{proof}
We first show that $\sqrt{n}B_n$ and $\sqrt{n}C_n$ are stochastic processes which converge for $n \rightarrow \infty$ to a tight Gaussian process. For this limit process we show that it exhibit a small fluctuation near zero. We then prove the claim by exploiting the consistency of the Fr\'echet sample mean and it bootstrap version. 

\textit{Step 1. Derivation of limit process.}  Define the function class $$\mathcal{G} \coloneqq \Big\{-1\Big(\cdot \in [-\pi, t-\pi]\Big), t\in [0, \pi)\Big\}\cup \Big\{1\Big(\cdot \in (t+\pi, \pi)\Big), t\in [-\pi, 0)\Big\} .$$ 
Notably, the Banach space $\ell^\infty(\mathcal{G})$ contains the probability measures $\mathbb{P}, \mathbb{P}_{n}$, and $\mathbb{P}_n^*$ for each $n \in \NN$. In particular, it holds for $t \in [0, \pi)$ that
$$\begin{aligned}
(\mathbb{P}_n - \mathbb{P})\big( -1(\cdot \in [-\pi, t-\pi ])\big) = B_n(t), \quad 
(\mathbb{P}_n^* - \mathbb{P}_n)\big( -1(\cdot \in [-\pi, t-\pi ])\big) = C_n(t)
\end{aligned} $$
and likewise for $t \in (-\pi, 0)$ we obtain
$$\begin{aligned}
(\mathbb{P}_n - \mathbb{P})\big( 1(\cdot \in (t+\pi, \pi))\big) = B_n(t), \quad 
(\mathbb{P}_n^* - \mathbb{P}_n)\big( 1(\cdot \in (t+\pi, \pi )\big) = C_n(t).
\end{aligned}$$
Hence, the empirical process $\sqrt{n}(\mathbb{P}_n - \mathbb{P})$ and the bootstrap empirical process $\sqrt{n}(\mathbb{P}_n^* - \mathbb{P}_n)$  can be identified in $\ell^\infty(\mathcal{G})$ with the (random) functions $\sqrt{n}B_n$ and $\sqrt{n}C_n$ on  $[-\pi, \pi)$. Notably, for each $n \in \NN$ the functions $B_n$ and $C_n$ are c\` adl\` ag on $[-\pi, \pi)$, i.e. right-continuous and limits from left exist. In the following we denote the space of such functions equipped with supremum norm by $D([-\pi, \pi))$.
Since $\mathbb{G}$ is the union of two Donsker classes (Van der Vaart \& Wellner, 1998, Example 2.5.4) it follows that $\mathcal{G}$ is also a Donsker class. Hence, $\sqrt{n} B_n$ converges weakly in $D([-\pi, \pi))$ to a tight centered Gaussian process $\mathbb{G}_B$. The covariance structure of $\mathbb{G}_B$ is characterized by
$$ \cov\lrBrack{\mathbb{G}_B(t) \mathbb{G}_B(s)} = \begin{cases}
 	\mathbb{P}(X \leq t - \pi) \wedge \mathbb{P}(X \leq s - \pi) - 	\mathbb{P}(X \leq t - \pi) 	\mathbb{P}(X \leq s - \pi) & \text{ if }  t, s \geq 0,\\
 	\mathbb{P}(X > t + \pi) \wedge \mathbb{P}(X > s + \pi) - \mathbb{P}(X > t + \pi) 		\mathbb{P}(X > s + \pi)& \text{ if }  t, s < 0,\\
 	\mathbb{P}(X \leq t -\pi) \mathbb{P}(X > s + \pi)  & \text{ if } t \geq 0 , s < 0.\\
 \end{cases}
 $$
 
 \textit{Step 2. Fluctuation of limit process near zero.} 
 In order to obtain concentration bounds for $\mathbb{G}_B$ we express it in terms of a standard brownian bridge. For this purpose, consider a standard Wiener process $\mathbb{W}$ on $[0,1]$ and define the Brownian bridge $\mathbb{B}$ by $\mathbb{B}(t) \coloneqq \mathbb{W}(t) - t \mathbb{W}(1)$. Defining the Gaussian process $\tilde{\mathbb{G}}_{B}$ on $[\pi, \pi)$ for $t \in [-\pi, \pi)$ by $$\tilde {\mathbb{G}}_{B}(t) \coloneqq \begin{cases}
 \mathbb{B}( \mathbb{P}(X \leq t - \pi) )   & \text{ if } t \in [0, \pi),\\
 \mathbb{B}( 1- \mathbb{P}(X > t+ \pi) )   & \text{ if } t \in [-\pi, 0).
 \end{cases}
 $$
we observe by a straight forward computation using $\EE[\mathbb{B}(t)] = 0$ and $\cov[\mathbb{B}(t)\mathbb{B}(s)] = \min(t, s) - t s$ for all $t,s \in [0,1]$ that $\tilde {\mathbb{G}}_{B}$ is centered and exhibits the same covariance structure as $\mathbb{G}_B$. Thus, $\tilde{\mathbb{G}}_B$ is a c\` adl\` ag modification of $\mathbb{G}_B$. Consequently, it follows for $0 <\delta < \pi$ and $L>0$ that 
\begin{align*}
	\mathbb{P}\lrCurlyBrack{ \sup_{t \in [0, \delta]} |\mathbb{G}_B(t)|\geq L } &= \mathbb{P}\lrCurlyBrack{ \sup_{t \in [0, \delta]} |\tilde{\mathbb{G}}_B(t)|\geq L } \\
	&\leq  \mathbb{P}\lrCurlyBrack{ \sup_{t \in [0, \delta]} \left|{\mathbb{W}}\Big( \mathbb{P}(X \leq t - \pi) \Big)\right|\geq \frac{L}{2} } + \mathbb{P}\lrCurlyBrack{ \sup_{t \in [0, \delta]} \left|\mathbb{P}(X \leq t - \pi){\mathbb{W}}(1)\right|\geq \frac{L}{2} }\\
 	& \stackrel{(*)}{\leq} \frac{2\EE\lrBrack{\left|{\mathbb{W}}\Big( \mathbb{P}(X \leq \delta - \pi) \Big)\right|}}{L} \,+ \, \frac{2 \mathbb{P}(X \leq \delta - \pi) \EE\lrBrack{\left|{\mathbb{W}}(1)\right|}}{L}\\
 	&  \stackrel{(**)}{=} \frac{2\sqrt{2} \Big(\sqrt{\mathbb{P}(X \leq \delta - \pi)} + \mathbb{P}(X \leq \delta - \pi)\Big) }{L\sqrt {\pi} }  \leq \frac{4 \sqrt{2}}{L \sqrt{\pi}} \sqrt{\mathbb{P}(X \leq \delta - \pi)}.
\end{align*}
Here, $(*)$ follows by Doob's inequality for sub-martingales \cite[Proposition 3.15]{G16} in conjunction with Markov's inequality and $(**)$ holds by the formula for expected values of half-normal distributed random variables.
Likewise, it follows by symmetry of the Brownian bridge $(\mathbb{B}(t))_{t \in [0,1]} \stackrel{\mathcal{D}}{=} (\mathbb{B}(1-t))_{t \in [0,1]}$ that 
$$\mathbb{P}\lrCurlyBrack{ \sup_{t \in [-\delta, 0]} |\mathbb{G}_B(t)|\geq L } \leq \frac{4 \sqrt{2}}{L \sqrt{\pi}} \sqrt{\mathbb{P}(X > -\delta + \pi)}.$$
As the unique Fr\'echet population mean is located at $\mu = 0$ the antipodal point exhibits no point mass. Hence, there exists for any $\eps>0$ a sufficiently small $\delta>0$ such that \begin{equation}\label{eq:ConditionGaussianProcess}
  \mathbb{P}\lrCurlyBrack{ \sup_{t \in [-\delta, \delta]} |\mathbb{G}_B(t)|\geq \frac{\eps}{2} } \leq \frac{\eps}{3}. 
\end{equation}

\textit{Step 3. Convergence of $\EE\lrBrack{H\big(\sqrt{n}B_n(\widehat \mu_n)\big)}$ to zero.}
W.l.o.g. choose $\eps\leq 1$. Further, define the function 
\begin{equation}\label{eq:Definition_h}
   h(f) \coloneqq 
\min\left( \max\left( \sup_{x \in
 [-\delta, \delta]}|f(x)| - \eps/2, 0\right), \eps\right),
\end{equation}
which is $1$-Lipschitz with respect to supremum norm on $[-\pi, \pi]$ and bounded by one since $\eps\leq 1$.
Notably, it holds for each $f \in D([-\pi, \pi))$ that
$$\eps 1\lrCurlyBrack{\sup_{x \in
[-\delta, \delta]}|f(x)| \geq \eps} \leq   h(f) \leq \eps  1\lrCurlyBrack{\sup_{x \in
[-\delta, \delta]}|f(x)| \geq \eps/2}$$
and thus \eqref{eq:ConditionGaussianProcess} asserts $\EE[h(\mathbb{G}_B)] \leq \eps^2/3.$
By weak convergence of $\sqrt{n} B_n$ to $\mathbb{G}_B$ in $D([-\pi, \pi))$ there exists some $N_1\in\NN$ such that for all $n \geq N_1$ holds $$\left| \EE\lrBrack{h(\sqrt{n}B_n)} - \EE\lrBrack{h(\mathbb{G}_B)}\right| \leq \frac{\eps^2}{3}.$$
Consequently, for $n \geq N_1$ we obtain that $$\mathbb{P}\lrCurlyBrack{ \sup_{t \in [-\delta, \delta]} |\sqrt{n}B_n(t)|\geq \eps } \leq  \frac{\EE\lrBrack{h(\sqrt{n}B_n)}}{\eps} \leq \frac{\EE\lrBrack{h(\mathbb{G}_B)}}{\eps} +  \frac{\left| \EE\lrBrack{h(\sqrt{n}B_n)} - \EE\lrBrack{h(\mathbb{G}_B)}\right|}{\eps}  \leq \frac{2}{3}\eps.$$
Furthermore, since $\widehat \mu_n \xrightarrow{\mathbb{P}} 0$ there exists some $N_2 \in \mathbb{N}$ such that for all $n \geq N_2$ holds $$ \mathbb{P}(|\widehat \mu_n | \geq \delta) \leq \frac{\eps}{3}.$$
Combining these assertions, we conclude for $n \geq \max(N_1, N_2)$ that  \begin{align*}
	\mathbb{P}(|\sqrt{n}B_n(\widehat \mu_n) | \geq \eps) &\leq \mathbb{P}(|\widehat \mu_n | \geq \delta) + \mathbb{P}\lrCurlyBrack{ \sup_{t \in [-\delta, \delta]} |\sqrt{n}B_n(t)|\geq \eps } \leq \frac{1}{3}\eps + \frac{2}{3}\eps = \eps,
\end{align*}
which yields for $n \rightarrow \infty$ that $\EE\lrBrack{H\big(\sqrt{n}B_n(\widehat \mu_n)\big)} \rightarrow 0$  and thus also in outer probability.

\textit{Step 4. Conditional convergence of $\EE[H\big(\sqrt{n} B_n(\widehat \mu_n)\big)\big| X_1, \dots, X_n]$ in outer prob. to zero.} We employ a similar strategy as in Step 3. Note that we cannot simply infer that $\EE[H\big(\sqrt{n} B_n(\widehat \mu_n)\big)\big| X_1, \dots, X_n] = H\big(\sqrt{n} B_n(\widehat \mu_n)\big)$ since $\widehat \mu_n$ might be non-unique. Let $\eps>0$, then it follows under $|\widehat \mu_n | < \delta$ and $\sup_{t \in [-\delta, \delta]} |\sqrt{n}B_n(t)|< \eps$ that $\sqrt{n} B_n(\widehat \mu_n)< \eps$ and thus $\mathbb{P}( \sqrt{n} B_n(\widehat \mu_n)\geq \eps \big|X_1, \dots, X_n)  =0.$
Hence, for $n \geq \max(N_1, N_2)$ we obtain that 
\begin{align*}
   & \mathbb{P}^*\lrCurlyBrack{\mathbb{P}\lrCurlyBrack{ \sqrt{n} B_n(\widehat \mu_n)\geq \eps \big|X_1, \dots, X_n} \geq \eps} \\
   \leq\;&\mathbb{P}\lrCurlyBrack{|\widehat \mu_n | \geq  \delta}+	\mathbb{P}\lrCurlyBrack{\sup_{t \in [-\delta, \delta]} |\sqrt{n}B_n(t)|\geq  \eps}\\
   +\;&\mathbb{P}^*\lrCurlyBrack{\{|\widehat \mu_n | < \delta\}\cap \{\sup_{t \in [-\delta, \delta]} |\sqrt{n}B_n(t)|< \eps\}\cap\left\{\mathbb{P}\lrCurlyBrack{ \sqrt{n} B_n(\widehat \mu_n)\geq \eps \big|X_1, \dots, X_n\}} \geq \eps\right\}}\\
   \leq \;&\frac{\eps}{3} + \frac{2}{3}\eps + 0 = \eps,
\end{align*}
which yields by Lemma \ref{lem:ConditionalConvergenceInProbability} that $\EE[H\big(\sqrt{n} B_n(\widehat \mu_n)\big)\big| X_1, \dots, X_n] \konvPStar 0 $ for $n \rightarrow 0$.

\textit{Step 5.  Conditional convergence of $\EE[H\big(\sqrt{n} B_n(\widehat \mu_n^*)\big)\big| X_1, \dots, X_n]$ in outer prob. to zero.} Given $\eps>0$ there exists by Lemma \ref{lem:BootstrapConsistencyFrechetMean} some $N_3 \in \mathbb{N}$ such that for all $n \geq N_3$ holds  
$$\mathbb{P}^*\bigg( \underbrace{\mathbb{P}\big(|\widehat \mu^*_n| \geq \delta \big| X_1, \dots, X_n \big) \geq \frac{\eps}{3}}_{\eqqcolon K_n^{(2)}} \bigg) \leq \frac{\eps}{3}.$$
Hence, for $n \geq N_3$ we see conditioned on $(K_n^{(2)})^c \cap \{  \sup_{t \in [-\delta, \delta]} |\sqrt{n}B_n(t)|< \eps\}$ that 
\begin{align*}
\mathbb{P}\lrCurlyBrack{ |\sqrt{n}B_n(\widehat \mu^*_n)| \geq \eps \Big| X_1, \dots, X_n } &\leq \mathbb{P}\lrCurlyBrack{ |\widehat \mu^*_n| \geq \delta \Big| X_1, \dots, X_n } + \mathbb{P}\lrCurlyBrack{  \sup_{t \in [-\delta, \delta]} |\sqrt{n}B_n(t)| \geq \eps \Big| X_1, \dots, X_n }\\
& \leq \frac{\eps}{3} + 0 < \eps.
\end{align*}
Consequently, it follows for $n \geq \max(N_1, N_3)$ that 
\begin{align*}
	& \mathbb{P}^*\lrCurlyBrack{\mathbb{P}\lrCurlyBrack{ |\sqrt{n} B_n(\widehat \mu^*_n)|\geq \eps \Big| X_1, \dots, X_n }\geq \eps} \leq \mathbb{P}^*(K_n^{(2)}) + \mathbb{P}\lrCurlyBrack{ \sup_{t \in [-\delta, \delta]} |\sqrt{n}B_n(t)|\geq \eps } 	\\
	& +\mathbb{P}^*\left((K_n^{(2)})^c \cap  \left\{ \sup_{t \in [-\delta, \delta]} |\sqrt{n}B_n(t)|< \eps \right\} \cap \left\{ \mathbb{P}\lrCurlyBrack{ |\sqrt{n} B_n(\widehat \mu^*_n)|\geq \eps \Big| X_1, \dots, X_n }\geq \eps \right\} \right) \\
	& \leq  \frac{1}{3}\eps + \frac{2}{3}\eps  +0 = \eps,
\end{align*}
hence by Lemma \ref{lem:ConditionalConvergenceInProbability} the assertion that $\EE[H\big(\sqrt{n} B_n(\widehat \mu^*_n)\big)\big| X_1, \dots, X_n]\konvPStar 0$ holds.

\textit{Step 6.  Conditional convergence of $\EE[H\big(\sqrt{n} C_n(\widehat \mu_n^*)\big)\big| X_1, \dots, X_n]$ in outer prob. to zero.} To prove this claim, we recall that the function class $\mathcal{G}$ is Donsker, thus by Van der Vaart \& Wellner (1998, Theorem 3.6.1) it follows for $n \rightarrow \infty$ that $$ \sup_{h \in \text{BL}_1(D([-\pi, \pi))} \left| \EE\lrBrack{h\Big(\sqrt{n}C_n\Big)\Big| X_1, \dots, X_n} - \EE\lrBrack{h(\mathbb{G}_B)} \right| \konvPStar 0. $$
Hence, for the previously defined function $h$ from \eqref{eq:Definition_h} there exists $N_4\in \NN$ such that for all $n \geq N_4$ holds 
$$ \mathbb{P}^*\Bigg(\underbrace{\left|\mathbb{E}\lrBrack{h(\sqrt{n}C_n)\Big| X_1, \dots, X_n }  - \mathbb{E}\lrBrack{ h(\mathbb{G}_B)}\right| \geq \frac{\eps^2}{6}}_{\eqqcolon K_n^{(3)}}\Bigg)\leq \frac{2}{3}\eps.$$
Consequently, for $n\geq N_4$ it follows conditioned on $(K_n^{(3)})^c$ by definition of $h$ and the choice of $\delta$, recall \eqref{eq:ConditionGaussianProcess}, that
$$
\begin{aligned}
\mathbb{P}\lrCurlyBrack{ \sup_{t \in [-\delta, \delta]} |\sqrt{n}C_n(t)|\geq \eps\Big| X_1, \dots, X_n }  &\leq \frac{\mathbb{E}\lrBrack{h(\sqrt{n}C_n)\Big| X_1, \dots, X_n }}{\eps} \leq  \frac{\mathbb{E}\lrBrack{ h(\mathbb{G}_B)}}{\eps} + \frac{\eps}{6}\\
& \leq  \mathbb{P}\lrCurlyBrack{ \sup_{t \in [-\delta, \delta]} |\mathbb{G}_B(t)|\geq \frac{\eps}{2}}  + \frac{\eps}{6} \leq \frac{1}{2}\eps. 
\end{aligned}
$$
Moreover, for $n \geq \max(N_2, N_4)$ it follows conditioned on $(K_n^{(2)})^c\cap(K_n^{(3)})^c$ that 
\begin{align*}
\mathbb{P}\lrCurlyBrack{ |\sqrt{n}C_n(\widehat \mu^*_n)| \geq \eps \Big| X_1, \dots, X_n } &\leq \mathbb{P}\lrCurlyBrack{ |\widehat \mu^*_n| \geq \delta \Big| X_1, \dots, X_n } + \mathbb{P}\lrCurlyBrack{  \sup_{t \in [-\delta, \delta]} |\sqrt{n}C_n(t)| \geq \eps \Big| X_1, \dots, X_n }\\
& \leq \frac{1}{3}\eps +  \frac{1}{2}\eps < \eps.
\end{align*}
Consequently, it follows that for $n \geq \max(N_3, N_4)$ that \begin{align*}
 &\mathbb{P}^*\lrCurlyBrack{\mathbb{P}\lrCurlyBrack{ |\sqrt{n}C_n(\widehat \mu^*_n)| \geq \eps \Big| X_1, \dots, X_n } \geq \eps}  \\
 &\leq \mathbb{P}^*(K_n^{(2)}) + \mathbb{P}^*(K_n^{(3)}) +  \mathbb{P}^*\lrCurlyBrack{ (K_n^{(2)})^c\cap(K_n^{(3)})^c \cap \left\{ \mathbb{P}\lrCurlyBrack{ |\sqrt{n}C_n(\widehat \mu^*_n)| \geq \eps \Big| X_1, \dots, X_n } \geq \eps \right\}} \\
 & \leq \frac{1}{3}\eps + \frac{2}{3}\eps + 0 = \eps,
 \end{align*}
which finishes the proof by Lemma \ref{lem:ConditionalConvergenceInProbability}.
\end{proof}

To prove that the terms in \eqref{eq:BS_Consistency_ErrorTermsNEW} involving the function $A$ tend to zero, we show that $\sqrt{n}\widehat \mu_n^*$ is conditionally tight given $X_1, \dots, X_n$. To this end, we required in our outline of proof that the Euclidean bootstrap sample mean $\sqrt{n}\overline X^*_n$ be conditionally tight. For convenience we give the following lemma, which we believe is part from mathematical folklore.

\begin{Lem}\label{lem:ConditionalTightnessBootstrapEuclideanMean}
For any $\eps>0$ there exist $M>0$ and $N\in \NN$ such that for all $n\geq N$,   $$
	\mathbb{P}^*\Big(\mathbb{P}\Big(|\sqrt{n}\,\overline X^*_n| \geq M \Big| X_1, \dots, X_n \Big) \geq \eps\Big) \leq \eps.$$
\end{Lem}

\begin{proof}	
	
Indeed, by tightness of $\sqrt{n}\,\overline{X}_n$  for any $\eps>0$ there exists some $M>0$ and $N_1 \in \NN$ such that $\mathbb{P}(|\sqrt{n}\,\overline{X}_n| \geq M) \leq \eps/4$ for all $n \geq N_1$. 
For this $M$ define the set $K_n^{(1)} \coloneqq \{|\sqrt{n}\,\overline{X}_n| \geq M\}$ and the function $h(x) \coloneqq \min( \max( |x| - M, 0),1)$ which is 1-Lipschitz and bounded for all $x \in \RR$ through
\begin{equation}
  1(|x|\geq M+1) \leq h(x) \leq 1(|x|\geq M),\label{eq:InequalityH}
\end{equation}
 thus $h \in \BL$. 
Hence, by \eqref{eq:InequalityH} we assert for $n \geq N_1$ that $\EE[h(\sqrt n \, \overline X_n)] \leq \mathbb{P}(|\sqrt n \, \overline X_n| \geq M) \leq \eps/4$. By consistency of the bootstrap for the Euclidean sample mean \cite[Theorem 23.4]{vdV00}  there exists some $N_2\in \NN$ such that for all $n \geq N_2$ holds
\begin{equation*}
	\mathbb{P}^*\Big( \underbrace{\left|
	\EE\lrBrack{h\lrCurlyBrack{\sqrt{n}(\overline X^*_n - \overline X_n)}\Big| X_1, \dots, X_n} -  \EE\lrBrack{h\lrCurlyBrack{\sqrt{n}\,\overline X_n}}
	\right|\geq \frac{\eps}{4} }_{\eqqcolon K_n^{(2)} } \Big) \leq \frac{\eps}{4}. 
\end{equation*} 
In particular, for $n \geq \max(N_1, N_2)$ we thus obtain conditioned on $(K_n^{(2)})^c$ that
\begin{align*}
	\mathbb{P}^*\lrCurlyBrack{ |\sqrt{n}(\overline X^*_n - \overline X_n)| \geq M+1 \Big| X_1, \dots, X_n}  &\leq  \EE\lrBrack{h\lrCurlyBrack{\sqrt{n}(\overline X^*_n - \overline X_n)}\Big| X_1, \dots, X_n} \\
	&\leq \EE\lrBrack{h\lrCurlyBrack{\sqrt{n}\,\overline X_n}} + \frac{\eps}{4} \leq \mathbb{P}(|\sqrt n \, \overline X_n| \geq M) + \frac{\eps}{4} \leq \frac{\eps}{2}.
	\end{align*} 
	Consequently, for  $n \geq \max(N_1, N_2)$ it follows conditioned on $(K_n^{(1)})^c\cap (K_n^{(2)})^c$ that
	\begin{align*}&\;\mathbb{P}\lrCurlyBrack{|\sqrt{n}\,\overline X^*_n| \geq 2(M+1) \Big|X_1, \dots, X_n} \\
			\leq &\;\mathbb{P}\lrCurlyBrack{|\sqrt{n}\,\overline X_n| \geq M+1\Big|X_1, \dots, X_n}  + \mathbb{P}\lrCurlyBrack{ |\sqrt{n}(\overline X^*_n - \overline X_n)| \geq M+1\Big|X_1, \dots, X_n}\\
	\leq &\;1\lrCurlyBrack{|\sqrt{n}\,\overline X_n| \geq M+1}  + \mathbb{P}\lrCurlyBrack{ |\sqrt{n}(\overline X^* _n- \overline X_n)| \geq M+1\Big|X_1, \dots, X_n}\\
	\leq &\; 0+ \frac{\eps}{2}= \frac{\eps}{2}< \eps,
	\end{align*}
	which yields for $n \geq \max(N_1, N_2)$ that 
	  \begin{align*}
	& \;\mathbb{P}^*\lrCurlyBrack{ \mathbb{P}\lrCurlyBrack{|\sqrt{n}\,\overline X^*_n| \geq 2(M+1) \Big|X_1, \dots, X_n} \geq \eps} \\
	\leq\; &  \mathbb{P}^*\lrCurlyBrack{ K_n^{(1)}} + \mathbb{P}^*\lrCurlyBrack{ K_n^{(2)}}+ \mathbb{P}^*\lrCurlyBrack{ (K_n^{(1)})^{c}\cap (K_n^{(2)})^{c}\cap \left\{\mathbb{P}\lrCurlyBrack{|\sqrt{n}\,\overline X^*_n| \geq 2(M+1) \Big|X_1, \dots, X_n} \geq \eps\right\}}\\
	\leq\; &  \frac{\eps}{4} + \frac{\eps}{4}+0 \leq  \eps
\end{align*}
and concludes the proof by Lemma \ref{lem:ConditionalConvergenceInProbability}.
\end{proof}

\begin{Prop}\label{prop:conditonalTightnessOfMuNSetting1}
	Assume the random element $X$ fulfills Assumption \ref{ass:antipodalDensityBounded}.
	With the function $A$ defined in \eqref{eq:DefFunctionA} we have  \begin{itemize}
		\item[(i)] $\EE\lrBrack{H\lrCurlyBrack{ \sqrt{n}A(\widehat\mu_n)}} \rightarrow 0$ and $\EE\lrBrack{H\lrCurlyBrack{ \sqrt{n}A(\widehat\mu_n)}\big|X_1, \dots, X_n} \konvPStar 0$ for $n \rightarrow \infty$.
		\item[(ii)] $\sqrt{n}\widehat \mu_n^*$ is conditionally tight given $X_1, \dots, X_n$, i.e. for any $\eps>0$ there exist $M>0$ and $N\in \NN$ such that $\mathbb{P}^*\lrCurlyBrack{ \mathbb{P}\lrCurlyBrack{|\sqrt{n}\widehat \mu_n^*| \geq M \big| X_1, \dots, X_n }\geq \eps}\leq \eps$ for all $n \geq N$, 
		\item[(iii)] and  $\EE\lrBrack{H\lrCurlyBrack{ \sqrt{n}A(\widehat\mu_n^*)}\big|X_1, \dots, X_n} \konvPStar 0$ for $n \rightarrow \infty$.
		
	\end{itemize}
\end{Prop}

\begin{proof}
{(i)} By assumption on the density $f$ near $-\pi$ it follows that $A(x) = o(x)$, i.e. for each $\gamma>0$ there exists a $\delta>0$ such that for all $|x| \leq \delta$ follows $|A(x)| \leq \gamma |x|$. 
Let $\eps>0$ and consider $M>0$ such that for all $n \in \NN$ holds  $\mathbb{P}(|\sqrt{n} \widehat \mu_n| \geq M) \leq \eps/4 $, by convergence in distribution of $\sqrt{n}\widehat \mu_n$ (Theorem \ref{them:CLT_Circle}) such an $M>0$ exists. Moreover, set $\gamma = \eps /(2M)>0$ and define $N_1 \coloneqq \lceil M^2/\delta^2(\gamma) \rceil\in \NN$, then it follows for all $n \geq N_1$ in case $|\sqrt{n} \widehat \mu_n| < M$ that $|\widehat \mu_n| < M/\sqrt{N_1}\leq  \delta(\gamma)$ and hence $$\sqrt{n}|A(\widehat \mu_n)| \leq \sqrt{n} \frac{ \eps }{2M} |\widehat \mu_n|  \leq \frac{\eps}{2}.$$
Consequently, for $n\geq N_1$ it holds by $H(x)= \min(2,|x|)$ that 
\begin{align*}
\EE\lrBrack{H\big(\sqrt{n}A(\widehat\mu_n)\big)}&= \EE\lrBrack{ H\big(\sqrt{n}A(\widehat\mu_n)\big) 1(|\sqrt{n} \widehat \mu_n| \geq M) }+\EE\lrBrack{ H\big(\sqrt{n}A(\widehat\mu_n)\big) 1(|\sqrt{n} \widehat \mu_n| < M) } \\
& \leq 2\frac{\eps}{4} + \EE\lrBrack{ H\Big(\frac{\eps}{2}\Big) 1(|\sqrt{n} \widehat \mu_n| < M) }\leq\frac{\eps}{2}+\frac{\eps}{2} = \eps,
\end{align*}
which implies $\EE\lrBrack{H\big(\sqrt{n}A(\widehat\mu_n)\big)} \rightarrow 0$ for $n\rightarrow \infty$. 
Moreover, conditioning on $\{\sqrt{n}|\widehat \mu_n| < M\}$ asserts for $n \geq N_1$ that $\mathbb{P}(\sqrt{n}|A(\widehat \mu_n)| \geq \eps \big| X_1, \dots, X_n) = 0$,  hence $$\begin{aligned} &\;\mathbb{P}^*\lrCurlyBrack{\mathbb{P}\lrCurlyBrack{\sqrt{n}|A(\widehat \mu_n)| \geq \eps \Big| X_1, \dots, X_n} \geq \eps } \\
\leq &\; \mathbb{P}^*\lrCurlyBrack{\sqrt{n}|\widehat \mu_n| \geq  M} + \mathbb{P}^*\lrCurlyBrack{\{\sqrt{n}|\widehat \mu_n| < M\} \cap \left\{ \mathbb{P}\lrCurlyBrack{\sqrt{n}|A(\widehat \mu_n)| \geq \eps \Big| X_1, \dots, X_n} \geq \eps \right\}} \leq \frac{\eps}{4} + 0 \leq \eps,\end{aligned}$$ 
which proves by Lemma \ref{lem:ConditionalConvergenceInProbability} Assertion (i).

{(ii)} Showing this assertion relies on Relation \eqref{eq:ConnectionBS_SampleMeanFrechetMean} which states
\begin{equation}\label{eq:CharacterizationOfBootstrapMean}\overline X^*_n= \big(1-2\pi f(-\pi)\big)\widehat \mu_n^*+ A(\widehat \mu_n^*) +  B_n(\widehat \mu_n^*) + C_n(\widehat \mu_n^*). \end{equation} We employ conditional tightness of the quantities $\overline X^*_n$, $B_n(\widehat \mu_n^*)$, $C_n(\widehat \mu_n^*)$  given $X_1, \dots, X_n$. In particular, there exists by Lemma \ref{lem:ConditionalTightnessBootstrapEuclideanMean} for $\eps>0$ some $M'>0$ and $N_2\in \NN$ such that for all $n \geq N_2$ holds
$$\mathbb{P}^*\Big(\underbrace{\mathbb{P}\Big(|\sqrt{n}\,\overline X^*_n| \geq M' \big| X_1, \dots, X_n \Big) \geq \frac{\eps}{5}}_{\eqqcolon K_n^{(1)}}\Big) \leq \frac{\eps}{5}.$$ Additionally, by Proposition \ref{lem:Convergence_B_C_Functions} in conjunction with Lemma \ref{lem:ConditionalConvergenceInProbability} there exists some $N_3\in \NN$ such that for all $n \geq N_3$ holds $$ \mathbb{P}^*\Big(\underbrace{\mathbb{P}\Big(|\sqrt{n}\,B_n(\widehat \mu_n^*)| \geq \frac{\eps}{5} \Big| X_1, \dots, X_n \Big) \geq\frac{\eps}{5}}_{\eqqcolon K_n^{(2)}}\Big) \leq \frac{\eps}{5}, \quad \mathbb{P}^*\Big(\underbrace{\mathbb{P}\Big(|\sqrt{n}\,C_n(\widehat \mu_n^*)| \geq \frac{\eps}{5} \Big| X_1, \dots, X_n \Big) \geq\frac{\eps}{5}}_{\eqqcolon K_n^{(3)}}\Big) \leq \frac{\eps}{5} $$
Furthermore, consider $\delta>0$ such that for each $x \in (-\delta, \delta)$ follows $|A(x)| \leq (1 -2 \pi f(-\pi)) x/2$ and select by Lemma \ref{lem:BootstrapConsistencyFrechetMean} some $N_4\in \NN$ large enough such that for $n \geq N_4$ holds  $$ \mathbb{P}^*\Big(\underbrace{\mathbb{P}\Big(|\widehat \mu_n^*| \geq \delta \Big| X_1, \dots, X_n \Big) \geq\frac{\eps}{5}}_{\eqqcolon K_n^{(4)}}\Big) \leq \frac{\eps}{5}.$$
Note, if $|\widehat \mu_n^*|\leq \delta$ it follows that the condition $(1 -2 \pi f(-\pi))|\sqrt{n} \widehat \mu_n^*| \geq L$, given some $L>0$, implies $$\sqrt{n}\big|(1 -2 \pi f(-\pi)) \widehat \mu_n^* + A(\widehat \mu_n^*)\big| \geq \frac{L}{2}.$$ 
Define $M \coloneqq 6\max(M', \eps/5)/(1-2\pi f(-\pi))$, then it follows conditioned on $\bigcap_{i =1}^{4}(K_n^{(i)})^c$ for all $n \geq \max(N_2, N_3, N_4)$ that
\begin{align*}
	&\mathbb{P}\lrCurlyBrack{|\sqrt{n}\widehat \mu_n^*|\geq M \Big| X_1, \dots, X_n } = \mathbb{P}\lrCurlyBrack{(1-2\pi f(-\pi))|\sqrt{n}\widehat \mu_n^*|\geq  6\max(M', \eps/5) \Big| X_1, \dots, X_n } \\
	\leq \;&\mathbb{P}\lrCurlyBrack{|\widehat \mu_n^*| < \delta, |\sqrt{n}(1-2\pi f(-\pi))\widehat \mu_n^* + \sqrt{n}A(\widehat\mu_n^*)|\geq 3\max(M', \eps/5)  \Big| X_1, \dots, X_n } \\
	& + \mathbb{P}\Big(|\widehat \mu_n^*| \geq \delta \Big| X_1, \dots, X_n \Big) \\
	\stackrel{\eqref{eq:CharacterizationOfBootstrapMean}}{\leq} \;&\mathbb{P}\lrCurlyBrack{ \sqrt{n}|\overline X^*_n - B_n(\widehat \mu_n^*) - C_n(\widehat \mu_n^*)|\geq  3\max(M', \eps/5) \Big| X_1, \dots, X_n } + \frac{\eps}{5} \\
	\leq \;& \mathbb{P}\lrCurlyBrack{ |\sqrt{n}\,\overline X^*_n|\geq  M' \Big| X_1, \dots, X_n } +\mathbb{P}\lrCurlyBrack{ |\sqrt{n}B_n(\widehat \mu_n^*)|\geq  \frac{\eps}{5} \Big| X_1, \dots, X_n }\\
	& +\mathbb{P}\lrCurlyBrack{ |\sqrt{n} C_n(\widehat \mu_n^*)|\geq   \frac{\eps}{5} \Big| X_1, \dots, X_n }+ \frac{\eps}{5} \leq \frac{4}{5}\eps< \eps.
\end{align*}
Consequently, it follows for all $n \geq \max(N_1, N_2, N_3)$  that 
\begin{align*}
	&\mathbb{P}^*\lrCurlyBrack{\mathbb{P}\lrCurlyBrack{|\sqrt{n}\widehat \mu_n^*|\geq M \Big| X_1, \dots, X_n }  \geq \eps} \leq \mathbb{P}^*\lrCurlyBrack{K_n^{(1)}} + \mathbb{P}^*\lrCurlyBrack{K_n^{(2)}} + \mathbb{P}^*\lrCurlyBrack{K_n^{(3)}} + \mathbb{P}^*\lrCurlyBrack{K_n^{(4)}}  \\
	&+\mathbb{P}^*\lrCurlyBrack{\bigcap_{i =1}^{4}(K_n^{(i)})^c \cap \left\{ \mathbb{P}\lrCurlyBrack{|\sqrt{n}\widehat \mu_n^*|\geq M \Big| X_1, \dots, X_n }  \geq \eps\right\}} \leq \frac{4}{5}\eps < \eps,
\end{align*}
which  yields by Lemma \ref{lem:ConditionalConvergenceInProbability} Assertion (ii).

{(iii)} Let $\eps>0$. By Assertion (ii) of this lemma there exists an $M>0$ and $N_5\in \NN$ such that for all $n \geq N_5$ holds $$ \mathbb{P}^*\bigg(\underbrace{\mathbb{P}\lrCurlyBrack{|\sqrt{n}\widehat \mu_n^*| \geq M\Big|X_1, \dots, X_n } \geq \frac{\eps}{2}}_{\eqqcolon K_n^{(5)}} \bigg)\leq \frac{\eps}{2}. $$
Let $\delta>0$ such that for all $|x| \leq \delta$ holds $|A(x)| \leq \eps |x| /(2M)$ and choose $N_6 \coloneqq \lceil M^2 /\delta^2\rceil $. Then it follows for all $n \geq N_6$ that the condition $\sqrt{n} |\widehat \mu_n^*| \leq M$ implies $|\widehat \mu_n^*|< M/\sqrt{N_6} \leq \delta$. Hence, under $\sqrt{n} |\widehat \mu_n^*| \leq M$ it holds for $n \geq N_6$ that $$\sqrt{n}| A(\widehat \mu_n^*)| \leq \sqrt{n} \frac{\eps}{2M}|\widehat \mu_n^*|  \leq \frac{\eps}{2}.$$
Consequently, conditioned on $(K_n^{(5)})^c$ it follows for $n \geq \max(N_5,N_6)$ that \begin{align*}
 &\mathbb{P}\lrCurlyBrack{\sqrt{n} |A(\widehat \mu_n^*)| \leq \eps\Big|X_1, \dots, X_n} \\
 \leq \;&	\mathbb{P}\lrCurlyBrack{\sqrt{n} |\widehat \mu_n^*| \geq M\Big|X_1, \dots, X_n}  + \mathbb{P}\lrCurlyBrack{\sqrt{n} |\widehat \mu_n^*| < M, \sqrt{n}| A(\widehat \mu_n^*)|\geq \eps \Big|X_1, \dots, X_n} \leq \frac{\eps}{2} + 0 < \eps. 
 \end{align*}
Concluding, we obtain for $n \geq \max(N_4,N_5)$ that \begin{align*}
	&\mathbb{P}^*\lrCurlyBrack{\mathbb{P}\lrCurlyBrack{|\sqrt{n}A(\widehat \mu_n^*) \geq \eps \Big| X_1, \dots, X_n }\geq \eps } \\
	\leq \;& \mathbb{P}^*\lrCurlyBrack{K_n^{(5)}}+\mathbb{P}^*\lrCurlyBrack{(K_n^{(5)})^c\cap \left\{\mathbb{P}\lrCurlyBrack{\sqrt{n}|A(\widehat \mu_n^*)| \geq \eps \Big| X_1, \dots, X_n }\geq \eps \right\}} \leq \frac{\eps}{2} + 0 \leq \eps,
\end{align*}
which yields Assertion (iii) by Lemma \ref{lem:ConditionalConvergenceInProbability} and finishes the proof.
\end{proof}

\section{Proofs for Moment Convergence of Fr\'echet Sample Means and Bootstrap Version}\label{app:MomentConvergence}

\paragraph{Proof of Proposition \ref{prop:UniformIntegrability}.}

Consider the function class $\mathcal{M}\coloneqq \{ m_\theta \coloneqq -d(\cdot, \theta)^2 \colon \theta \in \SSS\}$ and define $M \colon \SSS \rightarrow \RR,\theta \mapsto \EE[m_\theta(X)]$. By reverse triangle inequality it follows for all $x, \theta, \theta'\in \SSS$ that
\begin{equation}\begin{aligned}|m_\theta(x) - m_{\theta'}(x)| =&\; |-d(\theta,x)^2 +d(\theta',x)^2| = |d(\theta,x) +d(\theta',x)| |d(\theta,x) -d(\theta',x)|\\ \leq & \;2\pi d(\theta, \theta') \leq 2\pi | \theta - \theta'| \end{aligned}\label{eq:LipschitzPropertyMEstimator}
\end{equation}
which asserts that $M$ is continuous. Notably, $M$ coincides with the negative Fr\'echet function, hence by assumption $\mu= 0$  is the unique maximizer of $M$. By assumption  there exists $\delta >0$ such that $X$ has a continuous density $f$ on $[-\pi, -\pi+ \delta]\cup[\pi-\delta, \pi)$ which is bounded by $(1-\kappa)/(2\pi)$ for some $\kappa \in(0, 1]$. By Equation (3) in \cite{HH15} it holds for each $\theta \in [0, \delta]$ that
\begin{align*}
  M(\theta) - M(0) &= -\theta^2 + 4\pi \int_{-\pi}^{-\pi +\theta} (-\pi + \theta-x)f(x)dx 
  \leq  -\theta^2 + 4\pi \int_{-\pi}^{-\pi +\theta} (-\pi + \theta-x) \left(\frac{1- \kappa }{2\pi}\right)  dx \\
  &=-\theta^2 + \theta^2 \left(1-\kappa\right)   = -\kappa \theta ^2 = - \kappa d(0, \theta)^2 = - \kappa \theta^2.
\end{align*}
Likewise, it holds for each $\theta \in  [-\delta, 0]$  that
$$M(\theta) - M(0) \leq -\kappa d^2(0, \theta) = - \kappa \theta^2. $$
Since $\mu = 0$ is the unique maximizer of $M$ there exists by continuity of $M$ a sufficiently small $\eps>0$ such that $M^{-1}([M(0)- \eps, M(0)]) \subseteq [-\delta, \delta]$. Therefore, selecting $\kappa'>0$ sufficiently small such that $-\kappa' \theta^2 \geq - \eps$  for all $\theta \in [-\pi, -\delta]\cup [\delta, \pi)$ implies for all $\theta\in \SSS$ that
\begin{equation}
M(\theta) - M(0) \geq - \min(\kappa, \kappa') d(\theta, 0)^2 = - \min(\kappa, \kappa') \theta^2.\label{eq:LowerBoundMFunction}
\end{equation} 
Moreover, for a sample $X_1, \dots, X_n\iid X$ define  $M_n\colon \SSS \rightarrow \RR, \theta \mapsto\frac{1}{n} \sum_{i = 1}^{n}m_\theta(X_i)$. Then it follows by Lipschitz property \eqref{eq:LipschitzPropertyMEstimator} since the constant function $(2\pi)^{p}$ is integrable for any $p \geq 1$ with respect to the law of $X$according to \cite[Theorem 2.14.1 and Example 3.2.12]{vdVW96} that Assumption (2) in \cite{N10} is fulfilled for $\phi(\gamma) = \gamma$. The first assertion follows at once by \cite{N10} since the Fr\'echet sample mean $\hat\mu_n$ is a maximizer of $M_n$ and  therefore satisfies $M_n(\hat \mu_n) \geq M_n(0) \geq M_n(0)- 1/n$. The second assertion follows by $d(x,0) = |x|$ for each $x\in \SSS$,  Theorem \ref{them:CLT_Circle}(i) and \cite[Corollary to Theorem 25.12]{Billingsley1995}. \qed

\paragraph{Proof of Proposition \ref{prop:UniformIntegrabilityBootstrap}.}

Recall that the function class $\mathcal{M}=\{m_\theta\colon \theta \in \SSS\}$ from the proof of Proposition \ref{prop:UniformIntegrability} satisfies by \eqref{eq:LipschitzPropertyMEstimator}  that $|m_\theta(x) - m_{\theta'}(x)| \leq 2\pi d(\theta, \theta') \leq 2\pi |\theta-\theta'|$ for all $\theta, \theta'\in \SSS$, where $(2\pi)^{p}$ is integrable with respect to the law of $X$ for any $p \geq 1$. Hence, the first assertion follows at once from $d(x,y) \leq |x-y|$ for all $x,y \in \SSS$ and \cite[Theorem 2.2 and Remark 2.3]{K11}. 

For the second assertion we show for $p\geq 1$  that \begin{equation}\label{eq:OPConditionalExp}
    \EE\left[\left|\sqrt{n} (\widehat \mu_n^*- \widehat\mu_n)\right|^p \big| X_1, \dots, X_n \right]= O_P(1).
\end{equation}
Let $\epsilon>0$, then it follows for $M>0$ by Markov's inequality that \begin{align*}
    \Prb\left(\EE\left[\left|\sqrt{n} (\widehat \mu_n^*- \widehat\mu_n)\right|^p \big| X_1, \dots, X_n \right] \geq M \right) &\leq \frac{\EE\left[\EE\left[\left(\sqrt{n} (\widehat \mu_n^*- \widehat\mu_n)\right)^p \big| X_1, \dots, X_n \right]\right]}{M} \\
    &\leq \frac{\EE\left[\left|\left(\sqrt{n} (\widehat \mu_n^*- \widehat\mu_n)\right)^p\right|\right]}{M} \leq \frac{\sup_{n\in \NN}\EE\left[\left|\left(\sqrt{n} (\widehat \mu_n^*- \widehat\mu_n)\right)^p\right|\right]}{M},
\end{align*}
where the upper bound is finite by the first part. Hence, $M$ can be chosen large enough such that the upper bound is smaller than $\epsilon$, thus giving \eqref{eq:OPConditionalExp}.
The assertion then follows by \cite[Lemma 2.1]{K11} since $p\geq 1$ is arbitrary and 
$\sqrt{n} (\widehat\mu_n^* - \widehat \mu_n)$ converges weakly conditioned on $X_1, \dots X_n$ to the same limit as $\sqrt{n} (\widehat\mu_n - \mu_0)$ (cf. Theorem \ref{them:BS_CLT_Circle}), i.e. the distribution of $Z$. \qed

\end{document}